\documentclass[12pt,a4paper,twoside]{book}
\usepackage[italian,english]{babel}
\usepackage[utf8]{inputenc}
\usepackage[T1]{fontenc}
\usepackage{lmodern}
\usepackage{amsmath}
\usepackage{textcomp}
\usepackage{caption}
\usepackage{float}
\usepackage{graphicx}
\usepackage{array}
\usepackage[section]{placeins}
\usepackage{subcaption}
\usepackage{hyperref}
\usepackage{fancyhdr}
\usepackage{tabularx}
\usepackage{upgreek}
\usepackage{lineno}
\usepackage{multirow}
\usepackage{siunitx}
\usepackage{adjustbox}
\usepackage{rotating}

\begin{document}
	\begin{titlepage}
	\begin{center}
		{\LARGE{University of Ferrara}} \\[1ex]
	   	{\large{Physics and Earth Sciences Department}} \\[1ex]
	  	{\large{Master's Degree in Physics}} \\
		\vspace{1 cm}
		\includegraphics[height=6cm]{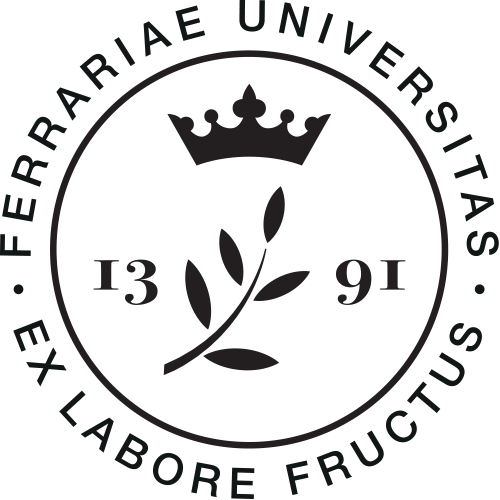}\\[1.5cm] 					
		{\LARGE\textbf{A Cylindrical GEM Inner Tracker}}\\
		{\LARGE\textbf{for the BESIII Experiment:}}\\
		\vspace{3mm}
		{\large\textbf{from Construction to Electronic Noise Studies}}\\[1.5cm]
	\end{center}
	\begin{flushleft}
		{\em Advisor}  \\
		{\em Dr}. {\sc Gianluigi Cibinetto}\\[1.5ex]
		{\em Co-advisor}  \\
		{\em Dr}. {\sc Ilaria Balossino} \\[3.5ex]
		\end{flushleft}
	\begin{flushright}
		{\em Candidate} \\
		{\sc Stefano Gramigna} 
	\end{flushright}
	\vfill
	\begin{center}
		{Academic Year 2019 - 2020}
	\end{center}
\end{titlepage}
 
	\pagestyle{empty}
	\pagenumbering{roman}	
  	\tableofcontents	
  	\listoffigures
  	\listoftables
  	\chapter*{Introduction}
\addcontentsline{toc}{chapter}{Introduction}
This thesis stems from the Italian project for the development of a CGEM detector to replace the innermost part of the drift chamber constituting the present inner tracker of the BESIII experiment at Beijing Electron Positron Collider II.

In the years of its operation, the innermost layers of the drift chamber have been showing signs of aging as a consequence of the prolonged exposure to the high luminosities achieved by the collider and the large beam-related background.
Its replacement with a newer kind of detector, with improved rate capabilities and better resistance to aging phenomena, has consequently become a priority for ensuring the continuous and effective performance of the experiment in the future, whose program has been extended by 10 years.

The upgrade, proposed by the Italian collaboration, is based on the GEM technology, a kind of MicroPattern Gaseous Detector outperforming in rate capability and aging robustness the previous generation of detectors that rely on wires for the amplification of the signal.
The detector will have to satisfy strict performance requirements while having to fit in the narrow space between the beam pipe and the BESIII detector.

The work described in this thesis began with a series of activities at LNF (Frascati National Laboratories) in summer 2019, where I participated in the construction of the innermost layer of the CGEM-IT. Thanks to an INFN research grant, I was able to continue working on the detector, I helped to build, at the Institute of High Energy Physics (IHEP), China’s biggest laboratory for the study of particle physics located in Beijing. During my three-months stay, I participated in the commissioning of the detector and to the installation of a setup for the study of electronic noise pick-up inside the BESIII experiment hall, near the interaction point. During my stay, and long after my return, I was responsible for the acquisition of data with the setup and for their analysis.

After an introduction describing the CGEM project and the context of its development, this thesis will follow the first part of the life cycle of a CGEM detector: its construction at LNF, its commissioning at IHEP and the noise studies needed to ensure its correct operation, under the conditions it will be subject to, with the real experimental conditions inside BESIII.

\bigskip

\noindent The thesis is organized as follows:

\begin{itemize}
	
	\item[] \textbf{Chapter 1} introduces BEPC-II, BESIII and the CGEM project. The working principles of a GEM detector are explained and an overview of the CGEM Inner Tracker and its dedicated electronics is provided.
	
	\item[] \textbf{Chapter 2} describes the construction of the innermost layer of the CGEM-IT: from the preparatory operations to the vertical assembly of the detector and its shipment.
	
	\item[] \textbf{Chapter 3} presents the initial phase of the commissioning of the detector after its arrival at IHEP. This began with the reception of the shipment and ended with its installation in a setup for the acquisition of cosmic ray data.
	
	\item[] \textbf{Chapter 4} describes the validation tests performed on the front end electronic boards before their installation on the detector. Details will be given about: the readout chain used for the tests and its main components; the technique adopted to assess the status of the electronics measuring the noise level, and the algorithm used.
	
	\item[] \textbf{Chapter 5} is focused on the noise test performed near the interaction point of BESIII. The aim of the test and the experimental technique are presented, together with a complete description of the setup. The data collected and the final results of the analysis are reported and discussed in detail.
	
\end{itemize}
    \pagenumbering{arabic}
    \pagestyle{fancy}
    \fancyhf{}
    \fancyhead[RE]{\small{\nouppercase{\rightmark}}}
    \fancyhead[LO]{\small{\nouppercase{\leftmark}}}
    \fancyhead[RO,LE]{\small{\thepage}}
    \setlength{\headheight}{14pt}
    \chapter{BESIII at BEPC-II and the CGEM-IT project}
\label{introchap}
The Beijing Spectrometer III (BESIII) \cite{bessite} is a general purpose detector located at the interaction point of the Beijing Electron Positron Collider II (BEPC-II). The experiment, in operation since 2009, is based in Beijing at the Institute of High Energy Physics (IHEP), China's largest laboratory for the study of high energy physics.
BEPC-II is a two rings $\mathrm{e^{+} e^{-}}$ collider running in the tau-charm energy region, between 2.0 and $\mathrm{4.9\, GeV}$. In 2016 it reached its design luminosity of $\mathrm{10^{33}\,cm^{-2}s^{-1}}$ at the center-of-mass energy of $\mathrm{3.78\,GeV}$, establishing a new luminosity record for accelerators in this energy range \cite{record}.

During its operation, the experiment collected about $\mathrm{30\, fb^{-1}}$ integrated luminosity at different energy points between 2.0 and $\mathrm{4.6\, GeV}$ including the world's largest samples of $\mathrm{J/\psi}$, $\mathrm{\psi(2s)}$ and $\mathrm{\psi(3770)}$. It provided a fundamental contribution to many fields of high energy physics: from hadron spectroscopy and the study of charmed hadron decays to the search for exotic states, like the first confirmed charged tetraquark, the $\mathrm{Z_c(3900)^\pm}$, in 2013 \cite{zcref}.

The experiment is arranged symmetrically around the collision point, with each of its five cylindrical subsystems, four detectors and a superconducting solenoid, layered around the beam pipe.
The innermost layers of the Multilayer Drift Chamber (MDC) constituting the present inner tracker of the detector have been suffering from a progressive aging process due to the high luminosity and beam related background. This led the BESIII collaboration to plan for its replacement with a new and improved tracking detector that could offer the increased radiation resistance required for prolonged operation in close proximity to BEPC-II interaction point. The detector proposed by the Italian collaboration is based on the Gas Electron Multiplier (GEM) technology, first devised by Dr. Fabio Sauli in 1997\cite{Sauli_1997}.

In this chapter, a brief description of the BEPC-II project and of the BESIII detector will be given together with an introduction to the CGEM upgrade.

\section{BEPC-II}
\label{bepici2}
\begin{figure}[h]
	\centering
	\includegraphics[width=.7\textwidth, keepaspectratio]{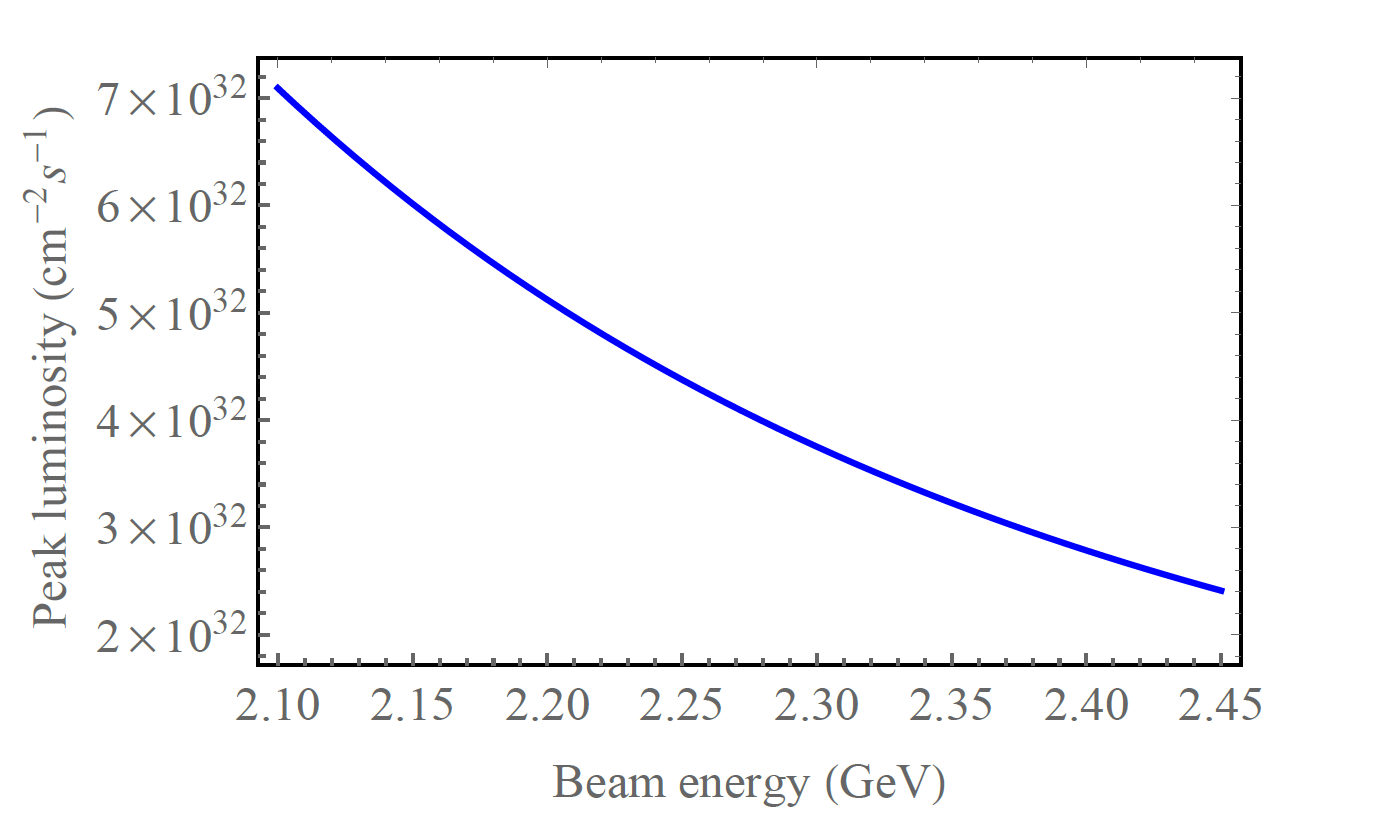}
	\caption[Estimate of BEPC-II peak luminosity in the energy region above $\mathrm{2.1\, GeV}$.]{Estimate of BEPC-II peak luminosity in the energy region above $\mathrm{2.1\, GeV}$ \cite{BESIII_2020}.}
	\label{lumigraph}
\end{figure}

The BEPC-II storage ring can operate as a double ring $\mathrm{e^{+}\,e^{-}}$ collider as well as a synchrotron radiation (SR) source. When it is operating as a collider, BEPC-II runs in the one-beam energy region between 1.0 and $\mathrm{2.45\, GeV}$ and its design luminosity is $\mathrm{10^{33}\,cm^{-2}s^{-1}}$ at $\mathrm{1.89\, GeV}$. Operating at energies above $\mathrm{1.89\, GeV}$ the peak luminosity is decreased, as estimated in the plot of figure \ref{lumigraph}, due to the limited power of the superconducting radio frequency cavities and difficulties in controlling bunch length and emittance. When operating as a synchrotron radiation source the accelerator is capable of achieving a current of $\mathrm{250\, mA}$ at the energy of $\mathrm{2.5\, GeV}$ \cite{BEPC-II_2009}.

The electron and positron rings of the accelerator are identical and cross each other at the northern and southern interaction points.
The northern interaction point hosts a bypass, that allows to connect the two outer half rings to form the synchrotron radiation ring, and the equipment used to measure the energy of the beams.
The southern interaction point hosts the BESIII detector together with all the equipment necessary for the joint operation of accelerator and detector. Here, a pair of special insertion magnets consisting of quadrupoles, dipoles and solenoids, is located. These serve to focus the beam, compensate the BESIII solenoid and, for operation in SR mode, as another bypass. Moreover, the southern interaction region hosts magnets, beam diagnostic instruments, vacuum pumps and cooling systems for both the beam magnets and the detector superconducting solenoid.

\begin{figure}[h]
	\centering
	\includegraphics[width=.7\textwidth, keepaspectratio]{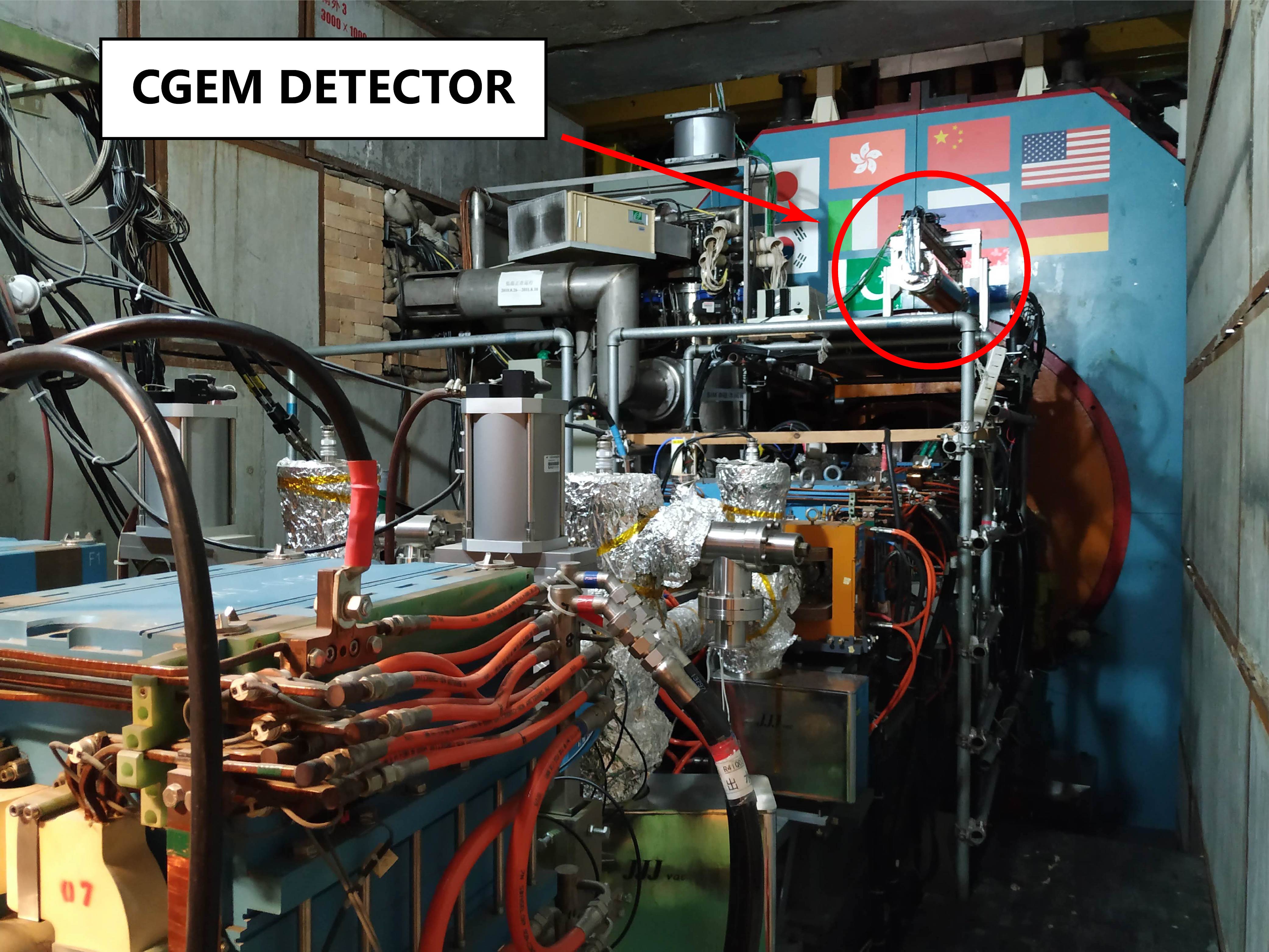}
	\caption[BESIII southern interaction point.]{Photograph of the instrumentation surrounding the southern interaction point, where the electronic pick-up noise studies described in chapter \ref{noisetest} of this thesis were conducted. The detector used for the tests is visible in the upper right corner, on top of a scaffolding.}
	\label{sipregion}
\end{figure}

A photograph of the instruments crowding the beam line at the southern interaction point is provided in figure \ref{sipregion}. This is the location that was chosen to conduct the electronic noise pick-up tests for the CGEM-IT detector that is part of this work and will be described in chapter \ref{noisetest}.

\FloatBarrier

\section{The BESIII Detector}
\begin{figure}[h]
	\centering
	\includegraphics[width=.9\textwidth, keepaspectratio]{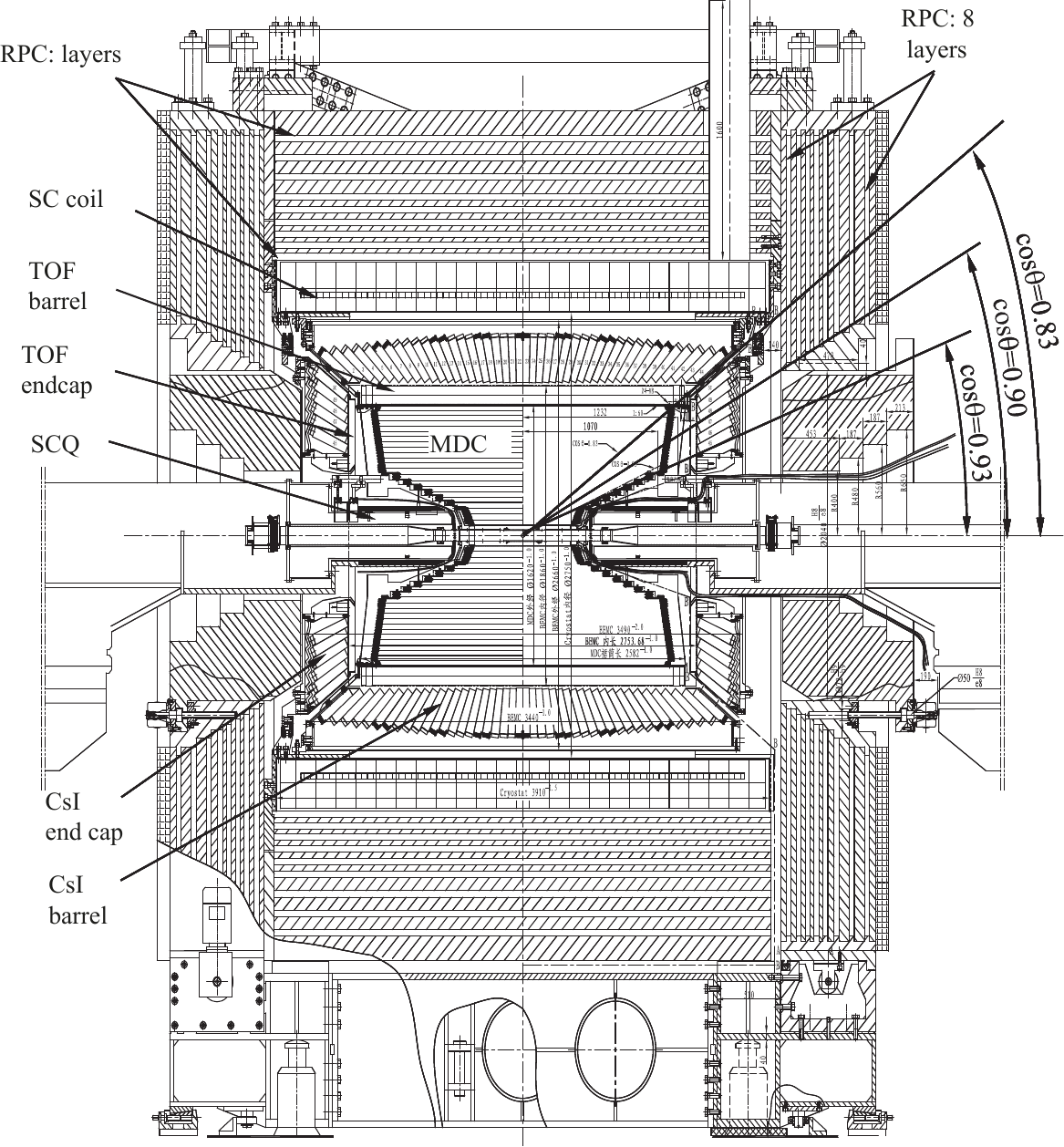}
	\caption[Schematic diagram of BESIII.]{Schematic diagram of the BESIII detector depicting its various subdetectors and the $\mathrm{1\, T}$ superconductive solenoid \cite{BESIII_2009}.}
	\label{besdiag}
\end{figure}

Figure \ref{besdiag} depicts the BESIII detector and its main subsystems. Particles produced in the interaction point encounter in the order: the multilayer drift chamber, the time-of-flight detector (TOF), the electromagnetic calorimeter (EMC), the superconducting solenoid and last, nested within the steel plates of the flux return yoke, the resistive plate chambers (RPCs) that constitute the muon identifier. The various subdetectors are briefly described in the following paragraphs while their design parameters are collected in table \ref{besdespar}.

\begin{table}[h]
	\centering
	\begin{tabular}{lrl} 
		\multicolumn{3}{l}{MDC}\\
		\hspace{4mm}Single wire $\mathrm{\sigma_{r\phi}}$ ($\mathrm{1\, GeV}$) & 130 & $\mathrm{\upmu m}$ \\
		\hspace{4mm}$\mathrm{\sigma_{z}}$ ($\mathrm{1\, GeV}$)& \textasciitilde2 & mm \\
		\hspace{4mm}$\mathrm{\sigma_{p}/p}$ ($\mathrm{1\, GeV}$)& 0.5 & \% \\ 
		\hspace{4mm}$\mathrm{\sigma_{dE/dx}}$ ($\mathrm{1\, GeV}$)& 6 & \% \\
		&&\\
		&&\\
		\multicolumn{3}{l}{EMC}\\
		\hspace{4mm}$\mathrm{\sigma_{E}/E}$ ($\mathrm{1\, GeV}$)& 2.5 & \%\\ 
		\hspace{4mm}Position resolution ($\mathrm{1\, GeV}$)& 0.6 & cm\\
		&&\\
		&&\\
		\multicolumn{3}{l}{TOF}\\
		\hspace{4mm}$\mathrm{\sigma_{T}}$ & & \\ 
		\hspace{8mm}Barrel ($\mathrm{1\, GeV/c}$ muons) & 100 & ps\\ 
		\hspace{8mm}End cap ($\mathrm{0.8\, GeV/c}$ pions) & 65 & ps \\ 
		&&\\
		&&\\
		\multicolumn{3}{l}{Muon Identifier}\\
		\hspace{4mm}No. of layers (barrel/end cap)& 9/8 &\\ 
		\hspace{4mm}Cut-off momentum & 0.4 & GeV/c \\ 
		&&\\
		&&\\ 
		Solenoid field & 1.0 & T\\
		$\mathrm{\Delta\Omega/4\pi}$ & 93 & \%\\
	\end{tabular}
	\caption[BESIII Design Parameters.]{BESIII Design Parameters \cite{BESIII_2020} \cite{BESIII_2009} .}
	\label{besdespar}
\end{table}

\FloatBarrier

\paragraph{Multilayer Drift Chamber}
The multilayer drift chamber provides track reconstruction together with momentum and dE/dx measurements for charged particles crossing its volume, allowing a first identification of the same. Data and simulations confirm that the dE/dx resolution achieved by the MDC allows $\mathrm{3 \sigma}$ $\mathrm{\pi/K}$ separation up to momenta of about $\mathrm{770\, MeV/c}$ \cite{mdcit}.

\begin{figure}[h]
	\centering
	\includegraphics[width=.6\textwidth, keepaspectratio]{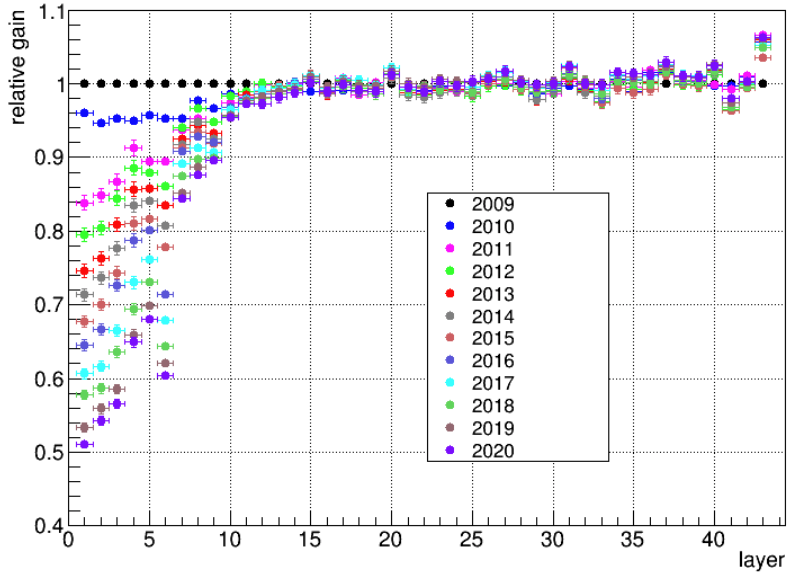}
	\caption[Relative gain loss of the MDC since its commissioning.]{MDC relative gain for each sensing layer in the years from its commissioning to 2020 \cite{note}.}
	\label{gainmdc}
\end{figure}

The MDC of the BESIII experiment comprises 43 layers of sensing wires and was built in two parts: the inner one houses the first 8 layers while the outer one accounts for the remaining 35. Once assembled together, the two parts share a common gas volume filled with a helium-hydrocarbon gas mixture in 60:40 proportions. The innermost of the two chambers is suffering from aging effects due to the presence of a beam-induced background with a hit rate up to $\mathrm{2\, kHz/cm^2}$ \cite{BESIII_2020}\cite{note}.
These problems arise from the deposition of insulating or conductive materials on the sensing wires and from the formation of a polymer deposit on the cathode. As the geometry of the electrodes and of the fields established between them is altered, gain and pulse height resolution are negatively affected, with a consequent decrease of the overall performance of the detector. Malter effect plays a role too, causing discharges unrelated to external irradiation inside the chamber \cite{Aging}. Due to these phenomena, since its commissioning, the first layers of the MDC have shown substantial gain losses, up to a maximum of about 46\% for the first layer \cite{BES30}. The progressive behavior of this performance decay is evident in the plot of figure \ref{gainmdc}. Despite the problems affecting the first layers, the MDC is still used to effectively collect quality data thanks to finely tuned HV adjustments and redundancy in the design.

\paragraph{Time-of-Flight System}
The barrel part of the TOF system is constituted by two layers of Saint-Gobain BC-408 scintillator bars directly connected to Hamamatsu R5924-70 photomultipliers. The end caps were upgraded in 2015, replacing the previous scintillator and photomultipliers configuration with a double layer of trapezoidal multi-gap resistive plate chambers (MRPCs). 

The TOF system allows to obtain the velocity of the particles produced in the interaction point by measuring the time elapsed between the collision and their passage through the detection layers. This information, when combined with the momentum measurement performed by the MDC, allows to identify charged particles, for which the system also provides a fast trigger signal.

\begin{figure}[h]
	\centering
	\includegraphics[width=.5\textwidth, keepaspectratio]{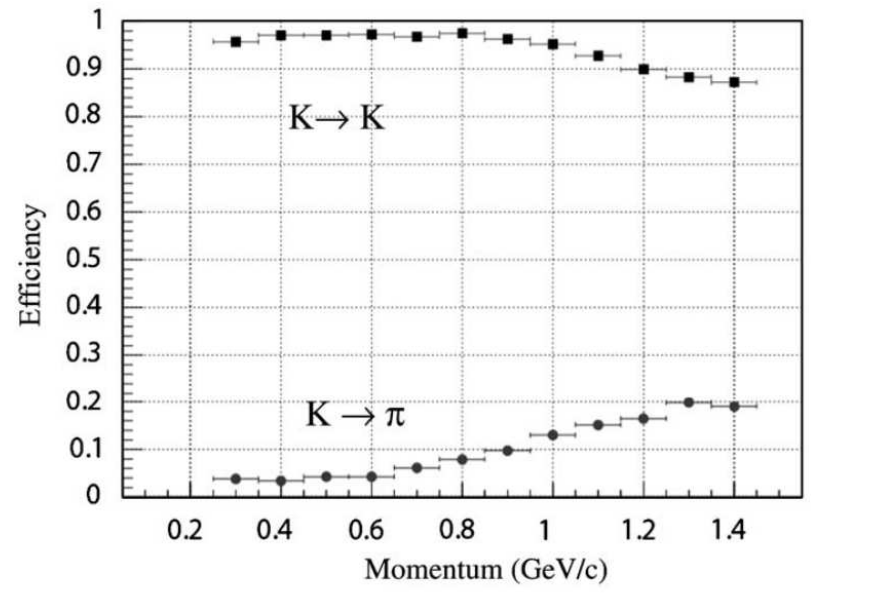}
	\caption[Efficiency of the TOF kaon identification and misidentification.]{Efficiency of the TOF kaon identification and the rate of their misidentification as pions, as a function of the particle momentum \cite{mdcit}.}
	\label{pikay}
\end{figure}

The TOF system, like the MDC, was designed to achieve good $\mathrm{\pi/K}$ separation; figure \ref{pikay} shows the efficiency of kaons identification and the rate of their misedintification as pions, attainable by the combining the data from the two systems, as a function of the particle momentum.

\paragraph{Electromagnetic Calorimeter}
The electromagnetic calorimeter consists of 6240 thallium activated cesium iodide crystals, each read by a single photodiode. It measures the energy of the particles and provides a trigger signal for the whole system.
For the scope of physics investigated by the BESIII experiment, it is essential to correctly identify radiative photons from the main charmonia, D mesons and neutral particles such as $\mathrm{\pi^0}$, $\mathrm{\eta}$, $\mathrm{\rho}$, etc.
The EMC must therefore be able to provide precise measurements for energies ranging from about $\mathrm{20\, MeV}$, for the most complicated decays, to the full one-beam energy for $\mathrm{e^+e^-\,\rightarrow\,\gamma\gamma}$.
In addition, the detector should be able to provide good e/$\mathrm{\pi}$ separation at momenta larger than $\mathrm{200\, MeV/c}$, as both these charged tracks produce similar showers inside the calorimeter \cite{giulio}.

\paragraph{Muon Detector}
The muon identifier is constituted by several layers of resistive plate chambers, inserted between the steel plates that provide the return flux of the axial magnetic field. This detector allows to separate muons from charged pions and other residual hadrons. Reliable muon identification expands the accessible physics channels, as at least one muon track is present in D meson decays, semileptonic decay of charmonia and the decay of $\mathrm{\tau}$.
The complete reconstruction of a muon track requires to associate hits from the muon detector with charged tracks in the MDC and energy measurements of the EMC. In addition, the merging of the data from the different detectors allows to lower the muon cut-off momentum \cite{giulio}.

\section{The CGEM-IT Project}
To replace the inner tracker of the BESIII experiment, the Italian collaboration has proposed to adopt the Gas Electron Multiplier (GEM) technology. GEMs are Micropattern Gaseous Detectors (MPGDs) and were first introduced in 1997 by Dr. Fabio Sauli \cite{Sauli_1997}. Since then, they have been adopted by many experiments, but their use to create a detector with cylindrical multiplication layers has been up to now limited to the realization of the inner tracker of the KLOE-2 experiment \cite{kloetdr}.

\subsection{GEM Operating Principles}
\begin{figure}[h]
	\centering
	\includegraphics[width=.4\textwidth, keepaspectratio]{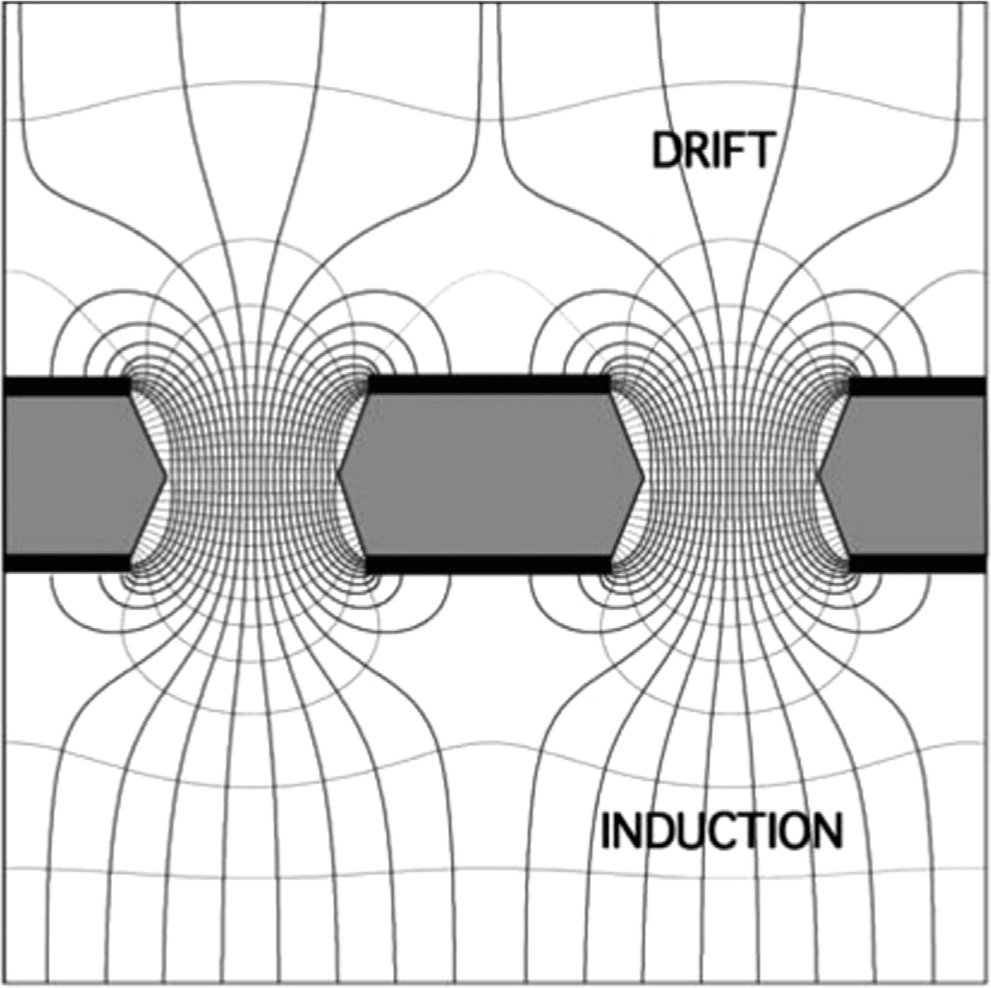}
	\caption[Field lines in the holes of a GEM foil.]{Representation of the field lines inside the holes of a GEM foil$\,$\cite{Sauli_2016}.}
	\label{gemfield}
\end{figure}

A Gas Electron Multiplier (GEM) is a thin insulating polymer foil, copper clad on each side and perforated by a large number of tiny holes by means of photolitographic techniques.
The thickness of the foil and the diameter of the holes are generally some tens of microns while the pitch is about 100 microns.
Applying a high voltage between the two faces of the foil, it is possible to generate an electric field inside its holes of the order of $\mathrm{10^4\, V/cm \,}$\cite{Sauli_1997}, which takes the geometrical configuration represented in figure \ref{gemfield}.

\begin{figure}[h]
	\centering
	\includegraphics[width=.6\textwidth, keepaspectratio]{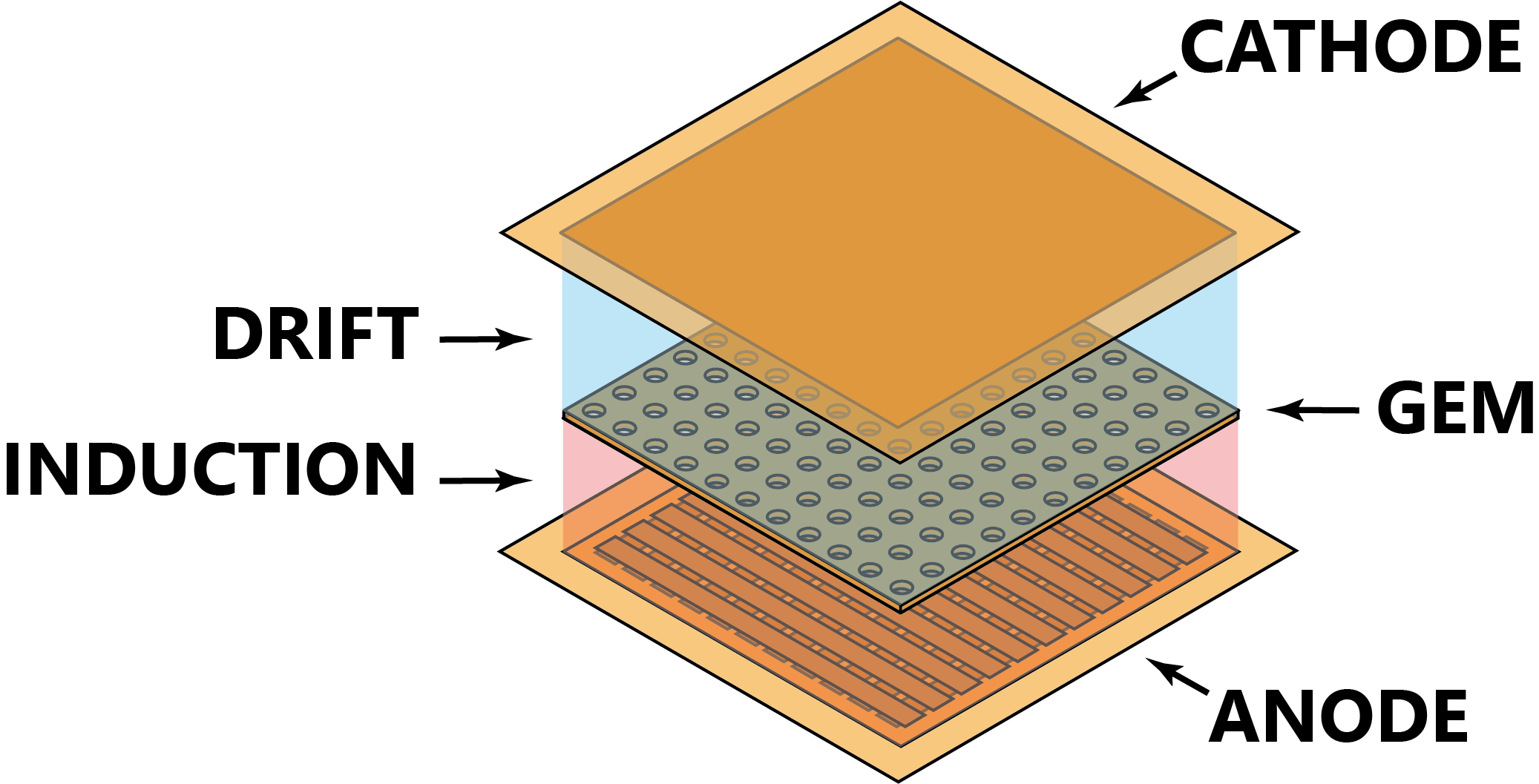}
	\caption[Schematic representation of a single GEM detector.]{Schematic representation of a single GEM detector.}
	\label{gemdet}
\end{figure}

A single GEM detector consists of: a cathode, an anode and a GEM foil, arranged in the configuration shown in figure \ref{gemdet} and separated by gaps of few millimeters.
The cathode is generally obtained by depositing a thin layer of copper on a polyimide foil. The anode, serving as the readout plane of the detector, is obtained etching strips or pads on an analogous copper clad foil using photolitographic techniques. The use of a polyimide substrate coated on both sides allows to realize a strip based anode with two-dimensional reading.
\pagebreak
Cathode, GEM and anode are enclosed in chamber filled with a mixture of two gases: a noble gas, favoring the ionization at the passage of a charged particle, and an organic gas, for quenching the avalanche originated by the multiplication.

By applying cascading potential differences between the electrodes in the system, it is possible to generate electric fields in the gaps that separate them.
The field generated between the cathode and the top face of the GEM foil is called drift field and serves to lead the electrons formed in the ionization toward the holes.
The field generated by the potential difference between the two GEM faces, confined inside the holes, is the strongest of the three and it is at the source of the multiplication process.
The field between the bottom face of the GEM and the anode is called induction field and serves to direct the electron avalanche resulting from the multiplication towards the anode.

A fast moving charged particle crossing the volume of the chamber ionizes the gas, which in turn releases electrons.
If these are produced in the region permeated by the induction field, they drift towards the anode and produce signals that are indistinguishable from the noise due to the lack of amplification.
Those generated in the region occupied by the drift field are instead led toward the holes, where they are rapidly accelerated and achieve enough energy to cause further ionization.
Electrons produced in these secondary ionizations get accelerated as well and the process assumes a cascading nature.
Part of the resulting electron avalanche is transferred to the anode by the induction field, where it is collected and produces an electric signal.
This is proportional to the overall charge of the avalanche and its characteristics are determined by capacitance and impedance of the chosen anode geometry.
A second part of the avalanche, which does not contribute to the formation of the signal, is instead guided toward the bottom face of the GEM by the conformation of the field near the holes and there it is reabsorbed. 

A single GEM detector can safely reach gains of the order of $\mathrm{10^3}$ \cite{Sauli_2016}. Stacking multiple GEM foils provides greater multiplication and allows for example a triple GEM detector to reach gains of the order of $\mathrm{10^4}$. Using multiple GEMs allows to operate them at much lower voltages, reducing the probability of a discharge \cite{Sauli_2016}.
In a multiple GEM detector the avalanche is transferred between the multiplication layers by transfer fields, generated by inducing a potential difference between the top and bottom faces of two adjacent GEM foils.

GEMs allow to overcome many of the drawbacks affecting the previous generation of tracking detectors, which rely on thin metal wires for signal collection.
Their spatial resolution, not limited by the macroscopic spacing separating the wires, depends from the microscopic pattern of the anode and can consequently reach values of tens of microns \cite{riccardo}.
The compact geometry of a GEM detector allows more efficient collection of the positive ions generated during the formation of the avalanche.
This, together with the increased spatial resolution, allows them to better resolve multiple tracks at a time and to remain effective at rates of the order of $\mathrm{10^6\, s^{-1}\,mm^{-2}}$ \cite{Sauli_2016}.
Another advantage deriving from the compact design of GEM detectors is the improvement in time resolution, which can be as low as $\mathrm{10\, ns}$ \cite{riccardo}. The signal produced is very fast because it is generated by the fast-moving electrons in the avalanche breaching the few millimeters separating the last multiplication layer from the anode.
Finally, the larger surfaces of its electrodes make GEM detectors less sensitive to aging effects with respect to their wire based counterparts, whose thin electrodes are more easily coated by the deposition of insulating materials. 

\subsection{The CGEM-IT Detector}
The Cylindrical GEM Inner Tracker (CGEM-IT) consists of three fully independent tracking layers, denominated L1, L2 and L3. Each of these is a complete cylindrical GEM detector, with three multiplication stages, capable of determining the position of a charged particle traversing its volume. The charge track is reconstructed extrapolating the particle trajectory from the position measurements performed by each of the three detectors. The adoption of GEM technology provides high counting rates and the resistance to aging effects needed for prolonged operation in the environment of BEPC-II collision point. Moreover, the increased resolution along the direction of the beam attainable by the CGEM detector allows a better reconstruction of the secondary vertexes. An argon-isobuthane gas mixture in proportions 90/10 was chosen according to the results of the preliminary analyses.

\begin{figure}[h]
	\centering
	\includegraphics[width=.7\textwidth, keepaspectratio]{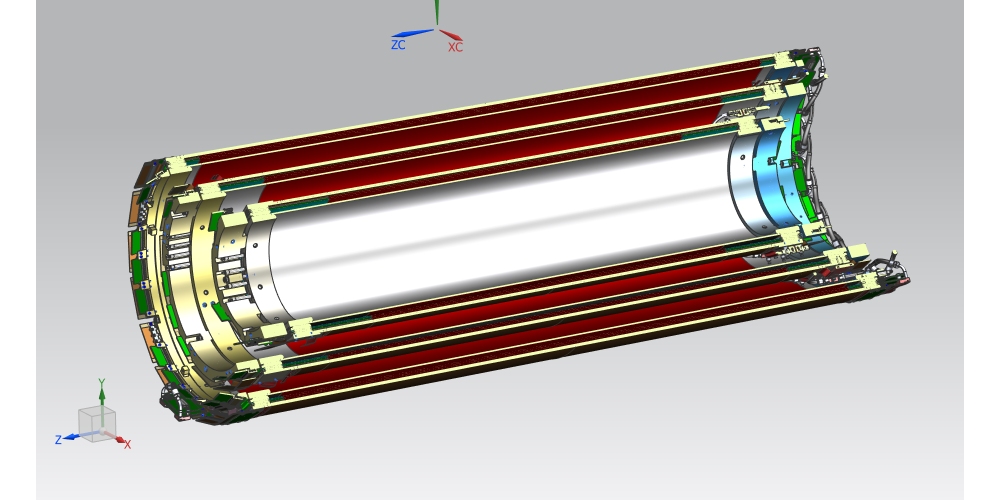}
	\caption[CGEM-IT Section]{Section of the three layers in their final configuration.}
	\label{cgemsection}
\end{figure}

A list of the requirements that the new detector has to satisfy can be found in table \ref{cgemreq} and a section of the three layers of the detector in their final configuration is represented in figure \ref{cgemsection}.

\begin{table}[h]
	\centering
	\begin{tabular}{lrl} 	
		$\mathrm{\sigma_{r\phi}}$ & $\leq$ 130 & $\mathrm{\upmu m}$\\
		$\mathrm{\sigma_{z}}$ & $\leq$ 1 & mm \\
		dp/p ($\mathrm{1\, GeV}$) & 0.5 & \% \\ 
		Material budget & $\leq$ 1.5 & \% $\mathrm{X_0}$\\ 
		Angular Coverage & 93 & \% $\times 4 \mathrm{\pi}$\\
		Rate capability & $10^4$ & Hz/cm$^2$ \\ 
		Minimum radius & 65.5 & mm\\ 
		Maximum radius & 180.7 & mm \\ 
	\end{tabular}
	\caption[CGEM-IT Requirements]{A list of the CGEM-IT upgrade requirements \cite{BESIII_2020}.}
	\label{cgemreq}
\end{table}

\FloatBarrier

Each layer consists of active elements, the ones including the electrodes that are powered to generate the fields, and passive structural elements that hold them together. The former include the cathode, the GEMs and the anode. The latter comprise the cylindrical structures strengthening anodes and cathodes and the Permaglass\footnote{Permaglass is a fiberglass reinforced epoxy resin with good mechanical characteristics and no outgassing issues.} support rings. The spacing between the cathode and the first GEM foil is $\mathrm{5\, mm}$ while all the other active components are separated by $\mathrm{2\, mm}$.

GEMs, anodes and cathodes are produced in planes at the CERN EST-DEM workshop\footnote{CERN EST-DEM is the Design and Manufacture of Electronic Modules (DEM) Group of the Engineering Support and Technology Division (EST).} and then given their final cylindrical shape when glued to the support rings during the construction of the detectors at the Frascati National Laboratories (LNF). The rings are used both as both spacers and supports for the gluing of the active elements of the detector. In addition, they host gas inlets and outlets and provide a stable base for the installation of the front end electronics.
The cylindrical structural components of the detector have undergone changes over the years of its development aimed to improve the overall robustness of the design \cite{mecreview}. The original Rohacell-Kapton
\footnote{Rohacell is a light polymethacrylimide based structural foam with good mechanical properties in a wide temperature range. Kapton is a polyimide film with exceptional temperature resistance capable of providing good insulation.}
 sandwich based design \cite{cdr}, used for the realization of L2, has been upgraded with a combination of Kapton, Honeycomb
\footnote{Honeycomb is a lightweight core material based on an hexagonal cell geometry. The one used in the construction of the detector is constitued by aramidic fibers held together by a resin.}
 and laminated carbon fiber meshes for both L1 and L3.

The new design has proven sturdier and safer to move with a limited increase to the overall radiation length of the detector.
The additional structural integrity is a fundamental requirement mandated by the international nature of the CGEM-IT project. The detector must in fact survive transcontinental shipping from LNF to the location of its final installation inside BESIII.

All the active elements of the detector are manufactured, through photolitographic techniques, on the same $\mathrm{50\, \upmu m}$ thick Kapton substrate, copper clad on one or both sides depending on the component. Each active element of the innermost detection layer is realized on a single sheet while, for the second and third layers, two sheets are joined together to form the final component. The dead zones represented by the junctions are aligned, so to minimize the loss of active area and angular coverage.

The anodic readout plane is built by etching strips on both the copper clad faces of a same Kapton foil and then applying this one on a Kapton layer $\mathrm{25 \, \upmu m}$ thick using an epoxy adhesive. The strips are arranged in two directions: the $\mathrm{570 \, \upmu m}$ wide X strips are aligned with the beam axis; the direction of the thinner $\mathrm{130 \, \upmu m}$ V strips is determined by a stereo angle which depends on the diagonal of the foils.  The pitch for both families of strips is $\mathrm{650 \, \upmu m}$. X strips provide the azimuthal coordinate while, combining the information from both X and V strips, it is possible to obtain the coordinate in the direction parallel to the beam.

The GEM foils are divided in separate HV sectors. This design has the benefit of limiting the capacitance of the structure and consequently the energy released in an eventual discharge. In addition, should any damage disable one of the sectors, the others can still operate independently, allowing to retain part of the original functionality. The bottom face of the GEM foil is divided into macrosectors. To each of these correspond ten microsectors on the top face. The GEMs of the three layers are divided into 4, 8 and 12 macrosectors respectively and so they house 40, 80 and 120 microsectors on their top surface.

The cathode is the simplest of the active elements. The initial copper coating is reduced to $\mathrm{3 \, \upmu m}$ in order to lower the radiation length of the detector. Similar foils constitute also the ground plane of the detector and the Faraday cage.

Details relative to the construction of a CGEM detector are provided in chapter \ref{construction} of this thesis together with a full stratigraphy of each of the three layers.

The front end electronics for the CGEM-IT detector are designed and realized by the INFN section of Turin while the back end electronics are developed by the INFN section of Ferrara. The former, installed on the detector and directly connected to its anode readout tails, consist of a series of Front End Boards (FEBs) each housing two TIGER chips, acronym of Torino Integrated GEM Electronics for Read-out \cite{tigerref} \cite{fabio}. The latter are instead based around the GEM Readout Card (GEMROC) Modules \cite{cpad}, constituted by a Field Programmable Gate Array (FPGA) card and a custom made interface board.

L1, L2 and L3 are equipped with 16, 28 and 36 FEBs respectively. Each TIGER can perform analog charge and time measurements on its 64 channels and digitize the results before transmitting them to the back end electronics.
For the first layer all the TIGER channels are connected to strips while, due to the different  anode geometries, TIGERs used for L2 and L3 read respectively 62 and 61 strips.

Each GEMROC handles the configuration of four FEBs and the organization of the incoming data stream. In addition, it monitors their operating parameters and handles the low voltage power distribution.

A more detailed description of the readout chain and its components is provided in chapter \ref{febtests}. 

The adopted design allows to operate the CGEM detectors in two modes: charge centroid and micro-Time Projection Chamber ($\upmu$TPC). Both of these are being used daily to collect cosmic ray data with two of the three layers, which are currently installed in a cosmic ray telescope setup in Beijing. These data allow to expand the knowledge of the detector and develop its dedicated physics and control software\cite{Farinelli_2020}.

    \chapter{Construction of a Cylindrical GEM Detector}
\label{construction}
The content of this chapter is mostly an account of my personal experience, acquired at the Frascati National Laboratories between June and August of 2019.
There, I participated to the realization of the innermost of the three layers of the CGEM-IT detector, working alongside the researchers and the technicians responsible for its construction.
I joined all but the very last phases of the construction as when they took place I was in China, thanks to an INFN research grant, for doing the works described in the last three chapters of this thesis.
The final assembly and sealing of the detector, as well as the preparation of its expedition, are still presented, to offer a complete review of the process.

\begin{figure}[h]
 	\centering
 	\includegraphics[width=.8\textwidth, keepaspectratio]{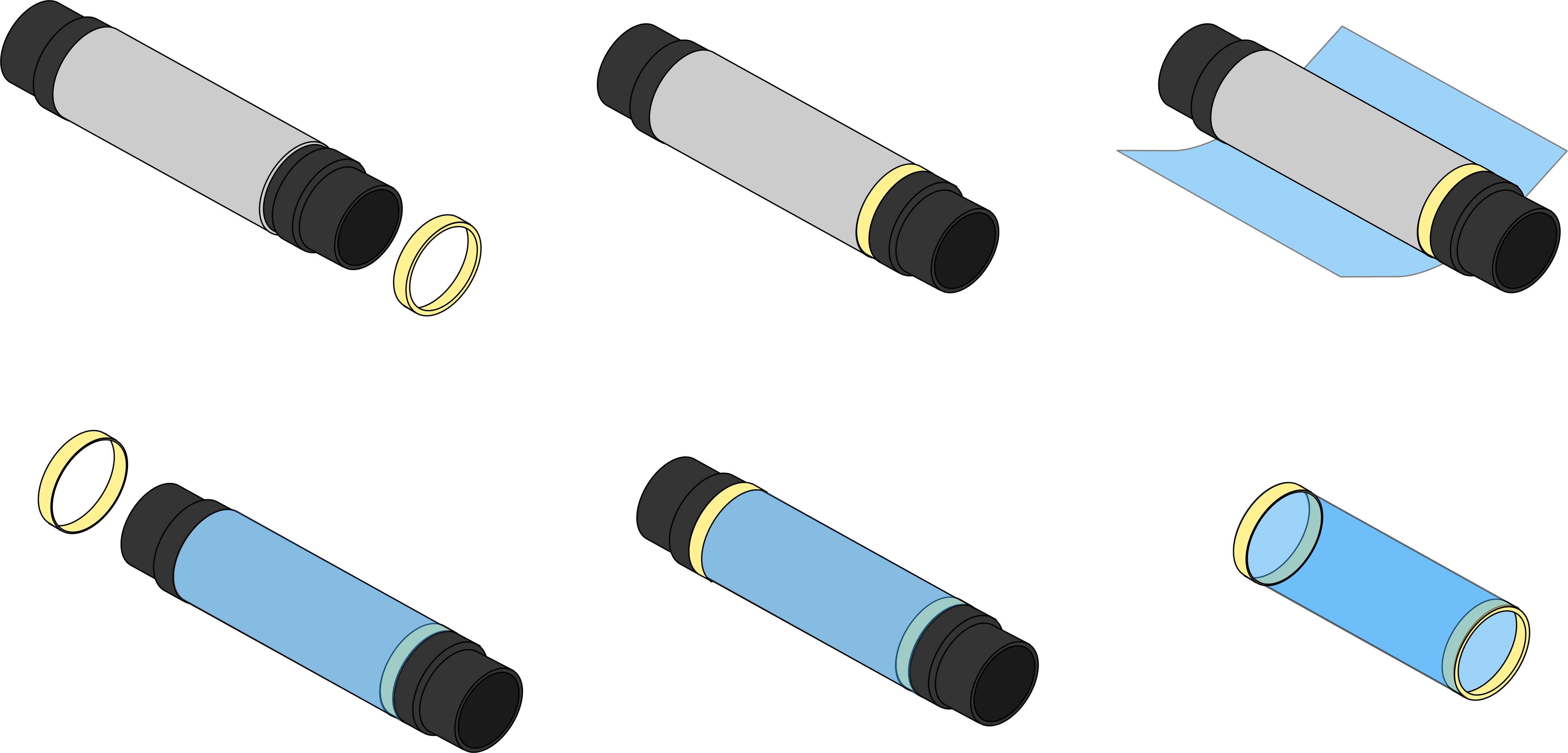}
 	\caption[Main steps of the procedure at the basis of the sublayer construction.]{Main steps of the procedure at the basis of the sublayer construction.}
 	\label{procedure}
 \end{figure}

Each of the three layers of the CGEM-IT detector is composed of five sublayers that are assembled together vertically. Each sublayer is built by wrapping a foil around a mold, which gives it its cylindrical shape. The molds can house a ring, to which the foil is glued, that will remain on the inside of the sublayer; the second ring will be glued on top of the foil. The main steps of this procedure, common to all the sublayers, are represented in figure$\,$\ref{procedure}. The anode and cathode sublayers comprehend cylindrical structural elements, also built by wrapping layers of material around the mold in an analogous way.
 
\FloatBarrier

\begin{figure}[h]
	\centering
	\includegraphics[width=.5\textwidth, keepaspectratio]{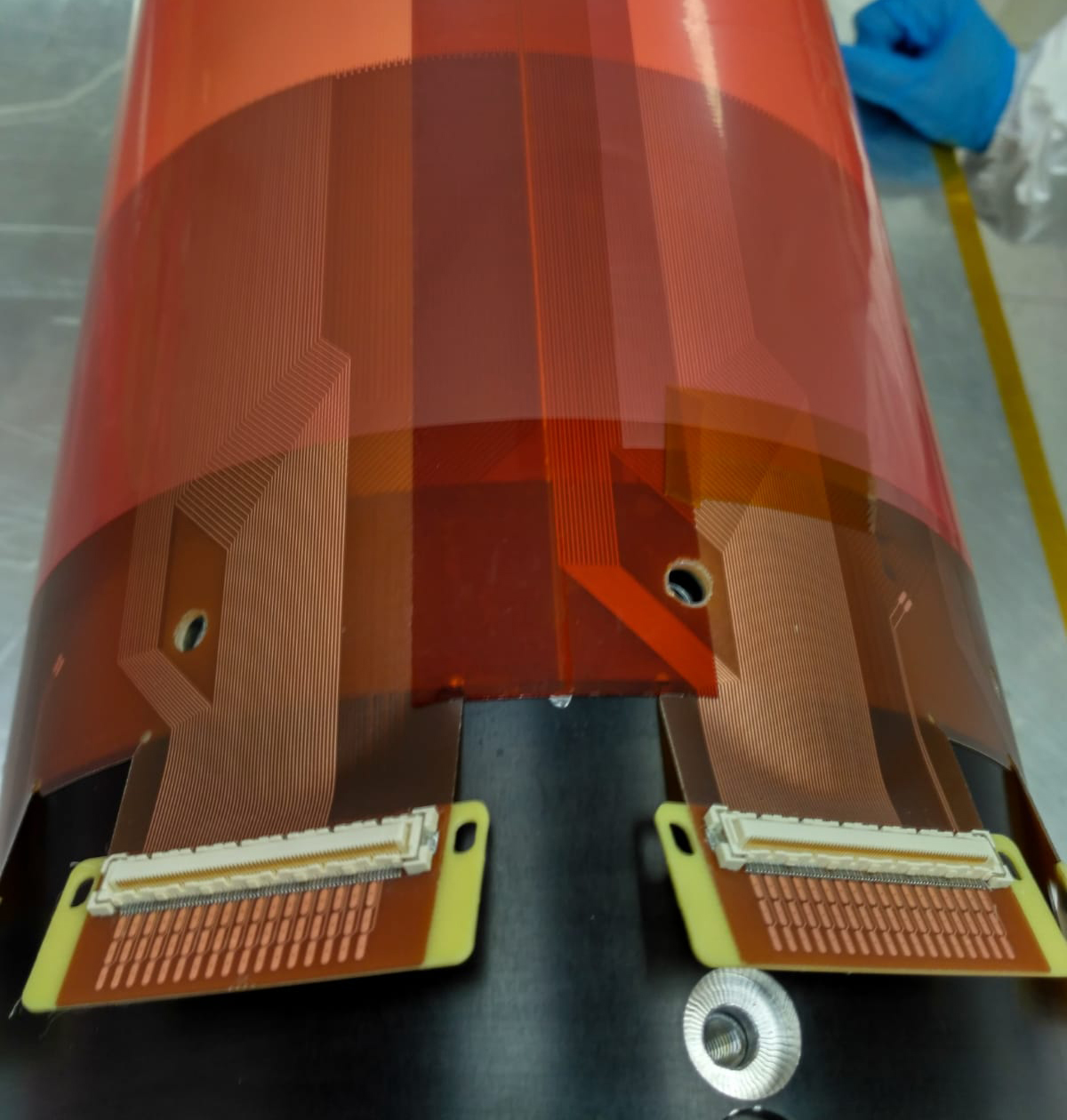}
	\caption[Photograph of the anode readout tails.]{Photograph of the anode readout tails taken during the construction of the anode sublayer, before the gluing of the outer ring.}
	\label{tails}
\end{figure}

Once the detector is assembled, the electrodes can be accessed through tails present at both ends of the rectangular foils. The anode foil of L1 has 16 readout tails, each carrying the signal from 64 strips; its GEM foils all have 4 HV distribution tails, with one channel for the macrosector and 10 for its corresponding microsectors; its cathodic foil has a single tail. Figure \ref{tails} shows the anode readout tails, on the mold, before the gluing of the outer rings.

\FloatBarrier

\pagebreak

The full stratigraphies of the three layers that compose the CGEM-IT are provided as a reference in tables \ref{strato1}, \ref{strato2} and \ref{strato3} in appendix \ref{stratos}, while their dimensions are provided in table \ref{dimensions}

\begin{table}[h]
	\centering
	\begin{tabular}{lrrr}
		Layer & \begin{tabular}[c]{@{}r@{}}Inner\\ diameter\\ (mm)\end{tabular} & \begin{tabular}[c]{@{}r@{}}Outer\\ diameter\\ (mm)\end{tabular} & Length (mm) \\ \hline
		1     & 153.8                                                           & 188.4                                                           & 532         \\
		2     & 242.8                                                           & 273.4                                                           & 690         \\
		3     & 323.8                                                           & 358.5                                                           & 847        
	\end{tabular}
	\caption[Dimensions of the layers.]{Dimensions of the three layers of the CGEM-IT.}
\label{dimensions}
\end{table}

Each topic is introduced by an overview of the procedures and a summary of the results. This is followed, with few exceptions, by an in-depth description of the operations.

\section{Preliminary and Ancillary Activities}
This section describes the operations needed to prepare the construction of the main components of the detector or are parallel to it. These include: the sourcing and quality assessment of the materials; the readying of the tools and of the workspaces; the definition of new procedures where previous ones had become obsolete due to changes in the design of the detector.

\subsection{Replacement of the Molds Vacuum System}
\label{vacsys}
In preparation to the L1 construction, it was necessary to replace the vacuum system of the molds used to build the sublayers. The molds of L1 had been used for building a previous iteration of the detector, after which their vacuum systems were decommissioned or partially dismantled.

The replacement of the piping of the L1 molds is hindered by their limited radius; the use of a particular piping arrangement helps to reduce the bends formed during connection. The new system must be tested through the construction of a vacuum bag around the mold, a procedure used also for the gluing of the components.

During my stay at LNF, I replaced the vacuum system of four out of the five molds used in construction.
All the systems were later tested and managed to reach the desired pressure values. 

\subsubsection{Description of the Molds}
\begin{figure}[h]
	\centering
	\includegraphics[width=.8\textwidth, keepaspectratio]{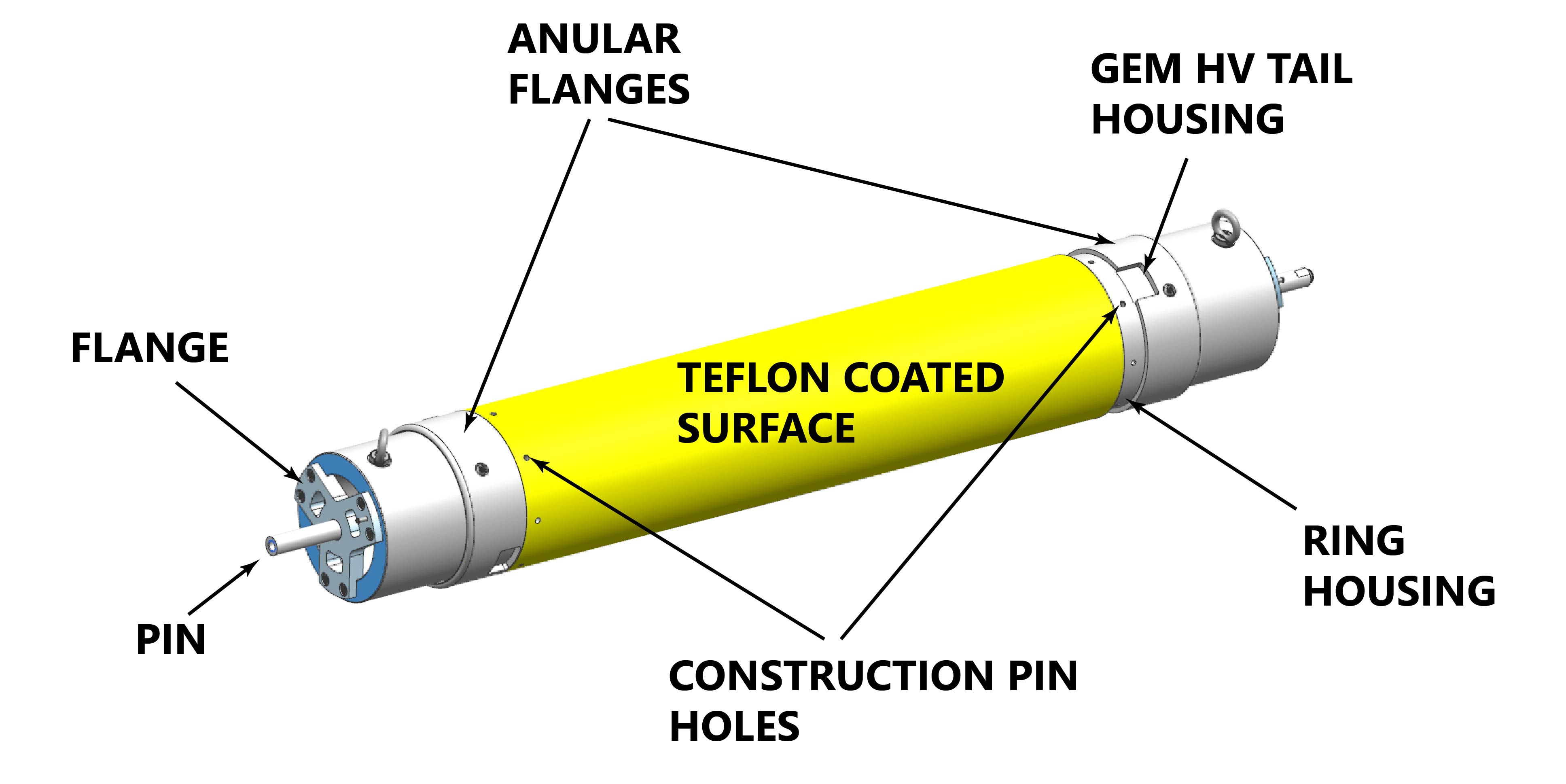}
	\caption[Technical drawing of one of the molds used in the construction.]{Technical drawing of one of the GEM molds used in the construction of the sublayers. The area in yellow represent the Teflon-coated surface that supports the foils during the gluing while the holes that transfer the suction are covered by the annular flanges.}
	\label{mold}
\end{figure}

A technical drawing of one of the Teflon-coated aluminum molds used in the construction is shown in Figure \ref{mold}. These are shaped like large pipes: the sublayer is built on their outer surface through a succession of vacuum assisted gluings while their inner cavity houses the pipes of the vacuum system. The vacuum is used to apply a homogeneous hydrostatic force to the parts to be glued, limiting the introduction of anisotropic stresses.

\FloatBarrier

The surfaces that support the foils were manufactured within strict geometric tolerances to prevent the formation of bends deriving from a loose fit. Their Teflon coating prevents accidental dripping of the epoxy glue from sticking to the mold and facilitates the release of the components during assembly. The other surfaces of the mold are instead anodized to prevent the release of conductive aluminum dust that could compromise the components. Each mold presents two set of holes at each end: one dedicated to the connection of the vacuum system and one for housing the pins used during assembly. This last set of holes must be patched while the mold is being used for gluing; the same patches are later pierced to allow the insertion of the pins during the final assembly.

Each mold is accompanied by a set of two aluminum annular flanges, which are anodized like the molds and have to be coated with anti-stick before the gluing begins.
These flanges cover the suction holes without closing them, forming a canal for the passage of the air.
One in each pair is designed to hold the inner ring in place during the construction of the sublayers. The annular flanges paired with the GEMs and cathode molds have housings to protect the HV distribution tails of the foils during gluing. 

At the edges of the molds, two alluminium flanges, housing pins, allow the mold to be mounted either on the rotating supports used for the gluings or on the machine used for the assembly of the detector.

\subsubsection{Description of the vacuum system}
The vacuum system is the piping that allows to depressurize the vacuum bags built around the outer surface of the mold during the construction process.
These pipes are attached radially to the walls of the mold cavity, through fast fittings inserted in the suction holes, and serve to transfer the suction from the pump to the inside of the bag.

\begin{figure}[h]
	\centering
	\includegraphics[width=\textwidth, keepaspectratio]{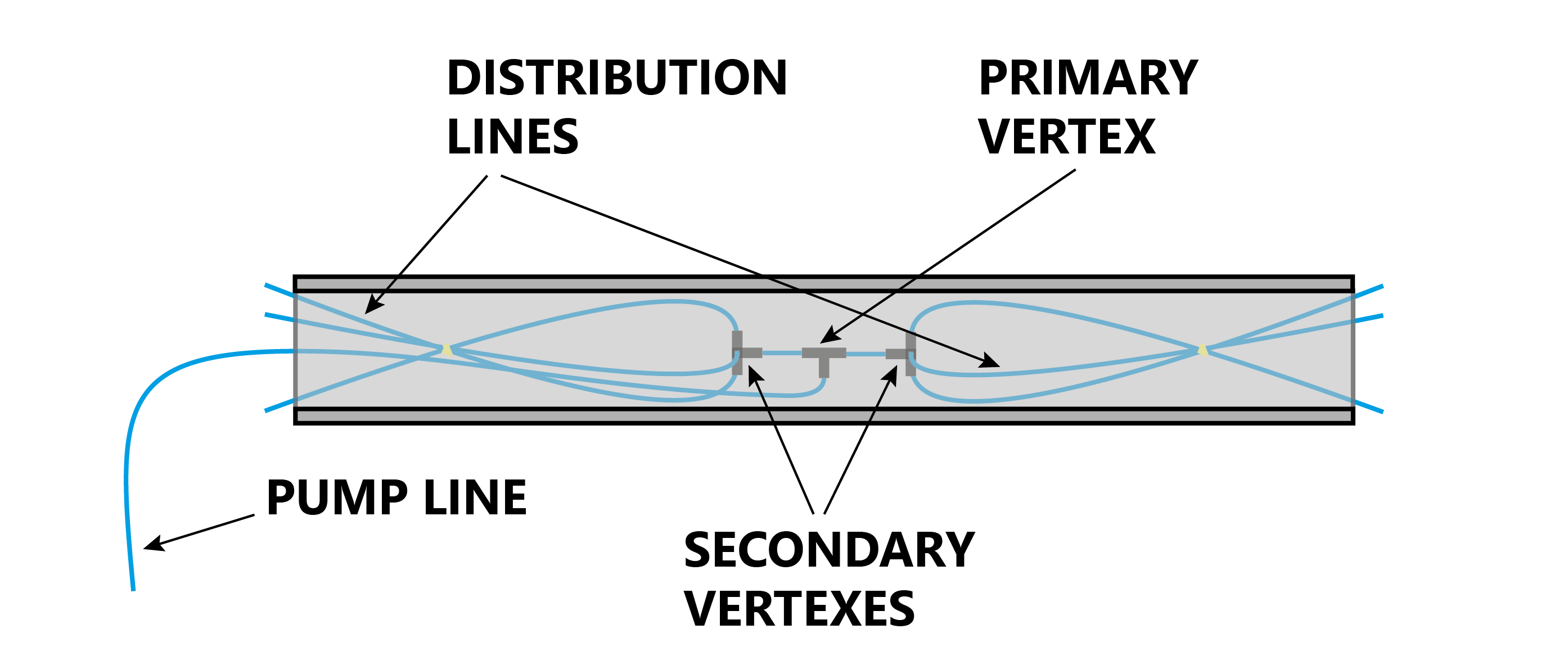}
	\caption[Schematic representation of the vacuum system.]{Schematic representation of the vacuum system and its fitting inside the L1 molds before the connection to the suction holes.}
	\label{pipes}
\end{figure}

In order to homogeneously distribute the suction, the piping should be as symmetric as possible with respect to the center of the mold. In addition, due to the limited space available on the inside of the L1 molds, the pipes have a tendency to develop tight bends.
To satisfy the symmetry requirements and prevent the choking of the pipes, a geometry like the one represented in figure \ref{pipes} has been adopted.

\FloatBarrier

The line from the pump is connected to a primary vertex at the center of the system. The pipes connecting the primary vertex to the two secondary vertexes, from which the six distribution lines depart, are kept as short as possible. This frees up more space for the distribution lines, which are so allowed to assume gentler curves. The three distribution lines at each side are made to cross and taped together, to fix the radius of curvature at the secondary vertex. The distribution lines are kept longer at insertion and then trimmed down to the most convenient length when they are connected to the suction holes.

The main difficulty presented by the installation of the L1 vacuum system is the cramped working space in which the operation must be performed.
The inner diameter of the molds does not allow to operate with both hands, so most of the work is completed from the outside, with the help of instruments like long pliers and dentistry mirrors.
Once the vacuum system is in place it must be tested, to verify the capability to reach pressure values below the millibar deemed necessary for a reliable gluing of the detector components.

\subsubsection{Vacuum Bag Preparation and Test of the Vacuum System}
\label{vacbag}
\begin{figure}[h]
	\centering
	\includegraphics[width=.7\textwidth, keepaspectratio]{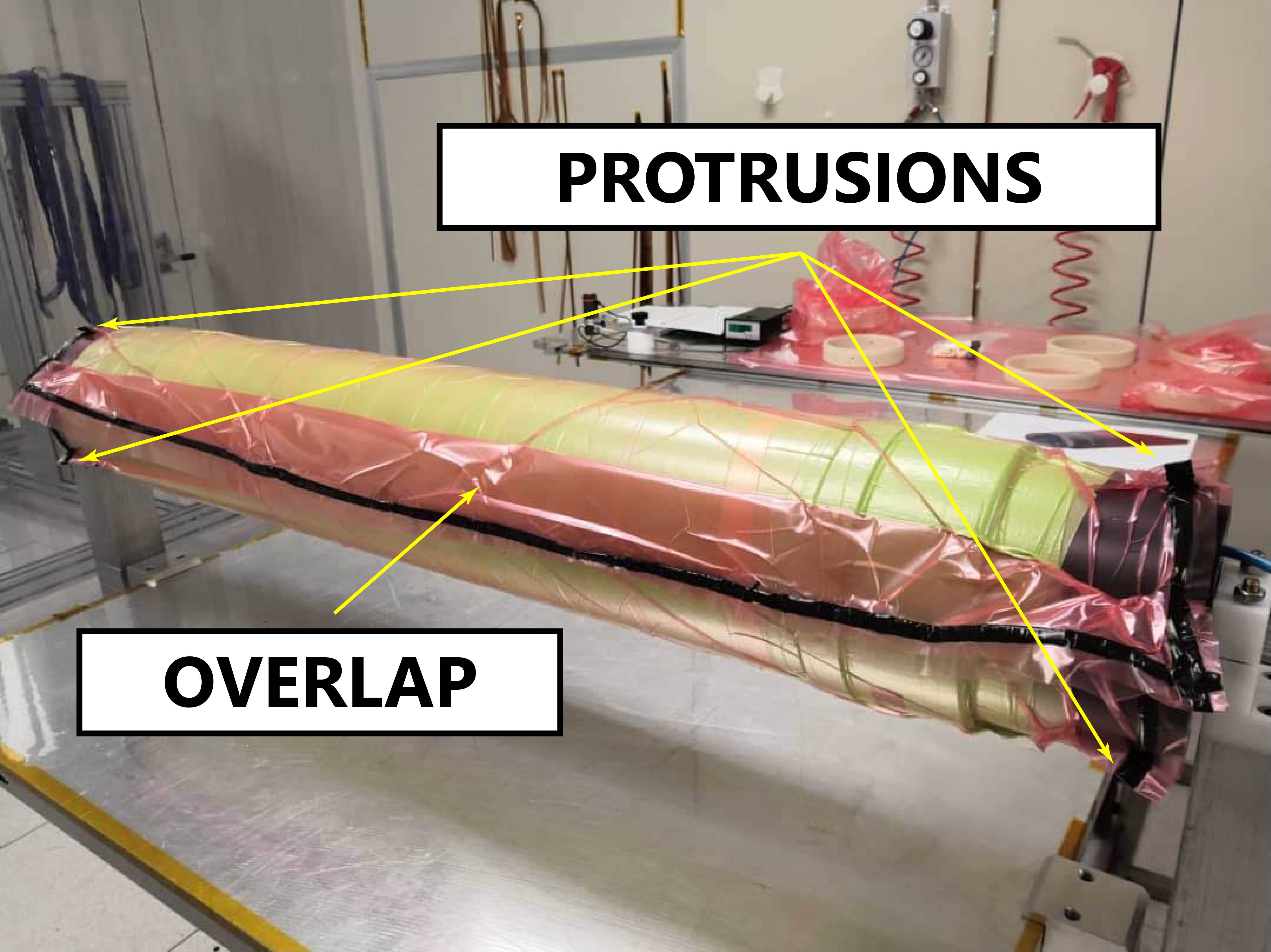}
	\caption[Photograph of a vacuum bag.]{Photograph of a vacum bag taken during a gluing. Both the pocket and the radial protrusions at the opposite ends of the mold are visible. The green peel-ply used is visible in transparency through the pink plastic of the vacuum bag.}
	\label{vacbagpic}
\end{figure}

To test this system a vacuum bag like the one shown in figure \ref{vacbagpic} must be prepared around the empty mold.

\FloatBarrier

The procedure for creating the vacuum bag, here described, is the same that is used many times during the construction; it represents the last step of the gluing of the various components of each sublayers.

Before building the bag, the surface of the mold is wrapped in peel-ply, a special textile that allows air to flow out of the bag through the suction holes by creating a permeable surface that spans the entire length of the component.
Then, a rectangular sheet of vacuum-bag plastic is cut from the roll and resized according to the mold dimensions.
One side must be as long as the mold while the other must be a bit longer than its circumference, to allow the edges to overlap and simplify the sealing of the bag along the length of the mold.
Sealant is applied to three out of the four sides of the rectangle.

With the help of a second operator the vacuum bag is attached to the outer rims of the mold, starting from 15-20$\,$cm from the edge.
This produces a hanging flap that will be later used to form the overlap.
The mold is then slowly rotated until the edges are completely sealed.

During the rotation three small protrusions are realized at each end by pinching together few centimeters of vacuum bag. This serves to compensate for the fact that the mold is smaller at its rims and larger in its central portion.
Once the bag covers the whole circumference of the mold, the flap is made to adhere with the excess portion of vacuum bag present at the opposite side. The overlap formed in this way both simplifies the sealing of the bag and keeps the sealant far from the delicate components during a real gluing.

Once the vacuum bag is closed and the seals have been checked a first time, the pump is turned on. While the pump is working, any air canals forming near the seals due to wrinkles in the vacuum bag must be smoothed out and closed.
The few remaining leaks are usually found by listening to the whistling sound they produce.
If there are no leaks in the vacuum system, the pressure, measured at the pump by a digital manometer, falls relatively fast. When it reaches values below the millibar, based on the experience of the previous constructions, the system is left running for several hours.

\subsection{GEM Quality Assurance}
GEMs are fragile and sensitive to dust, during their production process or their handling they can become damaged and rendered unusable.
For this reason, a spare is ordered for each foil. All the GEMs are tested on arrival, this allows to choose the best performing ones to be used for the detector.

The quality assessment of a GEM foil is divided in three parts: a first visual inspection of the foils, a series of capacitance and resistance measurements to verify the absence of contacts and, finally, a HV test for evaluating the GEM stability and performance at and above operating voltages.

I personally conducted the HV test of two GEM foils. One of them showed an absorption of current in correspondence of one of its microsectors, indicating the presence of a short-circuit, while the other one did not present any irregular behavior.

\subsubsection*{Detailed Description of the Testing Procedure}
\label{hvtest}
To protect the foils from dust that can enter the GEM holes, the shipping package can be opened only inside a clean room (CR) class 10000 or better\footnote{The cleanroom of LNF, where the detector was built, is a class 1000.} and only after being thoroughly cleaned with isopropyl alcohol and special CR-safe wipes.
After opening the package, the GEMs are visually inspected to spot any major faults present on their surfaces.
It is generally advisable to minimize the time a GEM foil is exposed to the air even inside a clean room and so, whenever not in use, they are kept covered and safe in the original package.

\begin{figure}[h]
	\centering
	\includegraphics[width=.7\textwidth, keepaspectratio]{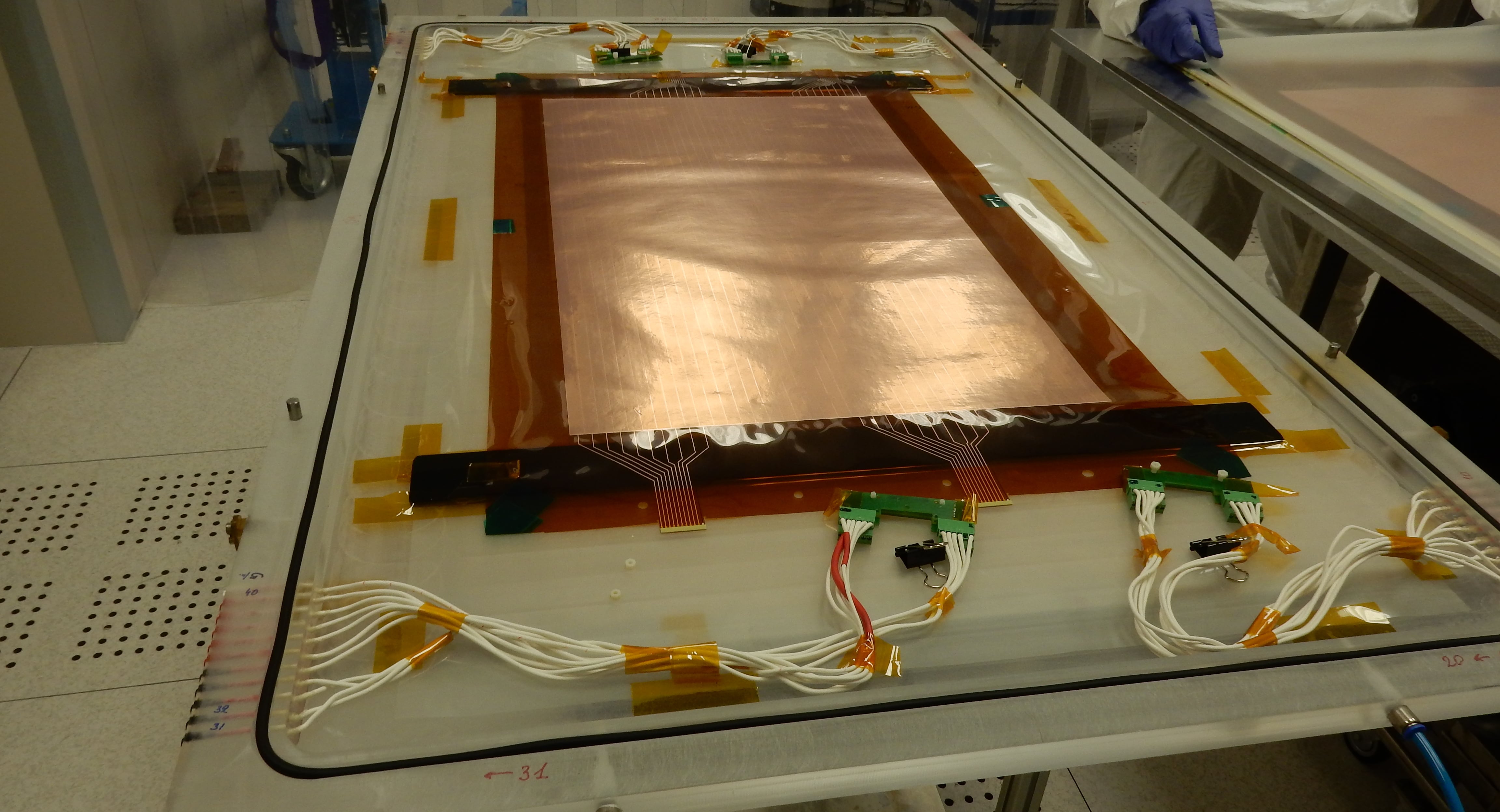}
	\caption[Photograph of the box used for the HV test of the GEM foils.]{Photograph of the box used for the HV test of the GEM foils.}
	\label{hvtable}
\end{figure}

The box used for the HV test of the GEM foils, represented in figure \ref{hvtable}, is equipped with a gas line and high voltage connections that allow to access and power the individual sectors from the outside, once the lid has been closed.
The GEM is placed inside the box and the connectors are clamped to the HV distribution tails of the foil.

\FloatBarrier

Before the proper HV test, a series of preliminary checks is performed on the GEM sectors, through the box connections, to diagnose possible problems with the box wiring and provide a first assessment of the GEM health status.
A multimeter is used to check if there are any contacts between the sectors and measure the capacitance between each microsector and its macrosector on the opposite side of the foil, which is then compared with the nominal value of 3.5$\,$nF.
These tests are repeated several times during the course of the construction: in particular after each GEM gluing and when the assembly is completed.

Having verified that the box connections are working properly, the lid is clamped shut and the chamber where the foil is enclosed flushed with nitrogen.
A hygrometer measures the humidity at the exit of the chamber.
When it falls below 8\% it is safe to apply a voltage to the GEM without prematurely inducing discharges.
A single macrosector and its corresponding ten microsectors are connected to the power supply and the voltage is raised in steps of 100$\,$V up to 400$\,$V, then in 50$\,$V steps up to 550$\,$V, and finally up to 600$\,$V in step of 10$\,$V. If the 600$\,$V mark can be reached without the GEM experiencing more than 1 discharge per minute, the voltage is maintained for 15 to 20 minutes. The number of discharges that occur at each voltage step and at 600$\,$V is registered, together with any other irregular behavior observed during the test. The measurement is then repeated on the remaining microsector groups.
Once all the GEM have been tested, any foil with evident damage or short-circuited sectors is immediately discarded and the candidates with lower discharge rate are chosen for the construction.

\subsection{GEM and Anode Resizing}
The sheets on which GEMs, anodes and cathodes are realized are larger, on arrival, than the components design dimensions, as the excess of polymer substrate is not removed before shipping. They have therefore to be resized using a machine especially designed for this purpose and capable of achieving a precision in the cut of the order of a tenth of a millimeter. Clean and precise cuts provide net borders that facilitate the precise gluing of the components. These often involve tiny overlaps that can be as small as just$\,$3 mm, with the glue being confined on a 2$\,$mm wide strip. Even a small error in the cut can complicate the gluing of the GEM overlap, one of the most delicate operations in the whole construction, and reduce the space available for the deposition of the glue, weakening the strength of the bond.

The cut is performed by aligning a guide ruler with markers present on the foil, through the use of microscopes, and then sliding a sled holding a scalpel blade in a single motion.

I was involved in both the preparation and the alignment for the resizing of the anode and of all three GEM foils. The final cut is instead always performed by an expert technician.

\subsubsection*{Detailed Description of the Resizing Procedure}
\begin{figure}[h]
	\centering
	\includegraphics[width=.9\textwidth, keepaspectratio]{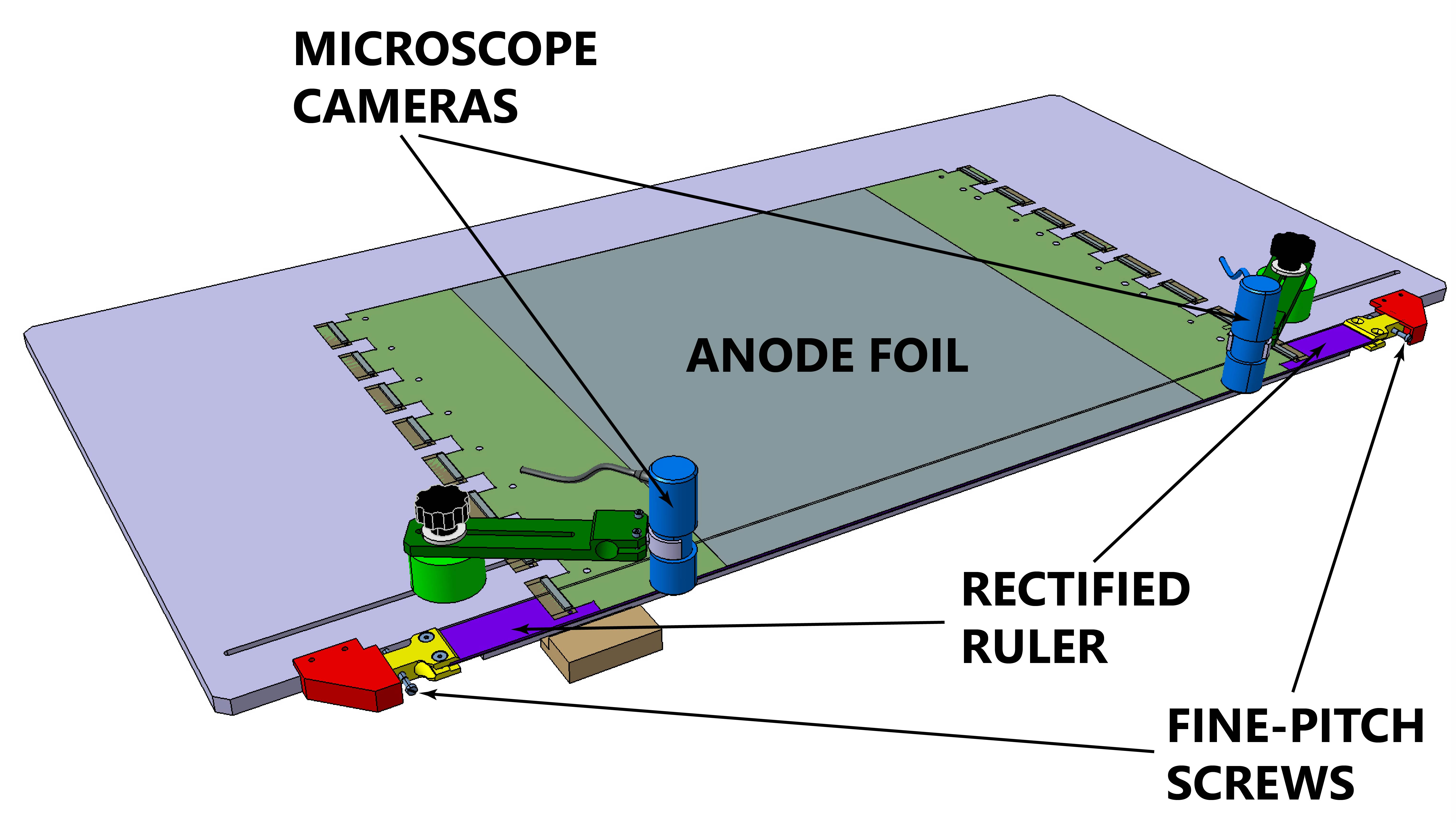}
	\caption[Drawig of the machine used for resizing the active elements of the detector.]{Drawing of the tool used for resizing the active elements of the detector. The microscope cameras are depicted in blue while the adjustable ruler is represented in purple. The foil on top of the plane of the machine is the anode readout.} 
	\label{guillotine}
\end{figure}

The machine used for the cuts is represented in figure \ref{guillotine} and consists of an adjustable rectified guide ruler, two microscope cameras to aid in the alignment and a sled which holds the scalpel blade. The sled has small wheels, fitted into a 3D printed body, and a holder that keeps the blade firmly in place. While the sled is moved across the table, the upward pointing blade grazes against the guide ruler.

\FloatBarrier

In preparation to the cut, the component is placed on the plane of the machine, roughly aligned with the adjustable ruler, taped in position, protected using one or more sheets of plastic material and weighed down to prevent movement. In order to evenly spread the force, a thin metallic bar is interposed between the component and the weights. The placement of this bar and of the lead bricks used as weights is performed with utmost care, to avoid nicking the delicate surfaces of the components.

The adjustable ruler is connected to the plane of the machine by two fine-pitch screws that can be tightened or loosened to modify its distance and inclination with respect to the component.
Small marks are present on the substrate to indicate where the cut should occur.
The alignment is achieved using the microscopes to observe the outline of the ruler through the semi-transparent Kapton substrate and then performing the necessary adjustmens to make it overlap with the marks.

The blade holder is then sled in a single movement along the length of the component, completing the cut. The edge is inspected with the microscopes in correspondence of the markers and with the naked eye along its length.

\subsection{Alignment of the Vertical Insertion Machine}
\begin{figure}[h]
	\centering
	\includegraphics[width=.5\textwidth, keepaspectratio]{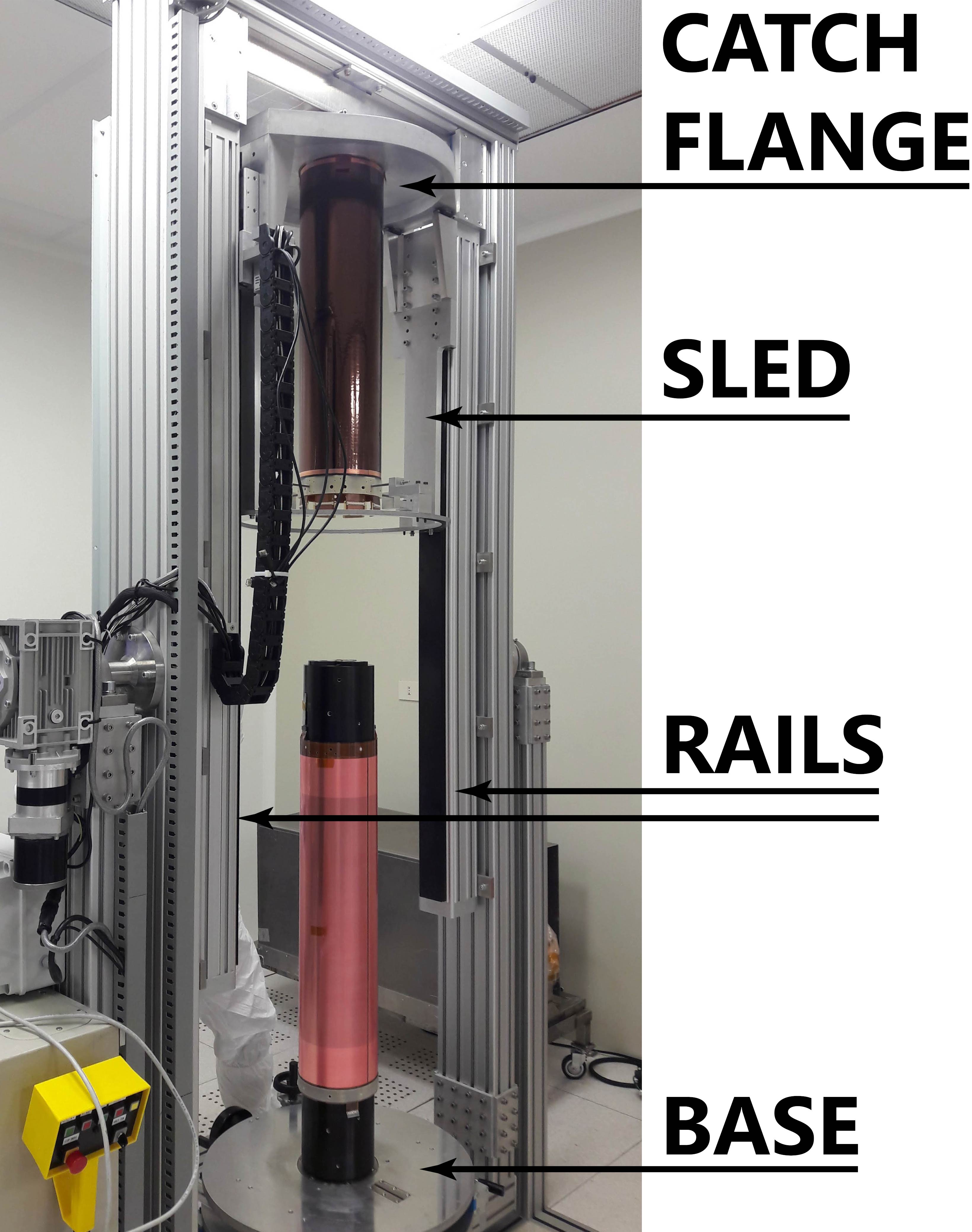}
	\caption[Photograph of the vertical insertion machine.]{Photograph of the vertical insertion machine during one of the phases of the final assembly.}
	\label{hourglass}
\end{figure}

Before the construction of the sublayers, the Vertical Insertion Machine (VIM) used for assembling the detector, photographed in figure \ref{hourglass}, must be aligned using the empty molds as reference.

\FloatBarrier

for the construction of L1, the alignment of the VIM was needed for the concerns raised by an earthquake that hit relatively close to the laboratories in the time period between the beginning of the works on L1 and the previous construction.

First, the misalignment must be evaluated for all the molds. This is performed using comparators, which slide on the surface of the molds, and a large caliper, to measure the relative distance between the mold and the catch flange of the VIM. The inclination of the VIM base is then adjusted until the misalignment falls within tolerance. The alignment must finally be confirmed for each mold.

I participated both in the moving of the molds and to the whole alignment process.
The VIM had indeed been thrown out of alignment as we measured 0.320$\,$mm of longitudinal misalignment and a maximum of 1.3$\,$mm in the XY plane for the anode mold and compatible values for the others. With few adjustments both values were brought within the tolerances and the success of the alignment was later confirmed for all the molds.

\subsubsection*{Detailed Description of the Alignment Procedure}
Each mold is positioned in the chuck at the bottom of the machine and two mechanical comparators are clamped to the sled that raises and extracts the sublayers from their mold. The comparators point in orthogonal directions and set to slide on the surface of the molds while the sled is raised or lowered. The maximum variation registered by the comparators is used as an indication to gauge the longitudinal misalignment. To measure the misalignment on the XY plane the catch flange is lowered to the same level as the upper rim of the mold. A large caliper is then used to take direct measurements of the relative position between the two.

The inclination of the base hosting the molds can be modified by tightening or loosening a set of three screws. Once an adjustment is made, the misalignment is measured again to evaluate the changes produced. Once a satisfactory result has been reached for a particular mold, all the others have to be checked as well in the new configuration. The tolerance is of 0.1$\,$mm for both the longitudinal misalignment and the one in the XY plane.

The whole procedure requires the operators to move the molds many times. Even though the L1 molds are the lightest of the three sets, they still require at least two operators and the use of a dolly to be safely moved around and placed inside the VIM.

\subsection{Cathode Structural Tests}
\label{cathtest}
Since the first L1 construction, the cylindrical structure that supports the cathode had to be redesigned to increase its rigidity. In place of a double Rohacell-Kapton sandwich, it was decided to use a single sandwich, realized using Honeycomb as core material and Kapton skins. To determine if the new structure had the desired characteristics, two tests were conducted. The construction of these samples was used as a test bench to define and perfect the procedures later adopted for the realization of the final cathode of the detector.

\subsubsection{First Cathode Structural Test}
\begin{figure}[h]
	\centering
	\includegraphics[width=\textwidth, keepaspectratio]{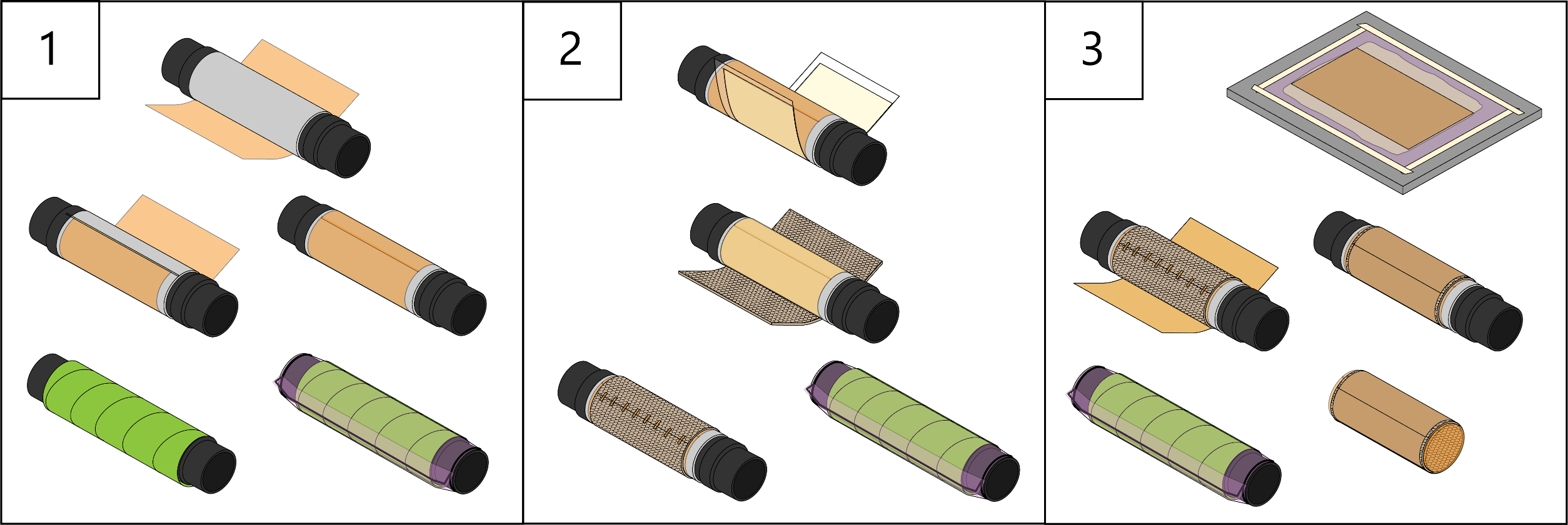}
	\caption[Main steps of the construction of the first sample.]{Main steps of the construction of the first sample.}
	\label{ikeasample1}
\end{figure}

The first cathode sample was realized around the cathode mold. Being just a test, no rings were used; the structure was built on top of a Kapton foil wrapped around the mold and closed by a glued overlap. On top of this were glued: a honeycomb layer 1.9$\,$mm thick and a copper clad kapton foil mimicking the cathodic foil. The passages of the construction, which involves a succession of cylindrical vacuum assisted gluings, are summarized in figure$\,$\ref{ikeasample1}

\FloatBarrier

\begin{figure}[!h]
	\centering
	\includegraphics[width=.7\textwidth, keepaspectratio]{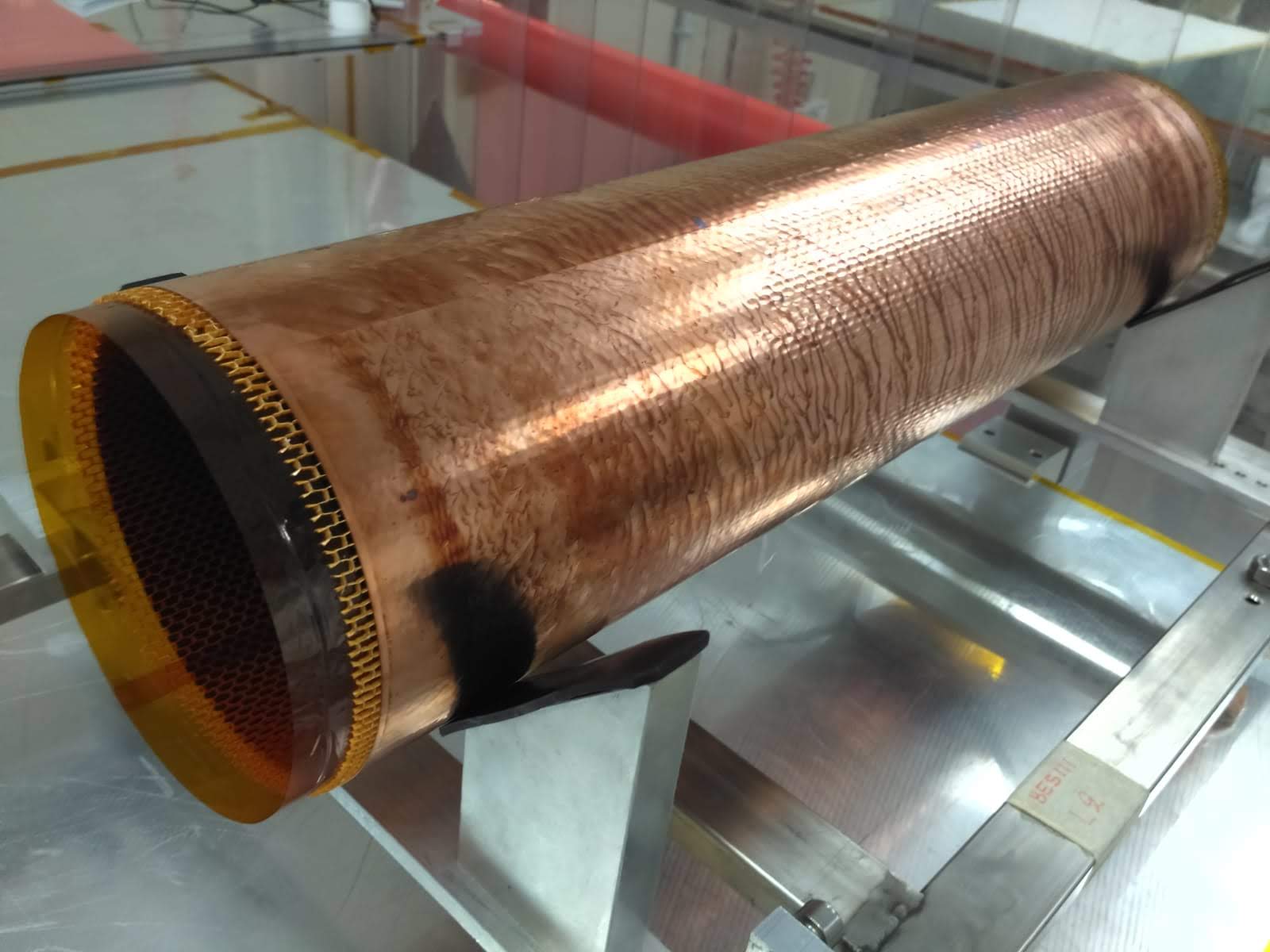}
	\caption[Photograph of the first cathode structure sample.]{Photograph of the first cathode structure sample. In this photograph is possible to observe the absence of dimples at the extremities, where the copper clad Kapton was not properly glued, and their depth at the center, where it was.}
	\label{sample1}
\end{figure}

I participated to all the phases of the construction and to the following evaluation of the sample produced. The sample was not satisfactory and another test was necessary. The copper clad Kapton was well glued at the center  of the sample but at the extremities air bubbles and detached areas could be observed. Moreover, the pressure of the vacuum bag against the Honeycomb ridges along the junction generated several spikes on the metal surface that could become a source of discharges in the final detector. The Honeycomb cells also created dimples on the cathodic plane that, although well within tolerances, are not ideal, especially compared to the smooth cathodic surface that is possible to obtain when gluing on a Rohacell-Kapton sandwich. Figure \ref{sample1} depicts the first sample and allows to observe the absence of dimples at the extremities where the copper clad Kapton was not properly glued. A second test was devised to fully define the procedures for the final cathode construction, and its structure.

\subsubsection*{First Sample Construction Procedure}
To provide a substrate on top of which to glue the Honeycomb panel, a $\mathrm{50\,\upmu m}$ thick Kapton sheet is resized and wrapped around the cathode mold, with an overlap of 1$\,$cm. A thin Mylar\footnote{Mylar is a strong and flexible polyester film.} strip is used to transfer epoxy glue to the overlap region. The overlap, once closed, is held in position by a sheet of plastic material wrapped around the mold and fixed using tape. The preparation of the vacuum bag retraces the one described in section \ref{vacbag}. 

As the manipulation of the Honeycomb tends to release dust particles, the operations were moved to a working area appositely prepared outside the clean room. Before being used, the Honeycomb panel has to be resized and, due to the stretchy nature of this material, all measurements and cuts must be made on the relaxed panel, taking care not to stretch it during the operations. A dry wrapping test is performed to verify the fitting of the resized panel on the mold. As the vacuum bag will compress the panel, a slightly oversize Honeycomb sheet is not a major drawback, the shape of the cell after gluing will be stretched along the longitudinal direction of the mold. An undersized sheet though, may lead to the opening of the junction and so has to be remade.

If the dry-test is satisfactory, the glue is transferred to the Kapton substrate already on the mold using a technique that is employed at many points in time during the construction of the final detector sublayers. A Mylar sheet is placed on a table and Araldite 2011\footnote{Araldite 2011 is a bicomponent epoxy adhesive. Due to outgassing issues, it is used when the surface of glue exposed to the gas is minimal.} epoxy glue is homogeneously spread on it with the aid of a Teflon roller. Two operators then wrap the Mylar sheet around the mold and press against its surface with a wipe in slow circular motions, to homogeneously spread the glue and remove residual air bubbles. Once the Kapton substrate has been completely coated, the Mylar foil is unwrapped from the mold and discarded.

With the glue now in place, the Honeycomb panel is once again wrapped around the mold. The junction is fixed in position using Kapton tape to prevent the sheet from unwrapping itself during the preparation of the vacuum bag. 

During this phase of the construction of the first sample, while the pump was in action and the glue was setting, the vacuum was lost due to a leak in the vacuum system. The system had been tested and had worked fine during the gluing of the previous Kapton substrate. The leak must therefore have occurred while moving the mold outside the clean room. The vacuum system had to be replaced and tested in a way akin to the one described in section$\,$\ref{vacsys} while the unfinished sample was on the mold.

Once fixed the vacuum system, the final step is to glue the copper clad Kapton foil to the Honeycomb, with a 1$\,$cm overlap. This cylindrical gluing is performed by first transferring the glue to the foil through a planar transfer. The glue is first spread on a table using a roller, the copper clad foil is laid flat on the glue, with the copper facing up, and massaged using a wipe. The foil, now covered in glue, is lifted from the table and wrapped around the mold. The overlap is closed and the gluing is completed by the preparation of a third vacuum bag.

\subsubsection{Second Cathode Structural Test}
\begin{figure}[h]
	\centering
	\includegraphics[width=.7\textwidth, keepaspectratio]{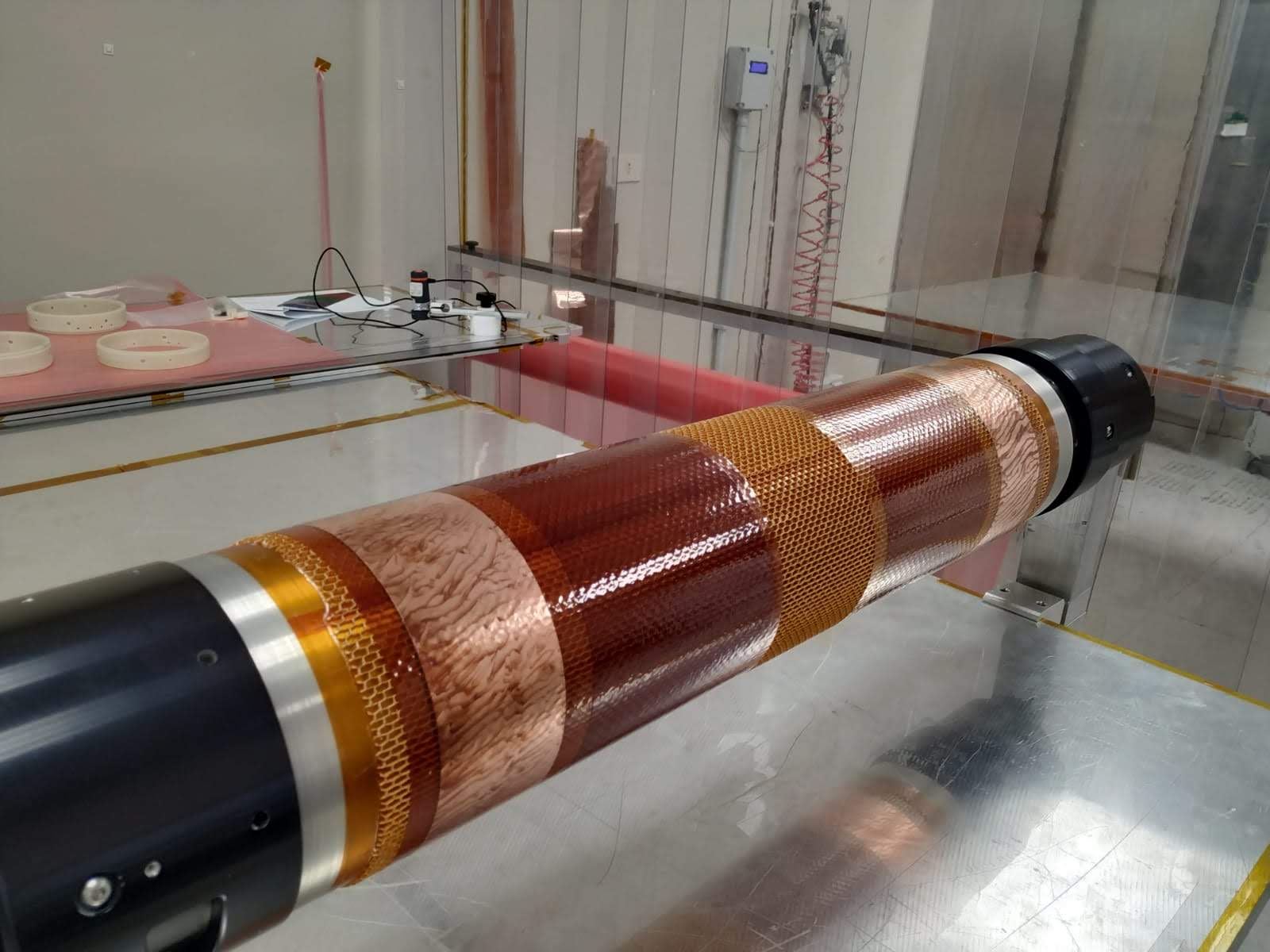}
	\caption[Photograph of the second cathode structure sample.]{Photograph of the second cathode structural sample, still on the mold. Shallow dimples are visible also near the margin of the strips, where the first sample presented evident signs of detachment.}
	\label{sample2}
\end{figure}

For a better evaluation of the behavior of the materials at the margins of the glued areas, where the first sample showed signs of detachment, we produced a second sample, which consisted of four separate strips on a common Kapton-Honeycomb substrate as shown by figure \ref{sample2}.

The preparation of this substrate retraces the one described in the previous paragraph, with the notable difference that this time the leftover ridges resulting from the resizing of the Honeycomb were trimmed away.

It was decided to add in between the Honeycomb and the copper clad Kapton mimicking the cathode a $\mathrm{25\,\upmu m}$ Kapton foil to favor the gluing between the much stiffer cathodic plane and the border of the Honeycomb cells. Two large strips of this material were glued on top of the Honeycomb. The copper clad Kapton layer strips simulating the cathodic plane were glued at the sides of the mold. The two central subsamples, instead, were obtained using $\mathrm{50\,\upmu m}$ thick Kapton in place of metallized Kapton as a final layer. This allows to see through the layers and evaluate the quality of the gluing in correspondence of the dimples and detached zones.

As for the first one, I participated in both the construction and the evaluation of the second sample. This test was much more succesful, all four samples showed no evidence of spikes and the dimples were much less pronounced.  The $\mathrm{25\,\upmu m}$ sheet closing the top of the Honeycomb cell allows the glue to pool and reduce the depth of the dimples. Overall rigidity and uniformity were also improved by a significant amount, thanks to the additional layer and the use of more epoxy adhesive. 

\section{Construction of the Detector Sublayers}
The five detector sublayers can be built in any order, as the procedures for their construction are independent from each other. The order of the steps is fixed and mandated by the design of the structural rings and by how the final assembly is performed. During my stay at LNF, I participated in the construction of all three GEM sublayers and of the cathode. For what concerns the anode sublayer, I was involved in the first cylindrical gluing of the readout plane and the final gluing of the ground plane. The remaining phases of the construction took place in Milan, at the company that provides the carbon fiber meshes for its support structure.

\subsection{Cathode Sublayer Construction}
\begin{figure}[h]
	\centering
	\includegraphics[width=\textwidth, keepaspectratio]{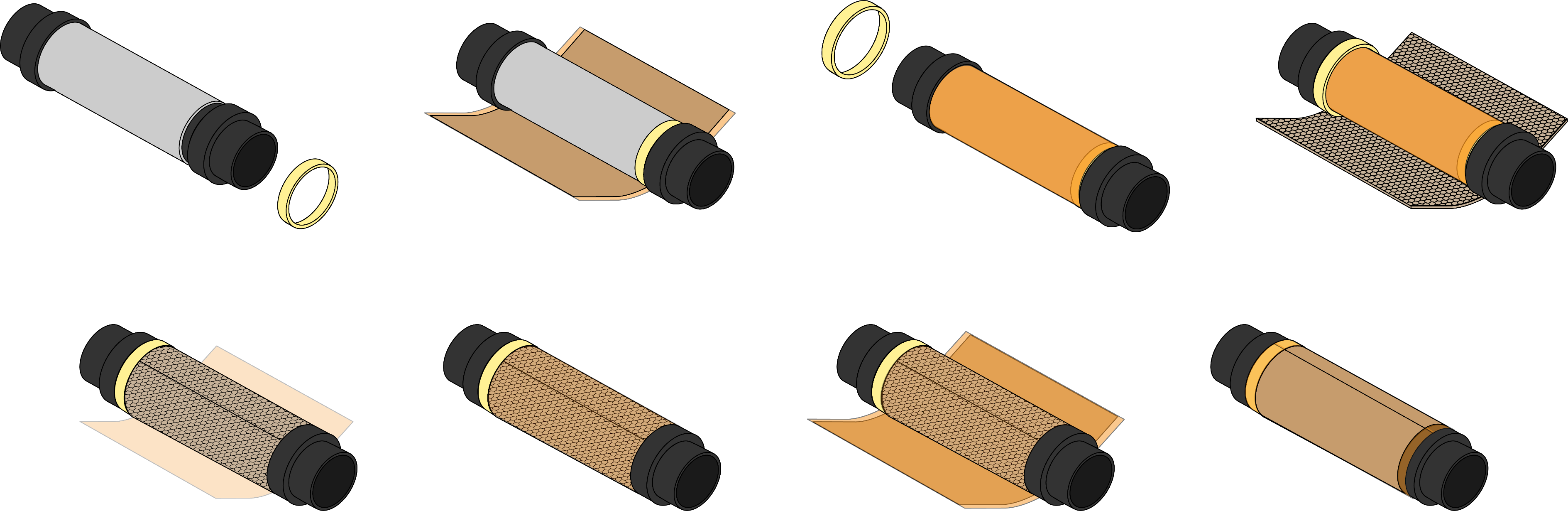}
	\caption[Main steps of the construction of the cathode sublayer.]{Main steps of the construction of the cathode sublayer.}
	\label{ikeacathode}
\end{figure}

The construction of the cathode sublayer starts with the gluing of the inner and outer rings to the Faraday cage of the detector. Atop the Faraday cage is built the cylindrical structural element, through a succession of cylindrical gluings. This support consists of a layer of Honeycomb 1.9$\,$mm thick and a $\mathrm{25\,\upmu m}$ Kapton sheet. The cathodic plane of the detector is finally glued on top of the structure and the outer ring. The main stepss of the cathode construction are summarized in figure \ref{ikeacathode}.

\FloatBarrier

I participated to all the phases of the cathode construction. Thanks to the techniques refined through the two tests described in section \ref{cathtest}, the cathode produced did not present any detached areas and the dimples caused by the honeycomb cells remained shallow.

\subsubsection*{Detailed Description of the Procedures}
The construction of the cathode involves the gluing of inner and outer ring to the Faraday cage, which remains outside the detector at inner radius. This represents a difference with respect to the construction of the other sublayers, where the rings are glued to the foil that gives the name to the sublayer: in the GEM sublayers the rings are glued to the GEM foil and in the anode sublayer to the anode readout foil. The cathodic plane, in this case, is instead glued on top of the outer ring and the cylindrical honeycomb sandwich built atop the cage.

The Faraday cage also constitutes one of the skins of the sandwich supporting the sublayer, whose core is the 1.9$\,$mm thick Honeycomb sheet. The other skin is a $\mathrm{25\,\upmu m}$ Kapton foil, as determined through the tests described in section \ref{cathtest}.

The construction of the cathode starts with the gluing of the Faraday cage on the inner structural ring. The ring is placed in its housing at one side of the mold, on one of the two annular flanges. Before the application of the glue, the Faraday cage is wrapped around the mold to check the alignment between the holes in the foil and those on the ring, together with the dimensions of the overlap. Similar dry-tests always precede the gluing of the foils and of the outer rings for the entirety of the construction.

If the dry-test does not raise any issues, after protecting the mold and the flange with a plastic sheet, glue is transferred to the ring, using a Mylar strip. At this point, the Faraday cage is wrapped around the mold, glue is transferred to one of the margins and the overlap is closed.
During the wrapping, the reference holes present on the foil are aligned to the ones on the ring using pins. Finally, the gluing is completed through the construction of a vacuum bag around the mold.

Once the epoxy has set and the mold has been freed from the bag, the glue can be transferred to the inside of the outer ring. Outer rings are designed with a cut in a point along their circumference, so they can be placed in position without removing the mold from its support. The ring is forced open, set in its designated position, at the free end of the Faraday cage, and another vacuum bag is prepared around the mold.

\begin{figure}[h]
	\centering
	\includegraphics[width=.7\textwidth, keepaspectratio]{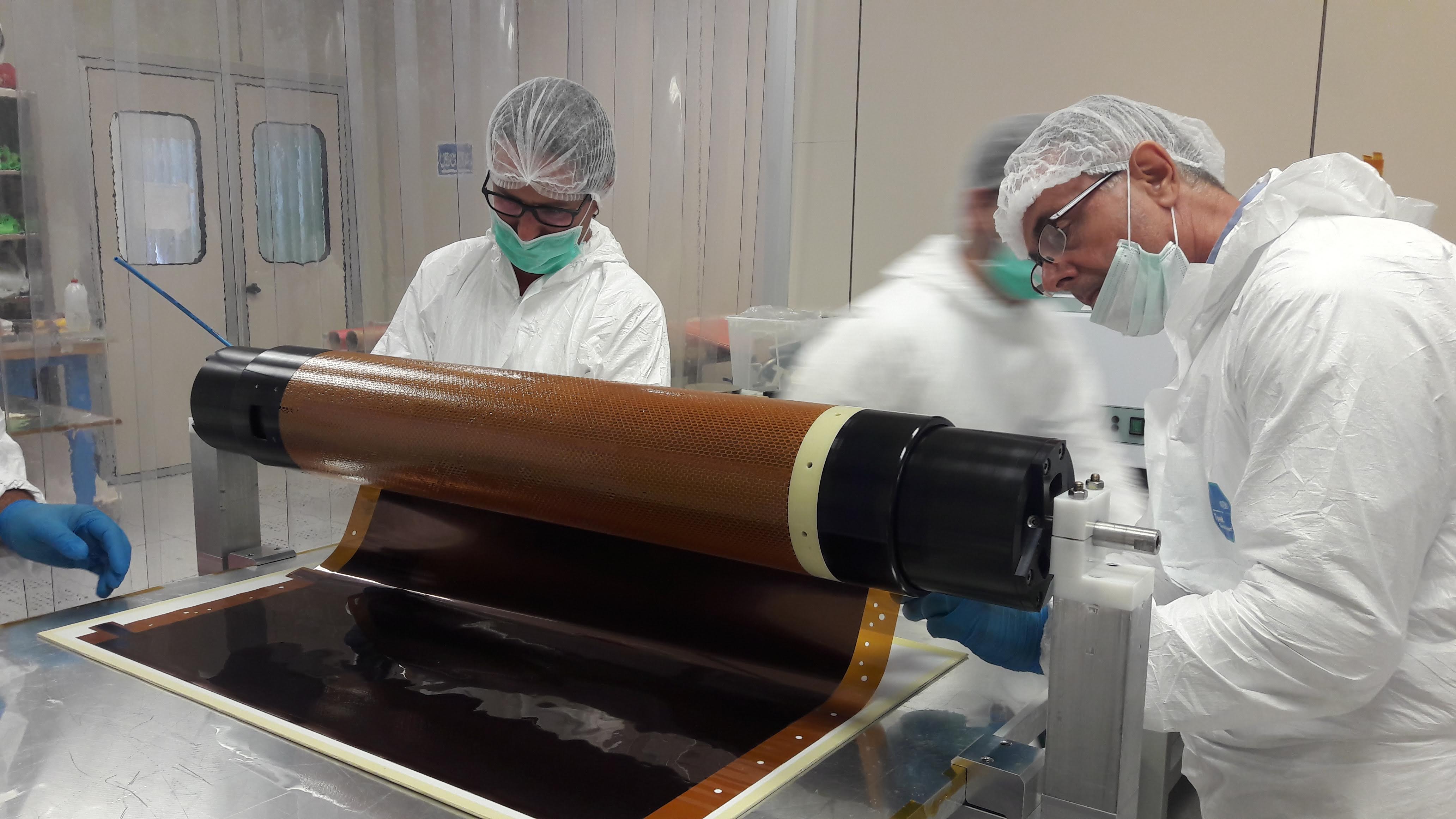}
	\caption[Photograph of the cathode construction.]{Photograph of the last step in the cathode construction: the gluing of the copper clad Kapton foil, constituting the cathodic plane of the detector, on top of the the $\mathrm{25\,\upmu m}$ Kapton layer and outer ring.}
	\label{cathodeconstr}
\end{figure}

At this point starts the construction of the cylindrical structural element. The procedures used are the result of adapting what was done for the construction of the samples to the more complicated design of the sublayer. The glue is transferred to the Faraday cage and the Honeycomb sheet is wrapped around the mold. The Honeycomb is put in contact with the ring on one side and fixed in position with Kapton tape, then it is pulled on the opposite side until it contacts the mold annular flange and is, once again, fixed in position. Some readjustment of the tape may be required to avoid inducing unwanted localized compression zones in the Honeycomb sheet. The junction is also closed with tape and, after checking that everything is firmly in place, a vacuum bag is prepared.
The $\mathrm{25\,\upmu m}$ layer was glued solely on top of the Honeycomb and not on top of the outer ring. Finally, the cathodic plane is glued on the ring and on the thin sheet covering the Honeycomb completing the sublayer. Figure \ref{cathodeconstr} shows this last step of the cathode construction. Overall, for the realization of the L1 cathode a total of five gluings is required, each one performed using a vacuum bag. 

\FloatBarrier

\subsection{GEM Sublayers Construction}
\begin{figure}[h]
	\centering
	\includegraphics[width=\textwidth, keepaspectratio]{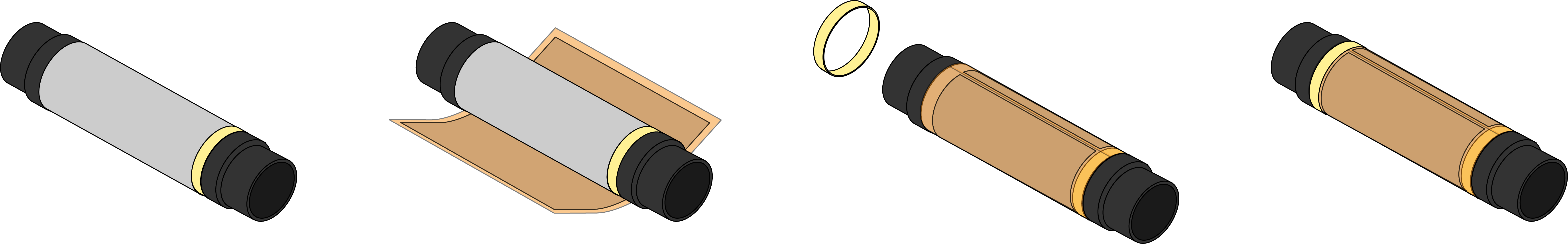}
	\caption[Main steps of the construction of the GEM sublayers.]{Main steps of the construction of the GEM sublayers.}
	\label{ikeagem}
\end{figure}

The GEM sublayers are realized by first gluing the foil on the inner ring and then gluing an outer ring at the opposite end. The gluing of the overlap that closes the foil is the most delicate operation of the whole construction and is performed by an expert technician with the help of at least two other operators.
The main steps of the construction of the GEM sublayers are summarized in figure \ref{ikeagem}.

I participated to the construction of all the three GEM sublayers of the L1 detector. No issues disrupted the flow of the operations and a series of connectivity and capacitance tests confirmed that the GEM foils were not damaged during the procedures.

\FloatBarrier

\subsubsection*{Detailed Description of the Procedures}
Like for the anode construction, each GEM foil must be glued together with an inner ring at one extremity and an outer ring at the other. The inner ring is initially set in a housing on the annular flange at one end of the mold. The GEM chosen for installation is removed from its package and a dry-test is performed in order to verify the alignments and practice the procedure before the application of the glue.
The presence of at least three operators is required for completing both the dry-test and the actual gluing: one of them controls the rotation of the mold while the other two guide the foil to prevent the formation of folds on its surface.

The operation starts with the foil resting flat on the table, on top of a plastic sheet that is used to help direct its movement without touching the active area. It is then delicately lifted and arched toward the mold, which is held above the table surface by a support that allows its rotation. The reference hole on the foil and the one on the inner ring are aligned and a pin is put in place to help hold the foil in the proper position during future maneuvers. A second reference pin is placed at the other side of the GEM foil. The mold is then rotated and the foil gently guided until about half of its length. At this point, one margin of the foil is facing upward. The final part of the wrapping is performed by lifting the other end of the foil and guiding it against the mold, with the help of the plastic sheet underneath it.

If the dry-test is deemed satisfactory, the GEM is unwrapped and returned to its package. Glue is transferred to the inner ring, using a Mylar strip, after protecting the housing and the mold by sliding plastic material underneath it. The GEM is unpacked again and glue is applied on a 2$\,$mm wide strip along its length using a thin spatula by an expert technician. The layer of glue must be gauged to guarantee a tight bond while also preventing any spillage over the nearby active area.
The procedure followed during the dry-test is then repeated and the longitudinal overlap is closed. At this point a sheet of vacuum bag is cut from the roll, resized, wrapped around the GEM foil and fixed in place by Kapton tape. The HV tails of the GEM are protected by mounting caps on top of their housings, the peel-ply is wrapped around the mold and the vacuum bag is prepared.

After at least 10 hours, corresponding to the setting time of Araldite 2011 at the temperature of the clean room, the GEM is removed from the vacuum bag and the gluing of the outer ring can begin. This, as already described, is placed in position with glue applied to its inner surface. Then, the last vacuum bag is prepared and the glue left to set under pressure. The last step is the sealing of the slit of the outer ring using Araldite 103\footnote{Araldite 103 is a bicomponent epoxy adhesive. Unlike Araldite 2011 it can be exposed to the gas and so it is used for the sealing of the detector.} epoxy glue, which is done with the help of a syringe after closing the extremities of the slit using Kapton tape.

At this point, the sublayer is ready, connectivity and capacitance tests analogous to those described in section \ref{hvtest} are performed, to check that the GEM was not damaged during the process, and finally everything is wrapped in protective material until assembly.

\subsection{Anode Sublayer Construction}
\begin{figure}[h]
	\centering
	\includegraphics[width=\textwidth, keepaspectratio]{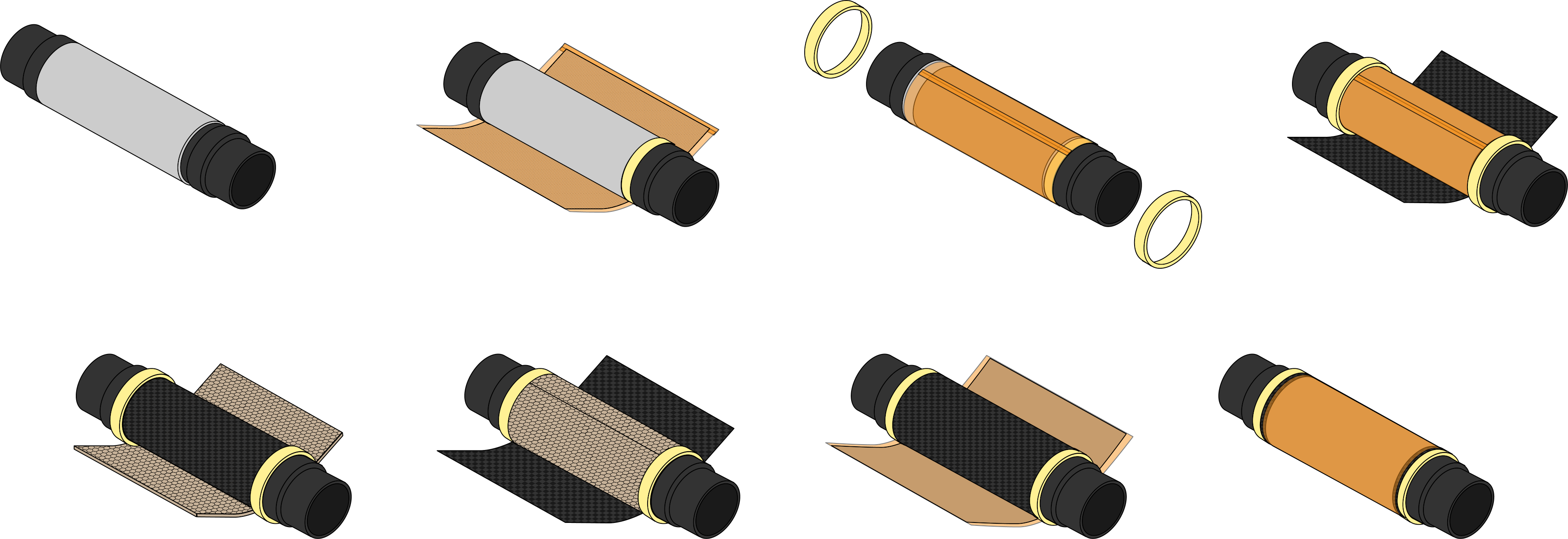}
	\caption[Main steps of the construction of the anode sublayer.]{Main steps of the construction of the anode sublayer.}
	\label{ikeanode}
\end{figure}

The construction of the anode sublayer begins with the gluing of the anode readout plane to an inner ring and two outer rings, in between which the cylindrical structural element will be realized. This structure consists of a Honeycomb core enclosed by two laminated carbon fiber meshes. The construction of the sublayer is completed by gluing on top of the outer carbon fiber skin the ground plane of the detector. The main steps of the construction of the anode sublayer are summarized in figure \ref{ikeanode}

I participated to the gluing of the anode readout plane and to the final gluing of the ground plane of the detector. The remaining steps of the construction were performed in Milan, at the company that produces the carbon fiber skins. The use of laminated carbon fiber meshes allows the application of the same construction techniques used for the other foils that compose the detector.

\FloatBarrier

\subsubsection*{Detailed Description of the Procedures}
The anode readout plane by design does not overlap, so a Kapton strip is glued at one of its sides to close the foil during the cylindrical gluing. As for the other sublayers, the anode is glued on a single inner ring but in this case the outer rings are two, one on each side. In between them, lies the carbon fiber and Honeycomb sandwich that constitutes the cylindrical support structure. The construction is completed by gluing the ground plane of the detector to the outer carbon fiber skin.

The Kapton strip for closing the anode readout foil is glued on the same vacuum table that is used for splicing together the active elements of the larger layers. The table has several sets of suction holes fitting the different dimensions of the different layers of the CGEM-IT, the ones not in use are patched with tape. The creation of the vacuum bag is analogous to the one performed on the molds but greatly simplified by the fact that everything happens on a flat surface.

\begin{figure}[h]
	\centering
	\includegraphics[width=.6\textwidth, keepaspectratio]{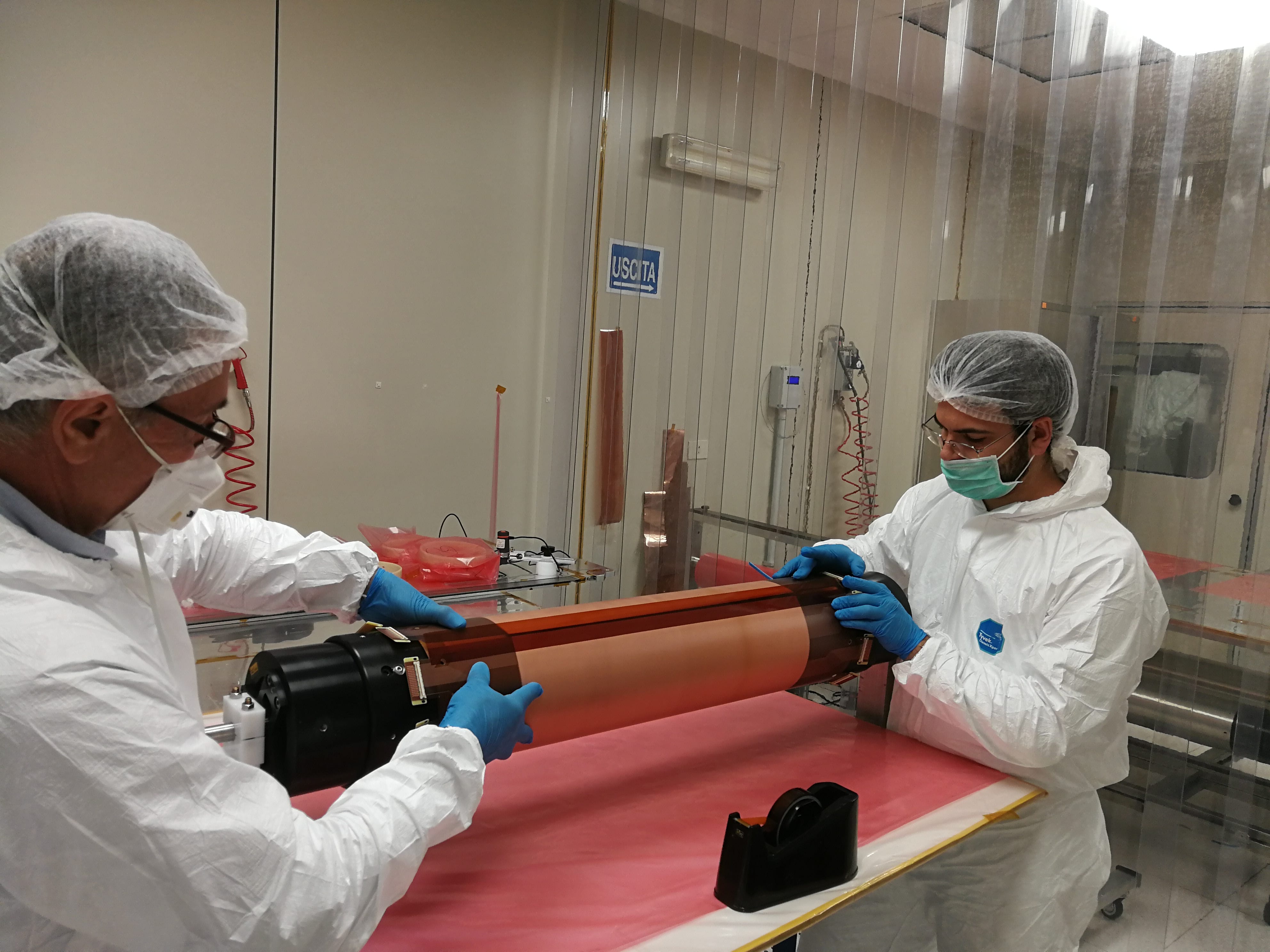}
	\caption[Photograph of the anode construction]{Photograph of a dry-test of the anode cylindrical gluing. The Kapton strip that is used to close the junction is clearly visible.}
	\label{anodeconstr}
\end{figure}

Once the strip is in place, glue can be applied to its free portion. At this point the glue is transferred to the inner ring, the anode foil wrapped around the mold housing the inner ring and the gluing completed with the preparation of a vacuum bag. Figure \ref{anodeconstr} depicts the dry-test that preceded the cylindrical gluing of the anode foil where the Kapton strip used for the junction is clearly visible.

\FloatBarrier

The remaining part of the anode construction, save for the final gluing of the ground plane, was not performed in Frascati but in Milan, at the Loson\footnote{Loson is a company specialized in the manufacturing of components realized with composite materials. The company produces the carbon fiber skins used in the construction of both L1 and L3. \url{https://loson.it/en/}} Headquarters. The mold with the anode readout was sealed, covered in protective materials and shipped to Milan.

The carbon fiber skins used for the construction must to be very thin, so to contain the radiation length of the detector. Because of this, they need to be lathed down to $\mathrm{60\,\upmu m}$ after being laminated in an autoclave. The resized carbon fiber skin can then be used in the construction employing the same techniques that are used for the other foil-like materials.

At this point the anode structure is built through a series of vacuum bag assisted gluings. The two outer rings are glued to the anode foil, the first skin is glued in between the two rings atop the anode, the Honeycomb layer on top of the first skin, and finally the last skin on top of the Honeycomb and above the extremities of the outer rings.

The anode sublayer, now almost complete, was then packed again and shipped to LNF for the gluing of the detector ground plane.
This gluing, being outside the detector, is the least delicate one, and was consequently done without the use of a vacuum bag. Peel-ply is tightly wrapped around the mold to provide some amount of pressure.

\section{VIM Assembly and Final Sealing}
The assembly of the detector is performed using the VIM (Vertical Insertion Machine), a custom CNC (Computerized Numerical Control) machine. The VIM raises and extracts the sublayers from their molds. The sublayers are extracted in order starting from the largest, the anode. Once two sublayers are glued together at one side the machine can rotate, allowing to perform the same operation on the other. When all the sublayers are glued together, the detector is sealed at both sides and removed from the machine. The final operations and tests are performed as the detector rests on a horizontal support.

\subsection*{Detailed Description of the Procedures}
Once all sublayers are ready, the final assembly can begin. The first mold inserted into the VIM is the one supporting the anode, which must be clamped in the chuck at the base of the machine.
During the assembly the VIM operates as a CNC machine; the movements are input manually by an operator and then executed automatically by the machine.
With the mold in place the catch flange of the VIM is lowered and fixed to the anode. 
As the flange is slowly raised, the anode is extracted from its mold. The empty mold is then removed from the VIM and substituted with the one of the largest GEM, G3. The flange, now holding the anode sublayer, is then lowered around the smaller G3 until the rings of anode and G3 are leveled. The relative position of anode and G3 is controlled many times during the whole procedure both by the operators and using a camera to check the alignment of pin holes present on the rings. Pins are used to connect the two layers and a Araldite 103 epoxy adhesive and silica microspheres mix is used to glue the two rings together. The quantity of microspheres to use has been gauged to penetrate predictably in the narrow space between the rings.

\begin{figure}[h]
	\centering
	\includegraphics[width=.6\textwidth, keepaspectratio]{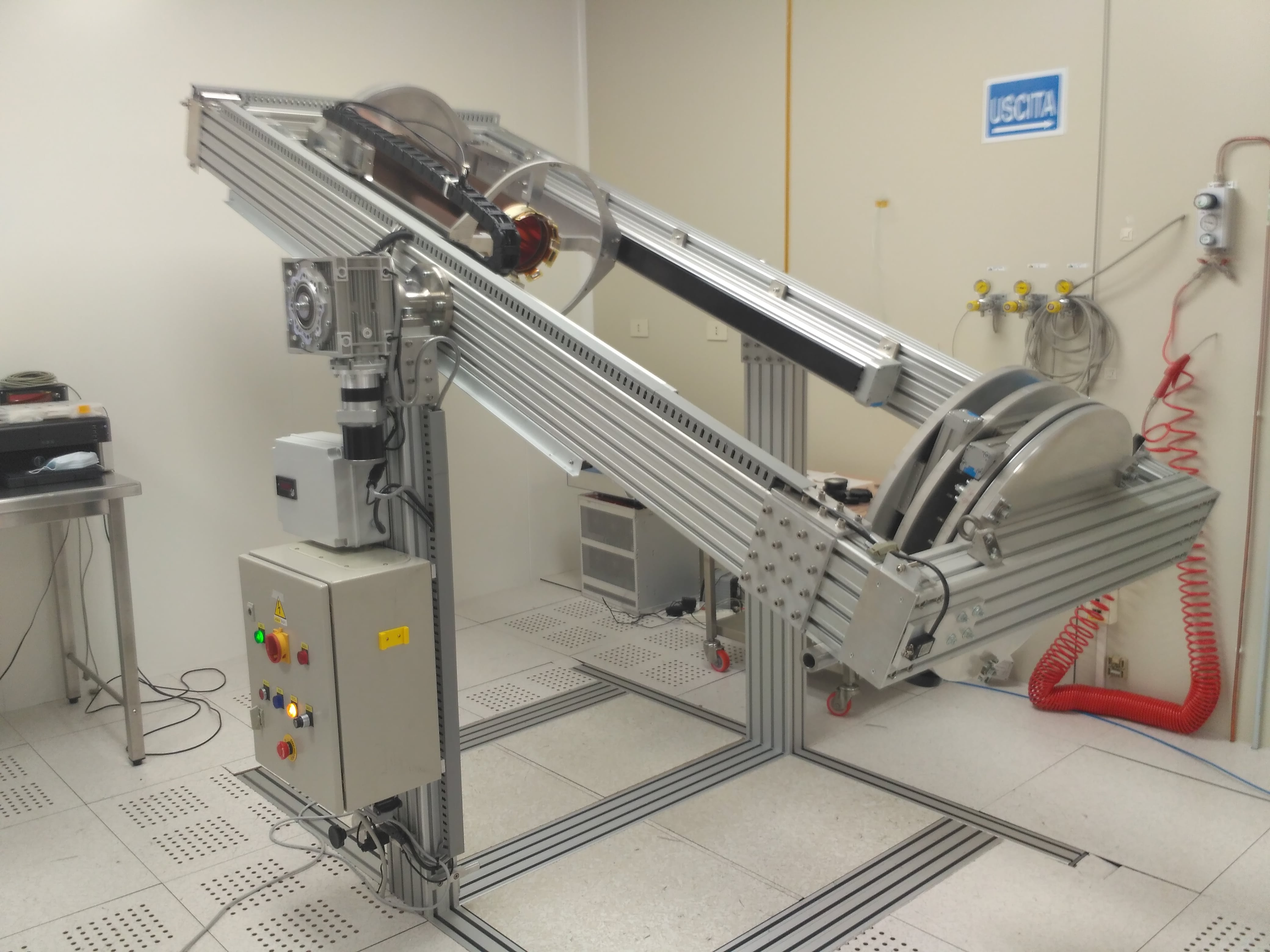}
	\caption[Photograph of the vertical insertion machine during a rotation]{Photograph of the vertical insertion machine during a rotation.}
	\label{rotation}
\end{figure}

After waiting the 22 hours necessary for the setting of Araldite 103 at the temperature of the clean room, the catch flange is raised once again and pulls G3, now glued to the anode, out of its mold. The mold is removed and the VIM slowly rotated by 180 degrees to allow access to the bottom rings as shown in figure \ref{rotation}. These rings are glued together with the same procedure used for the top ones and, after the glue has set, the VIM can be rotated again in its original position. This process is then repeated with all the remaining molds.

\FloatBarrier

Once the final cathode sublayer has been glued at both ends, its still missing second inner ring is installed. While the detector is still inside the VIM, both ends are sealed using Araldite 2011 and the gas connectors glued in place.
The detector is now solid enough to be manually removed from the VIM and placed horizontally on a crib where the pin holes are sealed.
At this point a pair of Permaglass service flanges are installed at both ends of the detector. These support the anode readout tails from below and protect both the HV distribution tails and the gas connectors. 
Another series of connectivity and capacitance tests is carried out with a multimeter to give a first assessment of the health status of the machine. The finished detector will later be connected to a HV power supply and turned on to verify if everything is working fine before the shipment. The procedure of this HV test is analogous to the one described in detail in section \ref{l1arrass}.

\section{Preparation for the Shipment}
Before being shipped, the most sensitive components of the detector must be protected. A vibration test was performed on a model of the detector at the university of Ferrara, to study the effect of the vibrations produced by the vehicles during transport. On the basis of this test, the padding materials used to protect the detector during shipment were chosen. The crate containing the detector was finally entrusted to a company specialized in international shipping of precious cargo.

The measures adopted allowed the successful delivery of the detector in Beijing, without any visible sign of damage. The detector was later subjected to a series of tests that confirmed the preliminary observations. These are described in section \ref{l1arrass}.

\subsubsection*{Detailed Description of the Procedures}
While inside the clean room the gas connectors must be closed with caps to prevent dust from entering the detector. The most exposed elements of the finished detector are its HV and anode readout tails, consisting of thin strips of Kapton substrate where the connections are etched.
These must be protected before shipping, as the bend induced by the glued rings make them rigid and therefore very fragile. The HV connections are protected by the service flanges while the anode readout tails are exposed, as they protrude from the outer rings. These were protected by a series of 3D printed ABS caps that were installed using the housings for front end electronics.
Two end caps were also 3D printed so to provide a surface for the padding material used in the shipping to slightly press against the detector from all sides and so prevent movement.

To study the stresses the detector would be subject to during the transport, vibration tests were performed at the university of Ferrara using a three axis vibrating machine. This instrument allows to reproduce the vibration profiles of the vehicles used for the transport, a truck and plane. After a mockup cylinder as large as L1 was used for these measurements, the combination of padding material that better damped the prevalent frequencies was adopted. The detector was then packed and entrusted to Montenovi\footnote{Montenovi is a company specialized in the shipping of artworks and other precious and/or fragile cargo. \url{http://montenovi.it/eng/}}, a company specialized in international shipping of precious and fragile goods.

    \chapter{L1 Commisioning at IHEP}
\label{commissioning}
This chapter reports a set of activities related to the implementation of L1 after its arrival in Beijing.

The chapter starts with the description of the decommissioning of the first-design L1 and the preparation of the setup for the new one. These activities have been part of my duties during my stay at IHEP.

Later, the operations performed to asses the good condition of the L1 after the shipment are presented together with the validation of its operational status before the installation in the cosmic ray setup. I took part only to the first part of these tasks since my fellowship ended few days after the arrival of L1.

The chapter also includes the description of the shift system that is in place for operating and monitoring the cosmic ray telescope setup from afar. These shifts allow to operate the detectors in safety and to continue collecting physics data while the access to the instrumentation is restricted by the ongoing global pandemic. The data are used to expand the knowledge of the detector and aid in the development of its software. I took part in these shifts since they were first required by the recalling of the team working at the CGEM-IT on site.

\section{Decommissioning of the first-design L1}
\label{decom}
Before the arrival of the L1 detector, it was necessary to extract from the cosmic ray telescope setup the first-design L1, which was partially damaged during shipment. The compromised detector would be later put to use, together with part of its electronics, for the realization of the noise measurements described in chapter \ref{noisetest} of this thesis.

\begin{figure}[h]
	\centering
	\includegraphics[width=.6\textwidth, keepaspectratio]{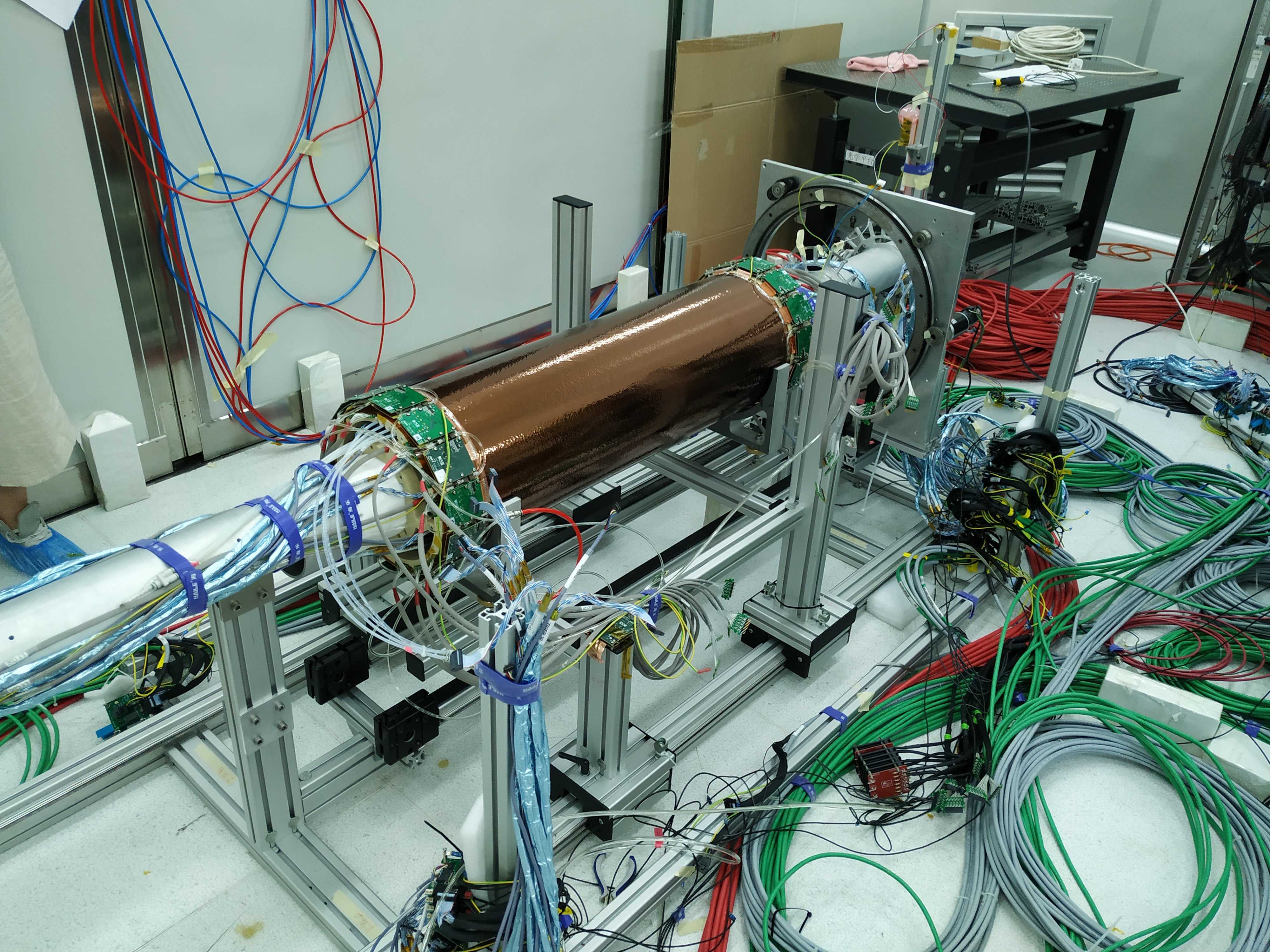}
	\caption[Photograph of the Cosmic Ray Telescope Setup.]{Photograph of the cosmic ray telescope setup taken before the extraction of L2.}
	\label{telescope}
\end{figure}

The cosmic ray telescope setup is represented in figure \ref{telescope}. This is arranged around the assembly support. Here the detectors are held from the inside with a stake. Rails built into the structure raising the stake from the ground serve to guide a sled that is used for inserting the detectors in the setup and extracting them. This sled can house a set of three cradles, which are sized according to the layers outer radii.

\FloatBarrier

At the time of the operations, the setup included the first-design L1 together with L2, as L3 had yet to be built. On the stake, the detectors are connected together through interconnection flanges, the same that will join them in their final installation inside BESIII. Each layer, when used in the cosmic ray telescope setup, is connected to: cooling system, high voltage (HV) power supply system, low voltage (LV) power supply system and readout chain, mimicking as much as possible the final configuration. Apart from the gas and cooling pipes, the other connections are realized through two sets of cables patched together using patch cards. The shorter cables, connected to the detector, are called Short Haul (SH) while the longer ones are called Long Haul (LH). This cabling configuration will allow to install inside BESIII the complete detector with the SH cables already connected. Once the three layers are assembled together the access to the connectors of the inner layers, where the cables have to be inserted, is in fact restricted by the outermost layer.

For the extraction of L2, the cooling system was disconnected from all the detectors.
To prevent the residual water in the pipes from leaking out and damaging the electronics, the inlets and outlets were plugged using caps.
The gas system was disconnected from L1 at this stage, as the detector would not be tested using the HV distribution.
Gas inlets and outlets were plugged as well to prevent dust from entering the detector.
The gas connections of L2 were instead left untouched, as part of the quality assurance protocol required to verify its status, through a HV test, during the extraction procedure.
The inner layer SH cables were disconnected from the patch cards and secured to the stake to prevent them from interfering during the extraction of L2.
LV and data SH cables of L2 were removed from the detector while the HV connections were momentarily left untouched in preparation of the test.

All layers are extracted in the same way. The cradle is slowly raised until it comes into contact with the outer surface of the detector and then it is fixed in position. At this point the detector is separated from the one inside it by removing the screws from the interconnection flanges.

After the separation, the status of L2 was verified through the LabVIEW interface handling the HV power distribution. To do this, all fields are turned on in order to verify the absence of abnormal current absorption that could indicate the presence of a short-circuit between different electrodes.

At the end of this test the SH-HV cables were disconnected from the patch cards and secured to the sled. The gas connection was interrupted and the gas inlet and outlet were connected together using a tube that was later secured to the sled. The layer was then carefully extracted pulling the sled and, once free from the stake, lifted on a second cradle secured to a table. Here it was reconnected to the gas system and to the HV power supply. After being flushed with the gas, it was once again tested through the HV distribution interface.

The first-design L1 was extracted in a similar manner but, in this case, all cables were removed and it was not tested using the HV system, as the noise measurements did not require it. The detector was then moved to a different location to perform the quality assurance of its readout electronics, described in chapter \ref{febtests}.

Apart from minor details regarding the cabling, the procedure used for inserting a layer in the setup retraces backwards the one for its extraction. In the particular case of the L2 insertion in the cosmic stand alone, to which I participated, more HV tests were performed at different stages of the insertion procedure.

\section{L1 Arrival and Health Assessment}
\label{l1arrass}
On its arrival at IHEP, the box containing L1 was brought inside the grey room housing the cosmic ray telescope setup. Once there, the box was opened, the protective padding was removed and L1 was placed on a cradle. The gas pipes used for closing the gas inlets and outlets during transport were removed and the layer was connected to the gas system to be tested for major leakages.

As none were found, the assessment of the detector status continued with a series of capacitance and resistance tests similar to the ones described in section \ref{hvtest}. These are performed first between sectors of the same GEM foil and then between sectors of GEMs facing each other. These measurements allow an internal diagnostic mapping of the detector: the capacitance values were found to be uniform and the resistance above the range of the instrument, as expected.

After these preliminary tests, the detector was connected to the HV power supply system in preparation of two HV tests. These are performed in sequence and follow the same procedure but using different voltage settings. Table \ref{hvalues} collects both the sub-nominal values used for the first test and the nominal operating values employed during the second one.

\begin{table}[h]
	\centering
	\begin{tabular}{llrr} 	
		Field    & Electrode    & Sub-nominal (V) & Nominal (V) \\ \hline
		Induction & G3B           & 800             & 1000        \\
		GEMs      & G1T, G2T, G3T & 200             & 270         \\
		Tranfer   & G1B, G2B      & 400             & 600         \\
		Drift     & Cathode       & 500             & 750   
	\end{tabular}
	\caption[Sub-nominal and Nominal voltage values for the HV tests.]{Sub-nominal and nominal voltage values used in the HV tests.}
	\label{hvalues}
\end{table}

First the voltage of each electrode is singularly increased in 50$\,$V steps up to the target value while the others are left at reference potential. This serves, as a confirmation of the preliminary checks, to rule out problems affecting the individual electrodes. The stability of the system as a whole is then verified in three consecutive steps. First the induction, transfer, and drift fields are turned on while the two faces of the GEM foils are instead kept at the same potential. Then, only the GEM fields are turned on and finally all fields are turned on together and left running for 12 hours.

Possible indicators of a problem that can be observed during the course of these tests are: the presence of frequent or large discharges and abnormally large current absorption. Discharges can be caused by a multitude of factors: impurities in the gas mixture used, problems in the HV connections, humid operating environment and others. A current absorption is caused either by a short-circuit affecting the GEMs, which somehow eluded the preliminary checks, or a contact somewhere else in the HV connections.

To determine the origin of the current absorption the cables of the affected sector are swapped with those of a healthy one. If the problem now appears in correspondence to the HV channel related to the new cable, then the problem is due to the detector. Vice versa the issue must be related to the previous cable or to the patch card it is connected to.

Shorts between the top and bottom face of the same GEM are rare and they can be, sometimes, cured through HV conditioning.
Current absorptions or discharges between two different GEMs are also very infrequent but they may point to some mechanical alteration of the internal structure and so they must be carefully evaluated.

I participated to the preliminary capacitance and resistance tests and to the subsequent HV cabling operations, the HV test were instead performed by my supervisor. L1 suffered no damage from the shipment and all sectors were found to be capable of operating at nominal values for prolonged periods of time.

\section{On-detector Electronics Installation}
\label{febinst}
Before its installation in the cosmic ray telescope setup, L1 had to be equipped with its front end boards (FEBs). These are described in detail in chapter \ref{febtests}. Each FEB consists of two parts: FE-1 and FE-2. FE-1 houses the TIGER chips, the copper pad for the soldering of the ground connection and the heatsink. FE-2 is mounted atop the first, houses connectors for LV and data SH cables and covers the delicate bonds of the chips. The two parts are installed separately and so the chip bonds are vulnerable during the whole operation.

\begin{figure}[h]
	\centering
	\includegraphics[width=.9\textwidth, keepaspectratio]{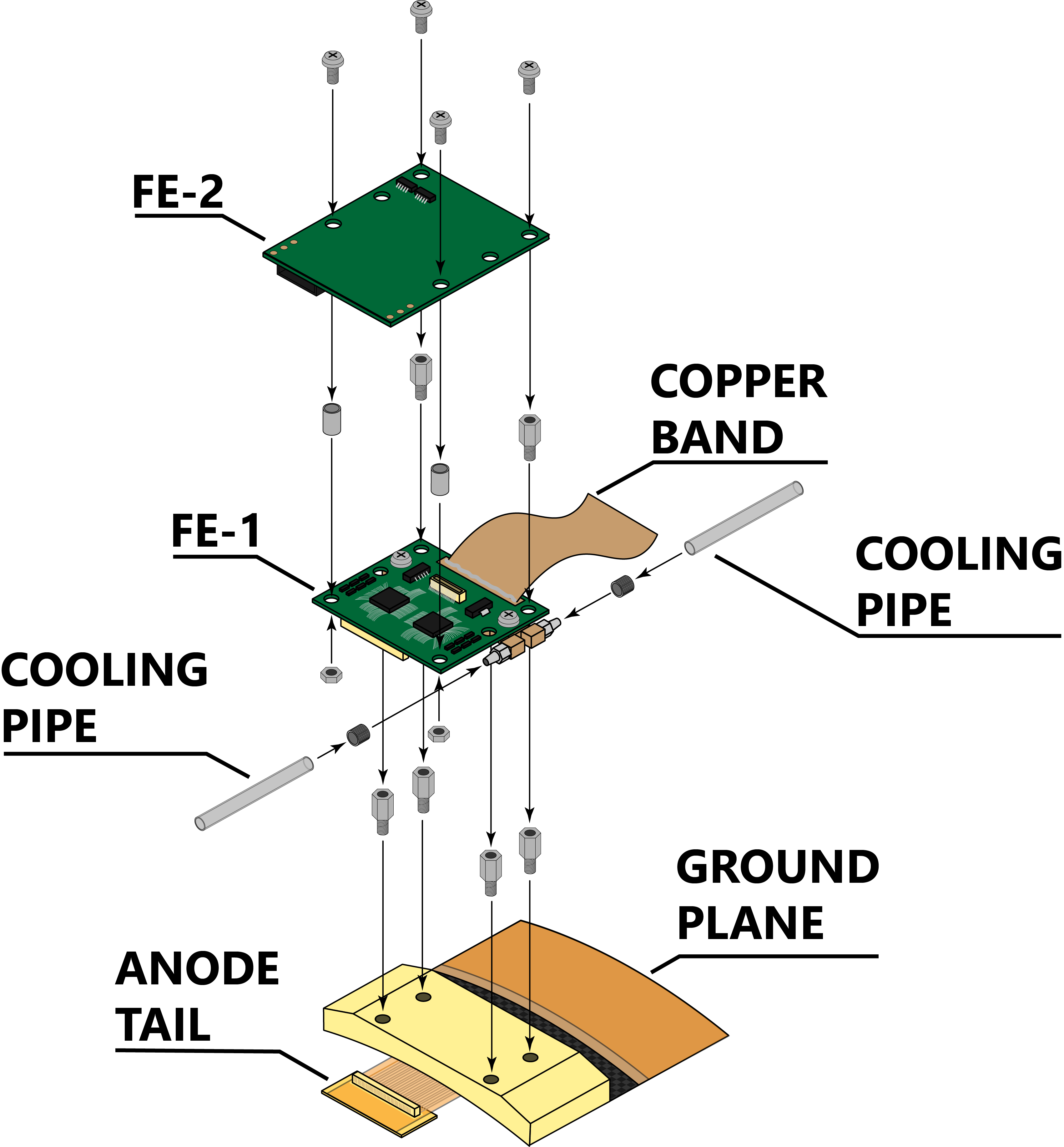}
	\caption[Schematic drawing of the installation procedure.]{Schematic drawing of the installation procedure.}
	\label{febmount}
\end{figure}

Before beginning the installation, it is necessary to touch some conductor and release any static charge accumulated. This serves to prevent discharges that could damage the electronics while handling the FEBs. Before FE-1 can be mounted on the anode ring, a copper band that allows its connection to the ground plane must be soldered to it. As the bonds are exposed, the soldering must be performed at low temperature and using good quality solder in order to avoid spitting.
Once the copper band is in place, FE-1 is installed on the outer anode rings using spacers and the anode readout tail is attached to it from below.
The pipes of the cooling system, which connect the FEBs in series, are also installed before mounting FE-2, as this eases the closing of the nut keeping them in place. Finally, FE-2 is installed atop FE-1 and fixed in position using screws. The last step of the installation process is the soldering of the copper band, already attached to the FEB, to the ground plane of the detector. A schematic drawing of the installation procedure is provided in figure \ref{febmount}.

At this point, the FEBs are ready to be cabled and connected to the cooling system to be tested.
The validation of FEBs and TIGERs is performed by measuring the noise level of their channels. Despite being part of the commisioning of a CGEM detector it is not described here as it is the subject of chapter \ref{febtests} of this thesis.

The installation of the FEBs requires a well coordinated team, as most of the operations I described require more than two hands. The most delicate steps are performed always by the same experts in order to reduce the risk of damaging the detector or the electronics.

\section{Remote Operation and Monitorning of the Cosmic Ray Telescope Setup}
Due to the eruption of the COVID-19 pandemic, the team of researchers working on site at the CGEM-IT was recalled, the last of its members leaving China in February 2020. Before their departure they rigged the cosmic ray telescope setup, now including L1 and L2, to be operated and monitored from afar.

All the data coming from the systems necessary to safely operate the detectors were then condensed in a user-friendly Grafana-based\footnote{Grafana is an analytics and interactive visualization web application. \url{https://grafana.com/}} interface, called \textit{shifter board} \cite{cpad}. This allows anyone with basic training to monitor the detectors.
Higher level interfaces, the same used when on site, remain at the disposal of the experts thanks to the possibility to remotely control all of the computers in the setup through the internet.

Each day has been divided in shifts that are covered on a voluntary basis by the members of the BESIII Italian collaboration. For each shift there are two experts and a shift leader. The experts are in charge of turning on and off the detectors, operating the data acquisition and solving the occasional problems affecting the setup. The shift leader coordinates the work of the experts relaying the information reported by the shifter. The positions of expert and shift leader require deep knowledge of the setup and its interconnected systems, therefore they can only be covered by the small group of researchers possessing the required expertise.

\begin{sidewaysfigure}[p]
	\centering
	\includegraphics[width=\textheight, keepaspectratio]{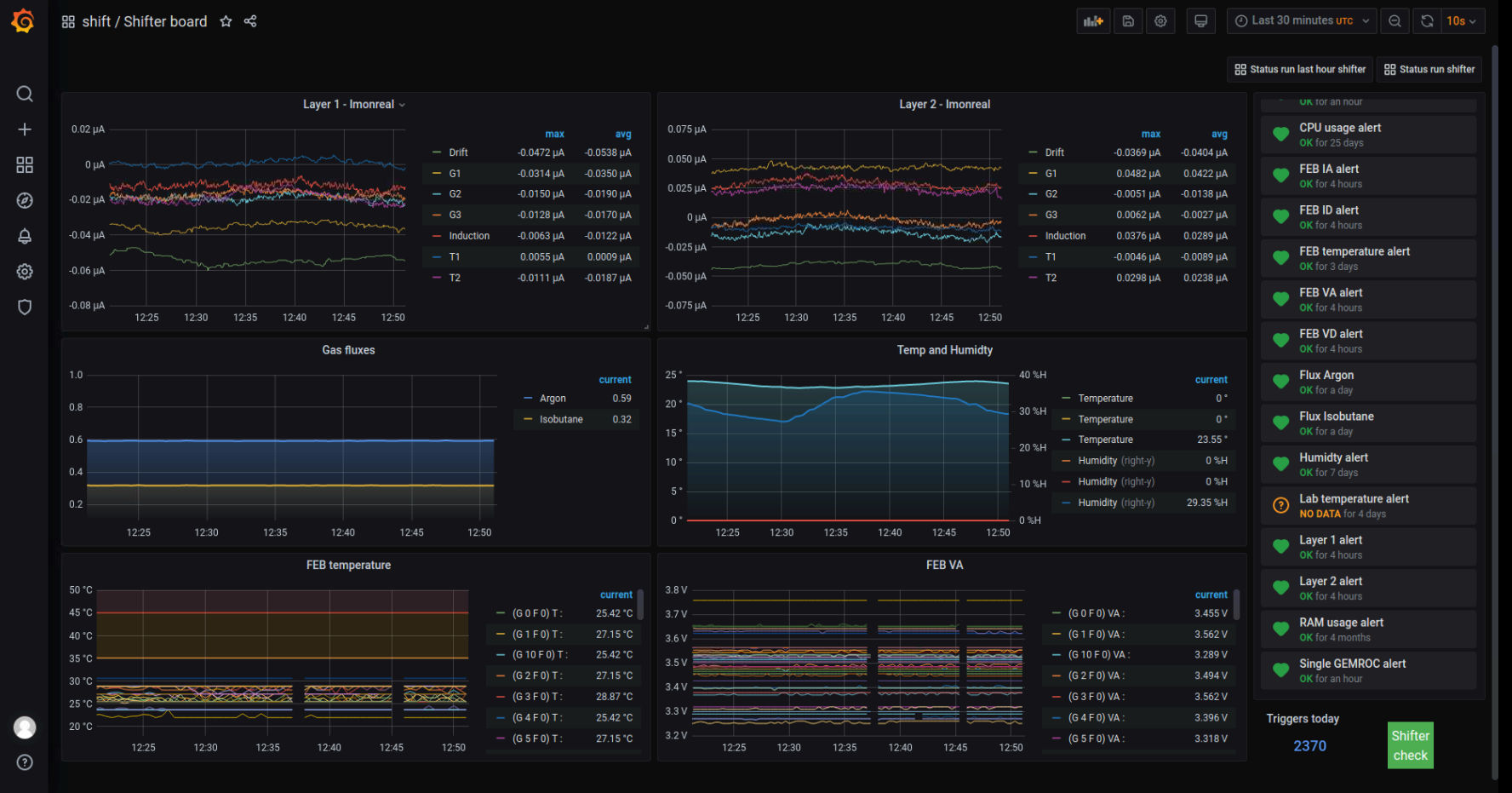}
	\caption[Screenshot of the shifter board main tab during a shift.]{Screenshot of the shifter board main tab during a shift.}
	\label{shiftboard}
\end{sidewaysfigure}

\begin{figure}[h]
	\centering
	\includegraphics[width=\textwidth, keepaspectratio]{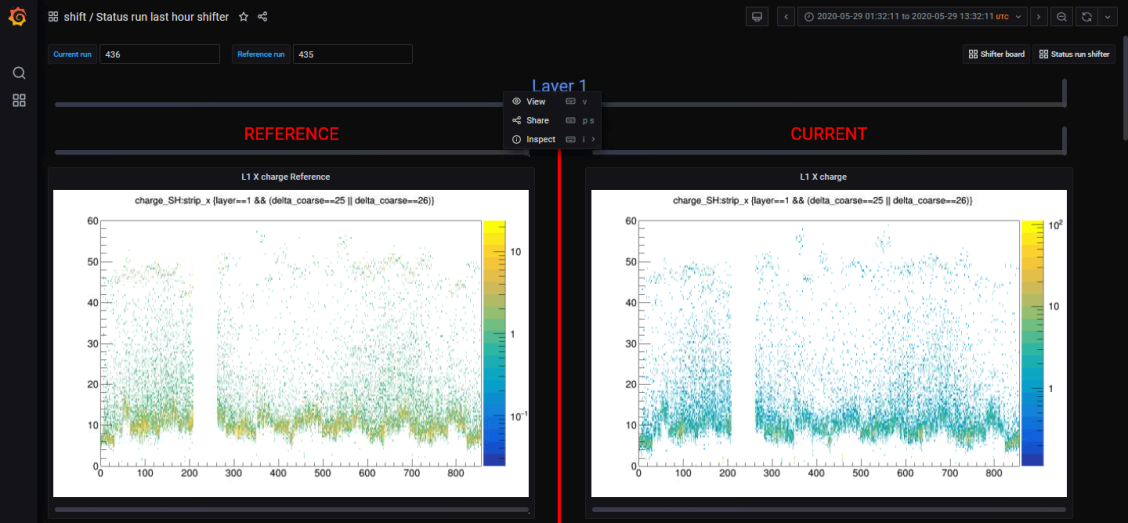}
	\caption[Screenshot of the shifter board tab allowing the comparison with a reference run.]{Screenshot of the shifter board tab allowing the comparison with a reference run.}
	\label{statusrun}
\end{figure}

During the course of a shift, the shifter is asked to monitor the setup through the shifter board, shown in figure \ref{shiftboard}, report any irregularities to the shift leader and note down the values of relevant parameters at regular intervals.
The interface shows plots of the following quantities: currents absorbed by the different electrodes of both detectors; relative mass flows of argon and isobuthane; environmental temperature and humidity; FEB temperatures and voltages. Together with these, the board provides
the number of triggers occurred during the day and a series of alarms regarding the status of acquisition and instrumentation. Pressure and flux of the water circulating in the cooling circuit can be monitored through a webcam pointed at the relative indicators.
The last check required involves the online data quality, accessible through a separate tab of the shifter board, shown in figure$\,$\ref{statusrun}. This tab reports a series of graphs providing an overview of the data collected during the current run and a chosen reference run. The shifter must compare the graphs and note down if the two present major differences.

\FloatBarrier

Monitoring the currents absorbed by the electrodes allows to evaluate the stability of the detectors. Oscillatory behavior of the currents in time is determined by the cycle of the grey room hosting the setup, which affects ambient temperature and humidity. Frequent or large discharges, above 100 nA, have to be noted down and notified to the shift leader. In some occasions the current absorbed by one of the electrodes has been seen to progressively increase. This may be due to particular humidity conditions, which may be affecting the HV connections, but the phenomenon disappeared by itself and was not possible to investigate it further.

The relative mass flows of the gases, measured as the amplitudes of the signal from the mass flowmeter, must remain at or near their nominal values: 0.32$\,$V for the argon and 0.6$\,$V for the isobuthane. These correspond to fluxes of 3.6$\,$l/h for argon and 0.4$\,$l/h for isobuthane. Alterations to the gas mixture indicate that one of the gas bottles may be almost empty. The rate at which the gas is consumed is known and so the swapping of the bottles is organized in advance and performed thanks to the help of the Chinese researchers working at IHEP.

The temperature and humidity readings consent to identify malfunctions of the grey room hosting the setup. This allows to turn off the detectors, before they become too unstable due to the increase in humidity affecting the HV connections, or the GEMROCs, which may overheat due to the increase in temperature.

The FEBs have to operate below 35$\,$°C with values above 45$\,$°C triggering an alarm on the shifter board. The voltages must instead remain within 3.2$\,$V and 3.7$\,$V. A rise in the temperature of multiple FEBs may indicate the presence of an issue with the cooling system.

Pressure and flux of the distilled water are monitored to see when the chiller tanks needs to be refilled. The cooling system is a closed circuit but a small leak near the UV lamp, used for sterilizing the water and preventing the formation of algae, requires periodic refilling.
The pressure must be between 0.200$\,$MPa and 0.300$\,$MPa while the flux must remain between 4.2$\,$l/s and 4.9$\,$l/s.
The refilling is done by the IHEP personnel and, as for the swapping of the gas bottles, it is scheduled in advance using known consumption rates. 
The temperature of the water is also displayed but it generally remains stable at 20$\,$°C. Recently this has been observed to raise up to 24$\,$°C during the course of a day, causing an increase in the FEBs temperature. This was due to the clogging of the air intakes of the chiller and it was solved thanks to the personnel working at IHEP.

The comparison plots are used to spot issues in the data acquisition. Unresponsive GEMROCs or FEBs create blanks in the current-run plots that are not matched in the reference ones. These problems can be fixed by the experts but may require the restart of the data acquisition.

The data acquired through the setup is used for expanding the knowledge of the detectors and for the development of the dedicated physics software. The shift system allows to continue advancing the project even under the current restrictions.

    \chapter{Quality Assessment of the TIGER Front End Boards}
\label{febtests}
This chapter reports the tests I conducted, with the assistance of the team of researchers working on the integration between the detector and its custom readout electronics, during my stay at IHEP.
The purpose of the tests was to verify the functioning of two batches of Front End Boards (FEBs) to be installed on the first and third layers of the detector.

The readout chain and the software employed for the tests are here described in detail.
The same equipment was also used to perform the noise studies described in chapter \ref{noisetest} of this thesis, although with minor variations driven by the different conditions in which the two measurements took place.

\section{Description of the Readout Chain}
\label{readout}
The FEBs housing the TIGER chips are designed to be mounted radially on the outer anode rings, from where they can be directly connected to the anode readout tails and to the detector ground.

Data and Low Voltage Patch Cards (DLVPCs) \cite{cpad} bridge together the Short Haul (SH) shielded cables connected to the FEBs on the detector and the Long Haul (LH) cables connected to the GEMROC, which are separate for data and Low Voltage distribution.

The GEMROC modules are connected to a low voltage power supply system by two sets of cables that carry the power needed by the GEMROC module itself and the four FEBs it controls.
Different power supply configurations are used according to the amount of FEBs and GEMROCs in the chain.
The system is always constituted by a combination of a CAEN SY5527 and/or SY4527LC mainframes and the necessary number of modules LV A2519 and A2517, which are used to power GEMROCs and FEBs respectively.
Each LV A2519 module can supply up to 8 GEMROCs while each A2517 module can supply 32 FEBs.

GEMROCs and power supply are controlled through a computer on the same Local Area Network (LAN). The LAN is set up through Ethernet connections and a switch. On the computer, the Graphical User Frontend Interface (GUFI) \cite{cpad}, an  internally developed suite of Python scripts that allows to handle the configuration of the FEBs and data acquisition, and the CAEN GECO interface, which is used to control the power supply system, are installed. GEMROCs communicate with GUFI using the UDP protocol while the GECO interface utilizes TCP/IP protocols.

\begin{figure}[h]
	\centering
	\includegraphics[width=\textwidth, keepaspectratio]{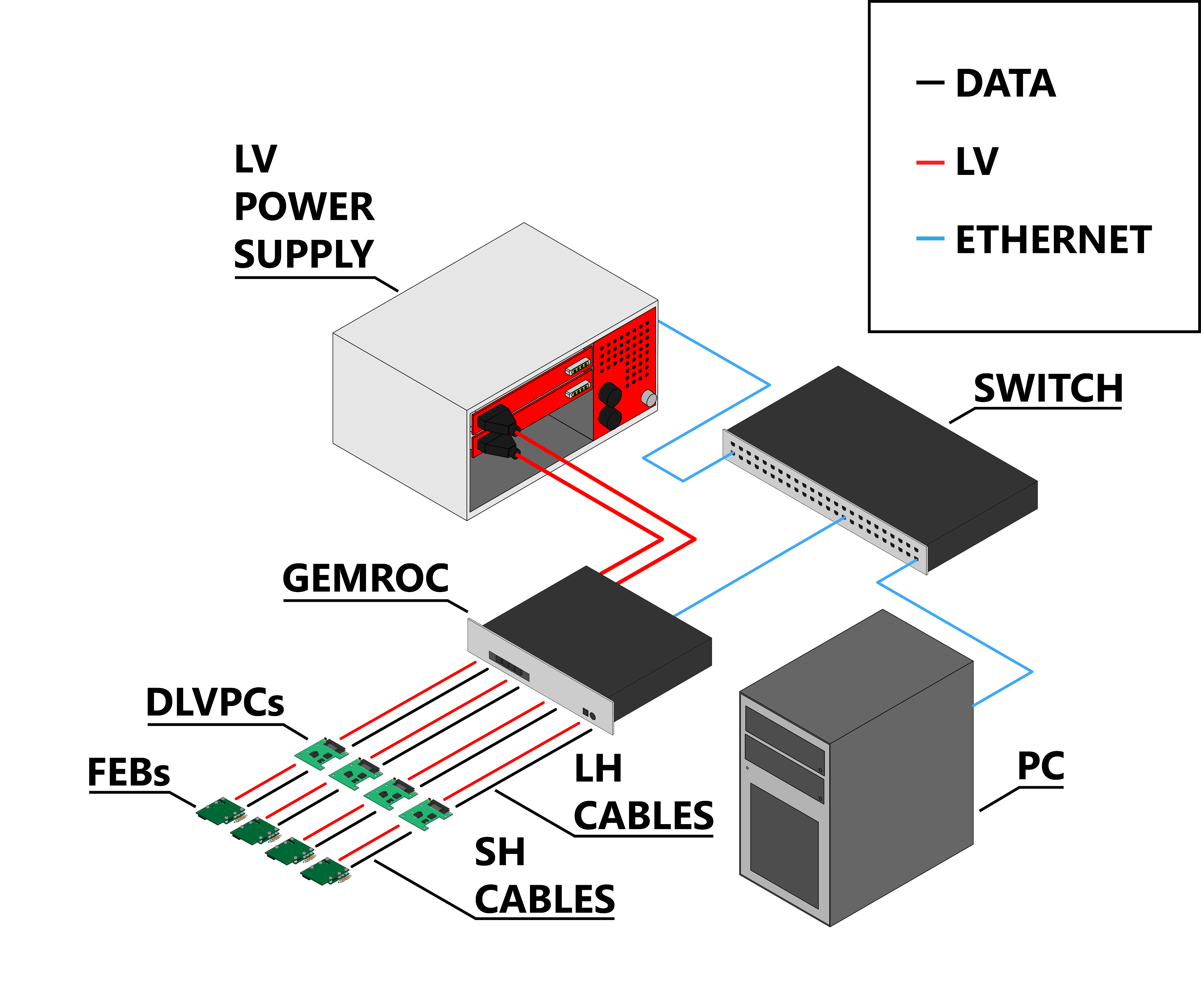}
	\caption[Simplified schematic representation of the readout chain]{Simplified Schematic representation of the readout chain including four FEBs. Cooling system and ground connections are not represented.}
	\label{readoutchain}
\end{figure}

A simplified schematic representation of a readout chain including four FEBs is provided in figure \ref{readoutchain}. Cooling system and ground connections are not represented.

\FloatBarrier

\subsection{Front End Electronics}
\begin{figure}[h]
	\centering
	\includegraphics[width=.7\textwidth, keepaspectratio]{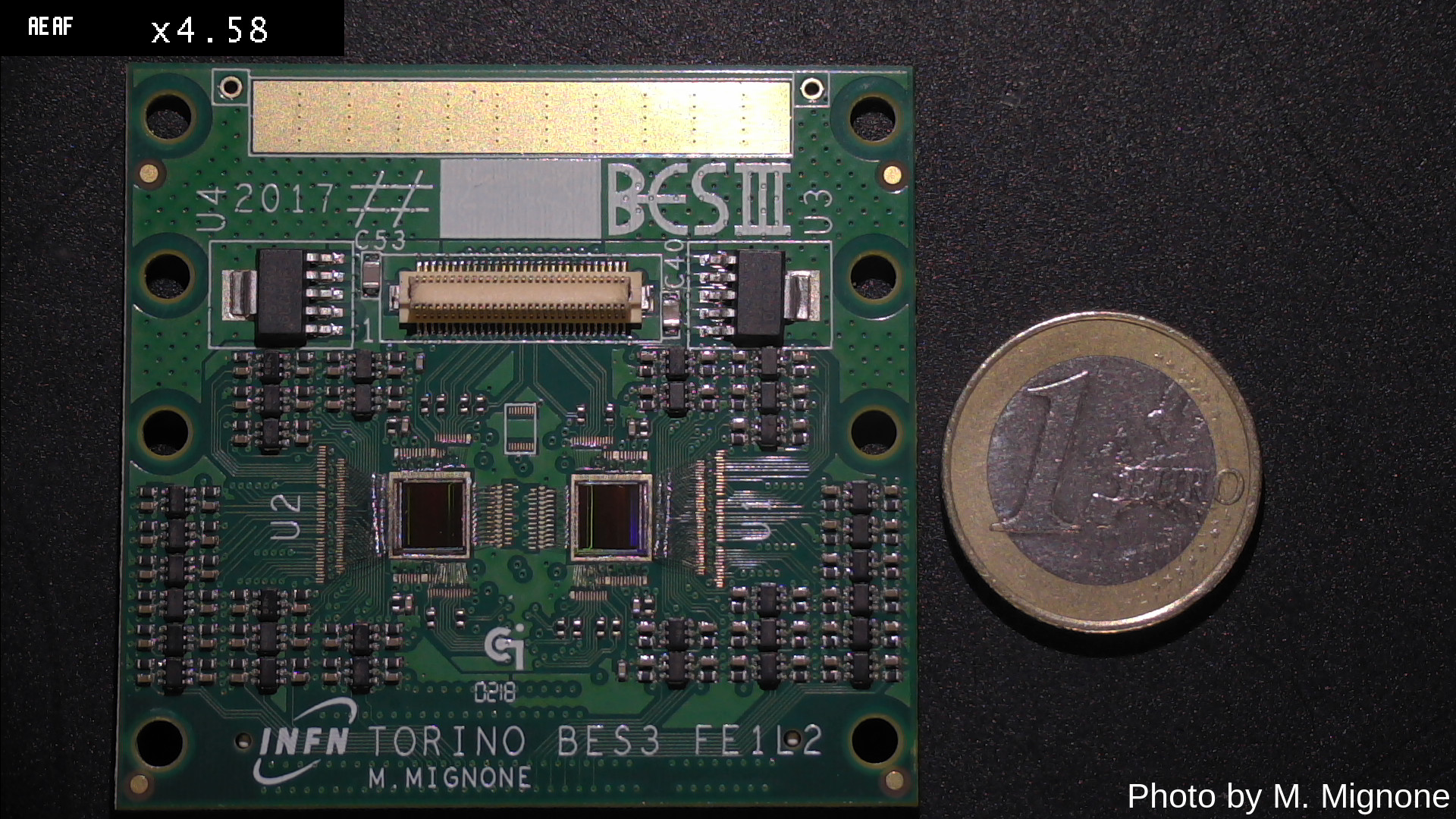}
	\caption[Photograph of the FEB part housing the TIGER chips]{Photograph of one of the FE-1 boards designed for L2 \cite{fabio}.}
	\label{febphoto}
\end{figure}

The front end electronic boards designed for the CGEM-IT detector are realized in two parts housing different components as described in section$\,$\ref{febinst}. Figure$\,$\ref{febphoto} shows the board housing the TIGER chips.
The FEBs of the first two layers have a different geometric configuration with respect to those of the third. L3 FEBs have a more compact design to fit within the tighter volume determined by the MDC and the surrounding instrumentation.

\FloatBarrier

The FEBs are connected to the anode readout tail from below and to the detector ground through a copper band soldered to FE-1.
The metal spacers keeping the two parts separated connect together their ground references.
LV and Data SH cables attached to FE-2 provide the necessary power and allow the FEB to exchange information with the GEMROCs.
As the FEBs include both analog and digital circuitry, a single power cable carries two different voltages.

\begin{figure}[h]
	\centering
	\includegraphics[width=\textwidth, keepaspectratio]{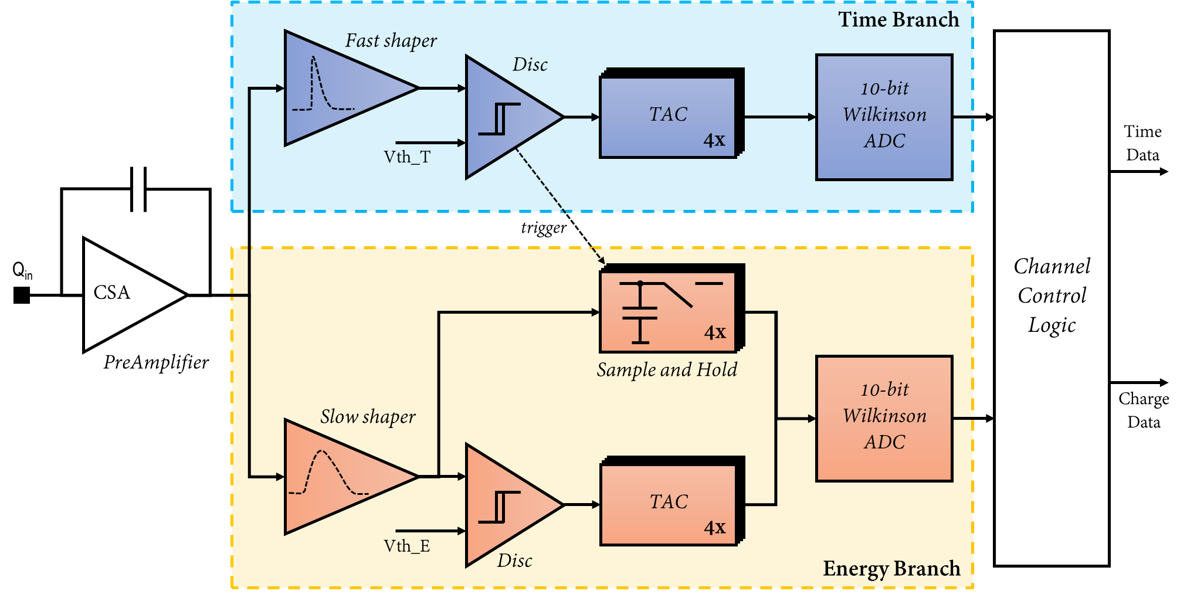}
	\caption[Schematic representation of the channel architecture,]{Schematic representation of the two electronic branches constituting each TIGER channel \cite{fabio}.}
	\label{archichannel}
\end{figure}

A single TIGER chip has 64 channels. The signals from the readout plane are amplified, duplicated and then fed to two different shapers, as shown in figure \ref{archichannel}. From this point forward the signals follow separate branches named T and E, dedicated to time and charge measurements respectively. The signal sent down the T branch has a much shorter rising time with respect to its counterpart. This serves to limit the error introduced on the time measurement by the jitter. The signal sent down the E branch, instead, presents a flatter peak that is more suitable to accurate charge integration.

\FloatBarrier

The water cooling the FEBs circulates in a copper heatsink mounted below the TIGERs. Heatsinks of neighboring FEBs are connected in series through polyurethane radiation resistant pipes. The distribution lines departing from the chiller are dimensioned to keep the serviced group of FEBs within the operative range of temperatures.

\subsection{Back End Electronics}
The back end electronics are constituted by DLVPCs and GEMROC modules. DLVPCs serve as a bridge between the SH and LH cables connecting FEBs and GEMROCs. Multiple DLVPCs can be assembled in stacks using similar metallic spacers that are used for the FEBs. As for the FEBs, this puts in contact their ground references. If more than a stack is present in the chain, their grounds are connected together through cables. The ground level for the whole system is then provided connecting one of them to the grounding strap.

A single GEMROC module handles 4 FEBs. The control unit of the modules is based on a commercially available FPGA card, which is connected to a custom made interface board.
GEMROCs organize the flux of data collected by the TIGER chips towards the computer used for the acquisition.
In addition, it distributes analog and digital voltages, sends the configuration signals for the TIGER chips and monitors their operating parameters. These include: two voltage levels for the analog and digital components, the currents absorbed by each and the temperature value provided by the FEBs internal sensors. 

\section{GUFI Interface}
The GUFI interface is a custom-made suite of Python scripts that allow to pilot the GEMROCs and, through them, to configure and monitor the FEBs, and the TIGER chips they house. In addition, the interface provides a set of debugging, calibration and preliminary analysis tools that are used to initiate and control the data acquisition.

The GUFI script, used to collect data for the tests described in this chapter and for the noise studies in chapter \ref{noisetest}, allows to perform a series of operations collectively called \textit{threshold scan} \cite{fabio}. The final result of this measurement is the assessment of the noise level of a channel. This information, together with the position of the baseline allows to set the threshold used for the acquisition of physics data.

The scan can also be used to evaluate the condition of the TIGERs, that is the presence of dead or damaged channels, through the analysis of the plots it produces.


\subsection{Threshold Scan}
\label{thresholdscan}
\begin{figure}[h]
	\centering
	\includegraphics[width=.8\textwidth, keepaspectratio]{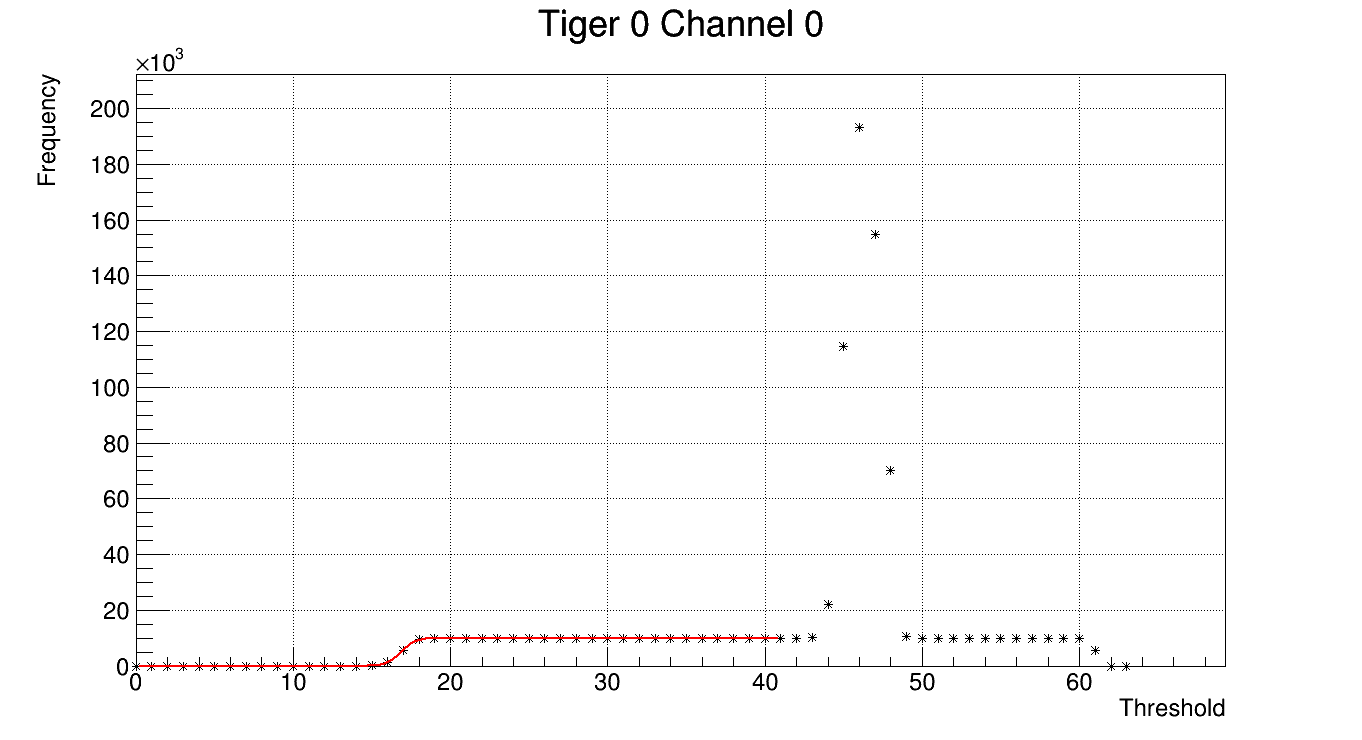}
	\caption[Example of plot obtained through a threshold scan]{Example of graph obtained by plotting the data collected during a threshold scan of a non pathological channel.}
	\label{goodts}
\end{figure}

To perform a threshold scan, a user configurable train of identical signals, called Test Pulse (TP), is generated and fed as input to the shaper of the chosen channels.
This allows to verify the noise condition of the channels through their response to a known signal. A specific TP amplitude is chosen for each TIGER through a configuration procedure described in section\ref{config}. Then, the discriminator threshold is progressively lowered, 63 times, in 0.5$\,$fC steps every tenth of a second from an initial value, depending on the configuration of each TIGER, above the TP amplitude. The number of threshold crossings for each step is recorded for each channel. Plotting these counts with respect to the threshold step of the measurement allows to obtain a graph like the one in figure \ref{goodts}.

\FloatBarrier

In these graphs, the lower value of the threshold step represents the higher value of the threshold setting. The counts are initially zero because the discriminator threshold is set above the TP amplitude. As the threshold approaches the peak of the pulse, the number of signals crossing the threshold progressively grows due to the distortion of the signal caused by the electronic noise. After the threshold surpasses the peak amplitude, the number of counts settles on a value that depends on the set TP frequency. 
Finally, when the the baseline is reached, the number of counts quickly rises and in some cases saturates the measuring capabilities of the chip, producing a tall peak like the one observable in figure \ref{goodts} or a truncated version of the same. The TP is symmetric with respect to the reference potential, therefore an inverted version of the signal that creates the first ramp causes the fall to the right of the baseline.
 
The ramp produced by the increase and following stabilization of the number of counts at the crossing of the TP amplitude can be fitted with an error function. The standard deviation obtained through this fit represents the noise level of the channel.

Figure \ref{goodts} is an example of the behavior displayed by a good, low noise channel. The absence of one or both structures, as well as their eventual deformation are indicators of a problem in the channel being examined. Examples of this pathological behaviors are presented in section \ref{monsters}.

The threshold scans on the T and E branches are performed separately. GUFI allows to perform the threshold scan automatically on groups of channels and save the resulting data on file. In this case, the scans are performed sequentially, a single channel at a time, according to the order defined by the cabling of FEBs and GEMROCs. The program also allows continuous data acquisition alternating between the two measurement branches.

GUFI can perform the fit and extract the noise level of the channels but, foreseeing the noise studies of chapter \ref{noisetest}, it was decided to develop a separate standalone algorithm to fit the raw data. This allowed to optimize the simple fitting algorithm for the small sample of known channels that have been repeatedly analyzed. The details of this algorithm I developed are provided in section \ref{algo}.

\subsection{TP configuration}
\label{config}
For a threshold scan to produce meaningful results, it is first necessary to center the TP in the acquisition window for all the TIGERs present in the readout chain.
The TP is well centered when both the structures produced while crossing the TP amplitude and the baseline are visible in the graphs shown by the GUFI, which are similar to the ones presented in figure \ref{goodts}.

A first threshold scan in launched on a few channels for each TIGER to asses the initial condition.
The graphs produced are used to inform changes to the TP configuration parameters, like the coarse and fine amplitude settings, together with the initial discriminator threshold.
Equal adjustments of the same parameters have shown to produce different effects for different chips, requiring the adoption of a trial and error approach.
After the adjustment of the parameters, a new threshold scan is launched on the same sample of channels and the plots are checked again. The process is repeated until a satisfactory centering of the TP within the acquisition windows is achieved. A final threshold scan is used to confirm the validity of the centering for all the channels to be tested.

The adjustments that grant the best results are different for each chip and so the procedure must be repeated every time the configuration of the readout chain is modified. For the purpose of evaluating the condition of the TIGERs a rough centering is sufficient as the graphs are double checked manually. For the automatized measurements performed for the noise tests of chapter \ref{noisetest} a finer configuration is required, as it grants more homogeneous plots that can be more easily analyzed through a simpler and faster algorithm.

\section{Threshold Scan Analysis Algorithm}
\label{algo}
The algorithm I developed aims to extract the noise level of a channel by fitting the graphs produced using the raw data collected in a threshold scan. For the noise test in chapter \ref{noisetest} the monitored channels were always the same, since the FEBs used were not replaced during the measurements. The code has been therefore optimized to provide quick and reliable results on large data samples of known channels.

For the tests described in this chapter, the algorithm was tuned to be more lenient and adapt to a wider range of situations. The graphs produced by the program were then inspected manually one by one while the output tables were used to double-check the observations.

\begin{figure}[h]
	\centering
	\includegraphics[width=.8\textwidth, keepaspectratio]{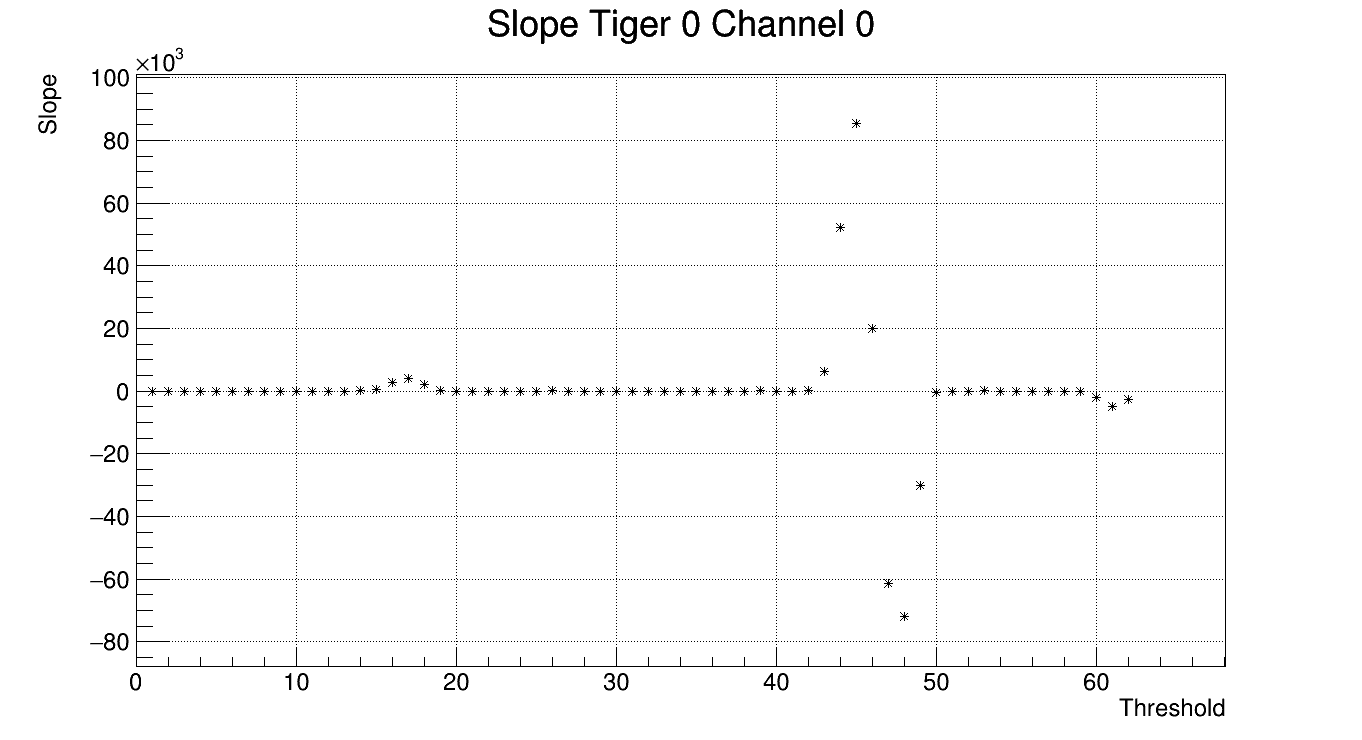}
	\caption[Slope graph]{Slope graph relative to the plot in figure \ref{goodts}. This is the plot used for the determination of the regions occupied by each structure.}
	\label{slope}
\end{figure}

The algorithm is written in C++ as a ROOT macro and utilizes the classes offered by the ROOT framework for both plotting and fitting. The first operation performed for each channel is the reconstruction of the plot described in section \ref{thresholdscan} from the raw data acquired by GUFI during the threshold scan. The program then produces a slope graph by averaging the number of counts at the threshold steps immediately before and after the one for which the slope is being computed. The slope graph corresponding to the plot in figure \ref{goodts} is shown in figure \ref{slope}.

\FloatBarrier

The slope graphs are used to determine, through cascading conditional statements, the borders of the regions where TP induced ramp, plateau and baseline peak are located. Performing this operation on the slope graph, instead of on the frequency one, has proven to be more reliable, as it allows to avoid mistakes introduced by fluctuations in the number of counts. Moreover, it allows to define non-adaptive filters that do not require to factor in some of the differences observed between the threshold scans of different channels.
The intervals obtained define where the fit should occur and they are used to provide a first identification of problematic channels. 

The recognition of pathological channels occurs through a sequence of conditional statements. These check the number of thresholds for which zero counts have been recorded and if all three structures have been identified. In case too many points are 0 or if one of the structures is missing, the channel is flagged and the fit is not performed.

The function used to fit the TP ramp is:
\begin{equation*} \label{fitfunction}
	f(x)=\dfrac{N}{2}\left(1 + erf\left(\dfrac{x-\mu}{\sigma \sqrt{2}}\right)\right)
\end{equation*}
where:
\begin{equation*}
	erf(x)=\dfrac{2}{\sqrt{\pi}}\int_{0}^{x}e^{-t^2}dt
\end{equation*}
is the standard C++ error function. This represents the cumulative distribution function of a normal distribution centered in $\mathrm{\mu}$ with standard deviation $\sigma$ and normalization N.

The borders of the TP ramp region and the average height of the plateau are used to compute a first rough estimate of the error function parameters to prime the fit: the initial mean value is set to the middle point of the TP ramp region, the standard deviation to its half-width and the height of the plateau is used for the normalization.

If the fit converges, its parameters are evaluated against known constraints. The normalization must fall within a narrow band around the plateau height and the mean must be contained within the ramp region. This serves to prevent the results from a converging but bad fit from being included in the final output.

During operation, these filters have proven to be in some measure redundant, but it was decided to leave all of them in place. This was to increase the system reliability against variations of the experimental conditions inside the BESIII experimental area, where the setup would be installed for the tests of chapter \ref{noisetest}.

If all the checks performed give a positive response, the standard deviation obtained through the fit and its error are written in the output table. Otherwise, next to the result of the suspect fit, an error message on the output file records which of the checks was not passed.

This process is repeated for all the channels included in the input file fed to the algorithm. For fast automatic execution, routines have been added allowing to feed a list of multiple data files and so to analyze large batches of data, automatically separating the two measurement branches. The generation of the plots can be disabled to drastically decrease computation time. This is the default operation mode used for the test of chapter \ref{noisetest}, where data taken in weeks or months of continuous acquisition were periodically analyzed in bulk.

\section{Execution of the Tests} 
The test of the two batches of FEBs took place under different conditions and with different setup configurations.

\begin{figure}[h]
	\centering
	\includegraphics[width=.6\textwidth, keepaspectratio]{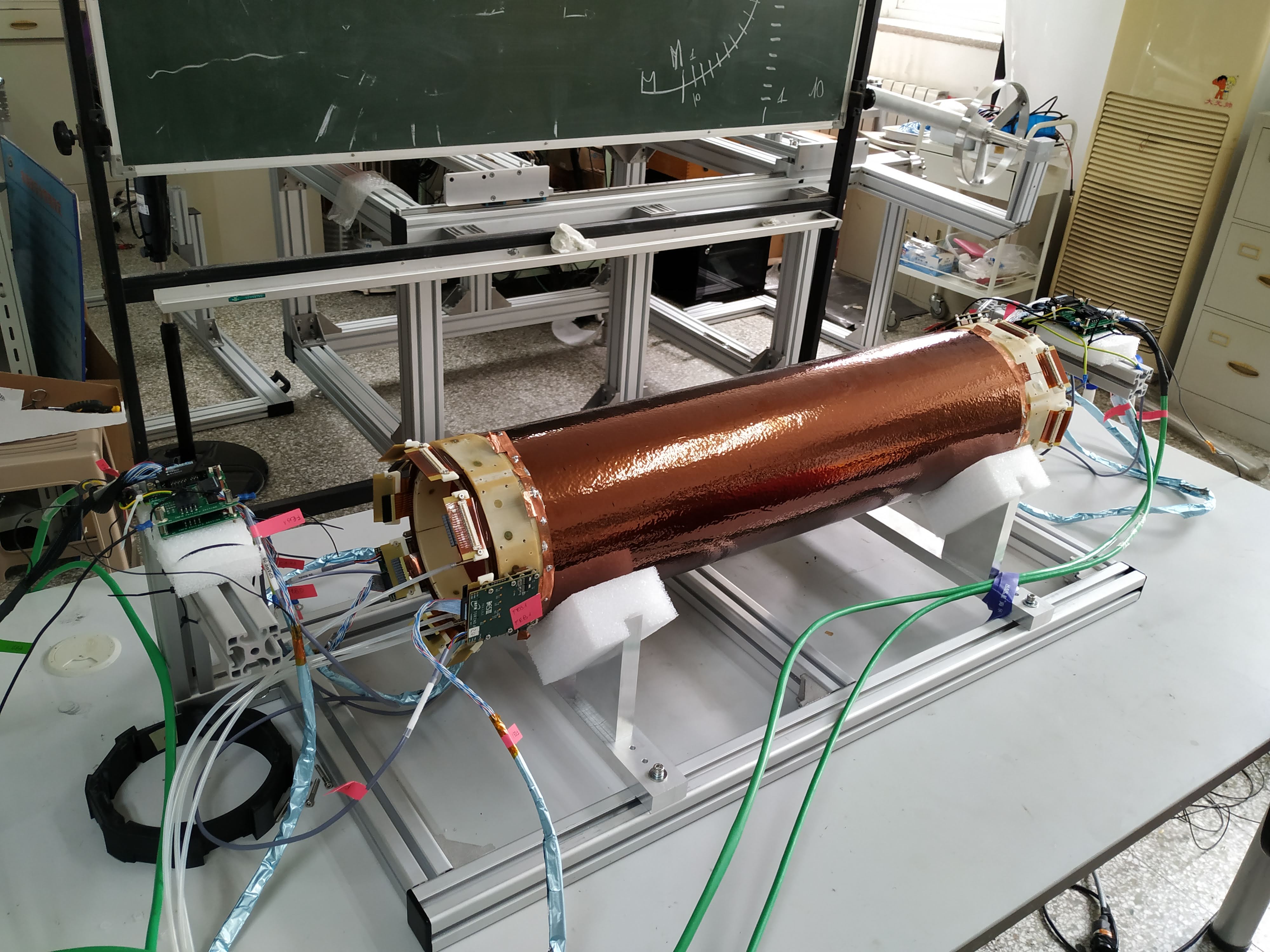}
	\caption[Photograph of the FEBs used in the noise measurements while being tested.]{Photograph of the FEBs used in the noise measurements, while connected to the first-design L1, during their test.}
	\label{fourfebs}
\end{figure}

The L1 FEBs were tested in a location separate from the one hosting the cosmic ray telescope setup and therefore without cooling. The higher temperatures the FEBs reach during the measurements cause an increase in the noise level but, as the purpose is the identification of dead or problematic channels and not the evaluation of the noise level per se, this does not hinder the test.
The four boards used for the noise measurements described in chapter \ref{noisetest} were tested while mounted to the first-design L1, as shown in figure \ref{fourfebs}, while the others were tested while placed on a table.
All the FEBs in this batch had already been used to collect data in the cosmic ray telescope setup and so preexisting documentation was available on their condition. The setup was made of a CAEN SY 5527 mainframe, a single LV A2519 module, a single A2517 module and a single GEMROC. This configuration allowed to test four FEBs at a time.

\FloatBarrier

The L3 FEBs were tested sharing both the power supply and cooling systems with the cosmic ray telescope setup. This introduced the need to optimize the operations in order to reduce the downtime of the cosmic data acquisition.
During the tests, the boards were disconnected from the detector and rested on a table. A CAEN SY4527LC mainframe was used and two GEMROCs were employed instead of one, allowing to perform a threshold scan on 8 FEBs simultaneously. This halved both the number of time consuming TP configurations and the times the chiller needed to be turned off to allow the replacement of the FEBs under test, with a consequent reduction of the downtime for the cosmic ray telescope setup. The cooling system was connected, using one of the free distribution lines available on location, to spot eventual leaks in the heatsinks of the L3 FEBs that were recently redesigned. 36 of the 41 L3 FEBs tested, the ones chosen for installation, had heatsinks installed and have been tested while connected to the cooling system; the remaining 5 were spares and were tested without it. 

Apart from minor variations dictated by the two different setups, the two tests were performed in the same way.
The set of FEBs to be tested was cabled according to the configuration described in section \ref{readout}. The TP of each TIGER was configured as described in section \ref{config} and a single threshold scan for each channel was launched through the GUFI interface. The data collected were used to produce graphs akin to the one in section$\,$\ref{thresholdscan} through the standalone algorithm I developed. The graphs were then quickly inspected and any irregular behavior was noted down. The log tables produced by the program were used to double-check the initial observations.

\section{Results of the test}
\label{monsters}
Apart from the few that were too damaged to establish communication with, all the FEBs mounted on the first-design L1 were tested. These tests confirmed and in some cases expanded the available documentation on their condition. The leading hypotheses are that the damages were sustained during operation, due to electrical discharges, or during transport.

A second series of tests involved 41 L3 FEBs that, apart from four of them, were never used before. Of all the FEBs tested, 35 were working perfectly, two had a single dead channel and the others presented either too many pathological channels to be used or could not communicate properly with the GEMROCs and cannot be configured. Three of the heatsinks presented minor leaks at the connectors or along the brazing and so were later replaced.

\begin{figure}[h]
	\centering
	\includegraphics[width=.7\textwidth, keepaspectratio]{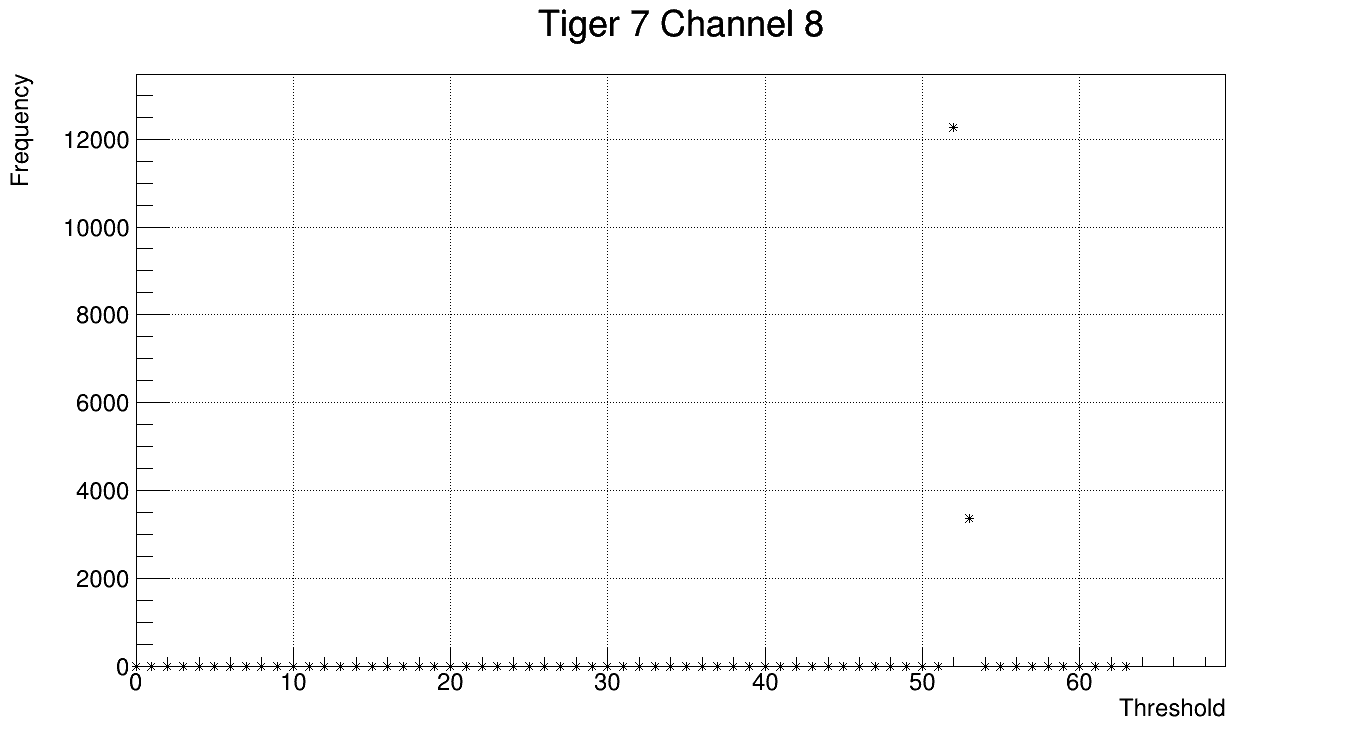}
	\caption[Threshold scan of a dead channel]{Example of graph produced by the threshold scan when the channel is dead. The ramp produced by the TP is absent and the baseline very narrow.}
	\label{deadchan}
\end{figure}

To determine the condition of the channels, the graphs produced in the threshold scan are inspected one by one in search of irregularities. A dead channel produces a graph like the one shown in figure \ref{deadchan}. Most of the points are zero, the ramp of the test pulse is not present and the baseline is limited to one or two points.

\begin{figure}[h]
	\centering
	\includegraphics[width=.7\textwidth, keepaspectratio]{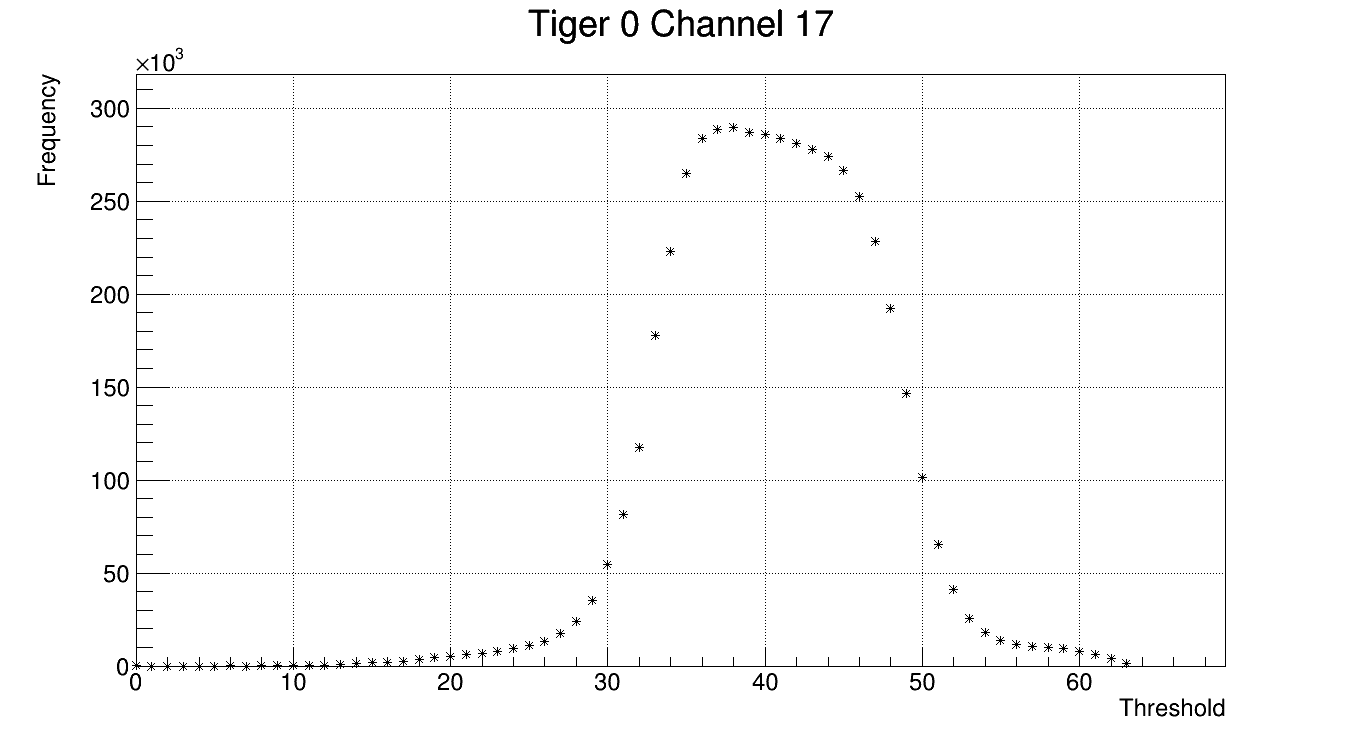}
	\caption[Threshold scan of a pathological channel]{Example of graph produced by the threshold scan on a channel displaying an irregular behavior. The baseline is so wide that it engulfs the ramp of the TP.}
	\label{slope5}
\end{figure}

Another kind of pathological channels produce graphs similar to the one in figure \ref{slope5}. In this case the baseline is very wide and engulfs the ramp produced by the TP. The number of counts saturates the measuring capabilities of the TIGER chips and therefore the curve appears truncated. There is still uncertainty on the origin of this behavior but, interpreting the graph, one can assume that the amplitude oscillations of the baseline must be comparable to the amplitude of the test pulse.

\FloatBarrier

    \chapter{Electronic Noise Studies in the BESIII Experiment Hall}
\label{noisetest}
This chapter describes a series of studies performed on noise data collected using a CGEM detector inside the BESIII experiment hall.
The purpose of the test is to verify the effect of the EM backgorund on the noise level of the front end electronics when operating close to the interaction point.
During my stay in China, I was directly involved in the preparation of the setup, which was the result of a combined effort by the researchers in charge of BESIII detector and those working on the CGEM-IT. After the installation, I was responsible for collecting the data with the setup and performing their analysis for about 11 months, until the test terminated.

The measurements began in the fall of 2019 and stopped in the summer of 2020. During this period, 4250 samples of data were collected in the different experimental conditions determined by the operation schedule of BEPC-II and BESIII. Now we plan to use the setup as a platform to aid the integration of the CGEM-IT within the Data Acquisition (DAQ) and slow-control software of the experiment.

In this chapter, the setup and the experimental technique are described in detail. An overview of the data collected in relation to BEPC-II operation schedule is provided. The analysis is performed separately on data batches taken during the different accelerator operation phases. The behavior of the relevant beam parameters during each of these is compared with the data. Finally, the results are compared with reference measurements to extract the variation in noise level corresponding to each phase.

\section{Aim of the Test and Experimental Technique}
The aim of the test is to measure the overall variation of the noise level under similar conditions to the ones that the CGEM-IT would face after its installation inside BESIII.
The primary factors that may contribute to the EM background close to the collision point were estimated to be: the BEPC-II magnets together with its auxiliary instrumentation; the presence of circulating beams; the operation of the BESIII detector and its superconducting (SC) solenoid and the collisions. These contributions are not entirely separable: BEPC-II must be operating for the beams to circulate and BESIII took data only with collisions.

The study was performed exploiting the schedule of the storage ring, which alternates periods of Synchrotron Radiation (SR) production and operation as a collider. While BEPC-II operates as a SR source, BESIII is off, only one beam circulates and there are no collisions. This phase allows to investigate the joint contributions of beam and accelerator. When BEPC-II operates as a collider instead, during the BESIII data taking, all the contributions are present.

In the collection of a data sample, the noise level of 512 channels is measured through a threshold scan, in the way described in section \ref{thresholdscan}. The noise level of the channels belonging to the same TIGER is then averaged to obtain the TIGER noise level, used in the analysis. Dead or compromised channels, for which it is impossible to measure a noise level, are excluded from the analysis. These were mapped at the beginning of the test and remained the same up to the last measurement.

Inside the TIGER, signals from different strips are handled by separate electronic pathways on a common substrate. All channels share the same design architecture but, as the process used in their production cannot be perfect, the behavior of each of them is unique. Moreover, these channels are connected to strips that are geometrically different on a macroscopic level. The X strips have the same length and they differ from V strips whereas the V strips differ also from each other. Because of these reasons, measurements taken on different channels should be treated as independent and studied separately.

This test aims to evaluate the overall variation of the noise level, not its absolute value on a channel-by-channel basis, as would be required before the collection of physics data. All the electronics channels are expected to react similarly to external EM noise. If immersed in a large EM background the noise level is expected to increase, while in a cleaner EM environment it is expected to fall.

Despite these considerations, the analysis of data collected through the T and E branch was kept separate to see if the electronics that distinguish them would show a different response to the same external conditions.

\section{Experimental Setup}
\begin{figure}[h]
	\centering
	\includegraphics[width=.9\textwidth, keepaspectratio]{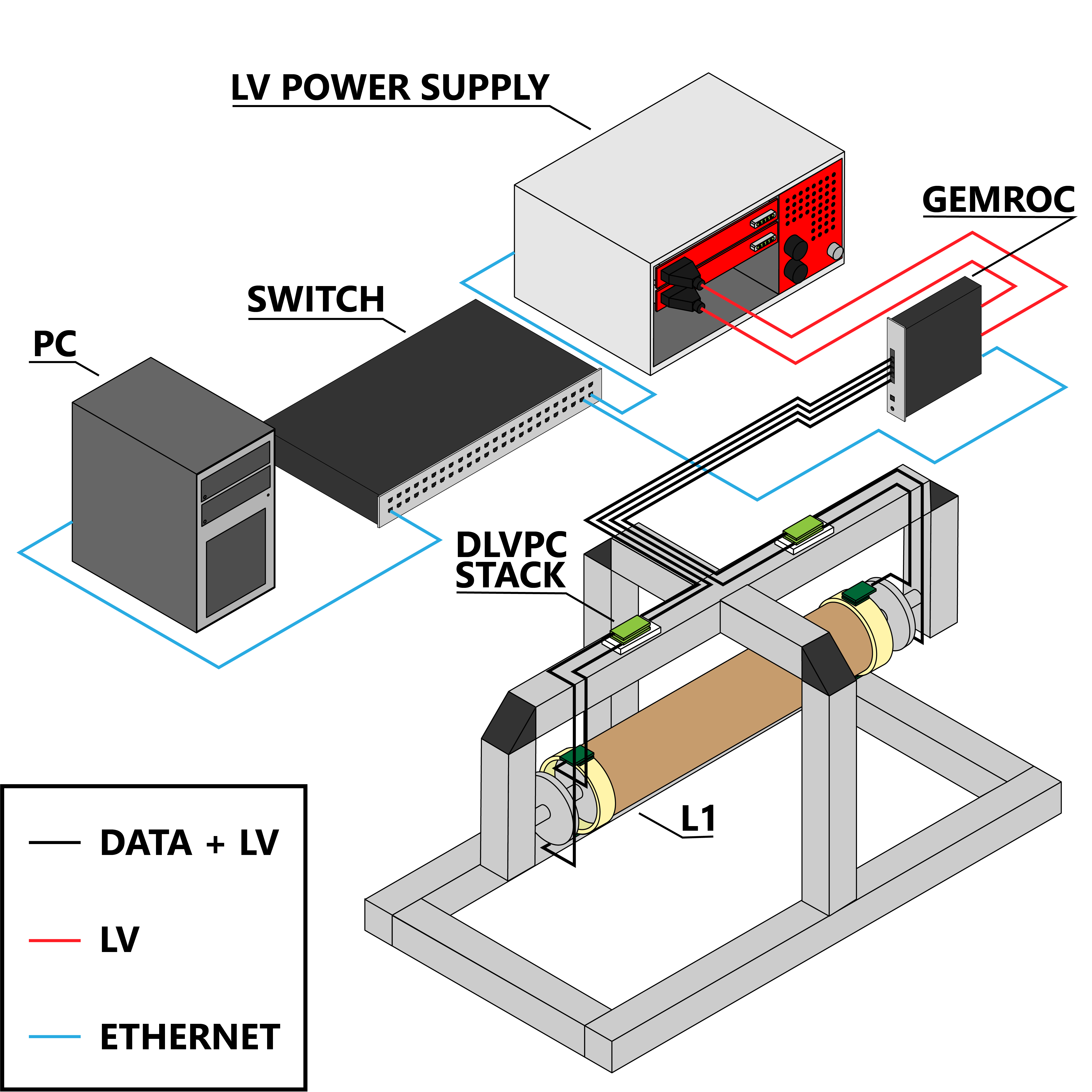}
	\caption[Schematic representation of the setup used for the noise measurements.]{Schematic representation of the setup used for the noise measurements in experiment hall.}
	\label{setup}
\end{figure}

A schematic representation of the experimental setup is shown in figure \ref{setup}.

The measurements were performed using the first-design L1 detector that was damaged during shipment. The detector is used passively, without a HV power supply since the aim of the test is the measurement of the sole electronics noise, not the one coming from physics events. In this configuration, while immersed in the EM background, its anode strips would be acting both as antennas and capacitor electrodes (the other being the ground plane).

On top of the detector four FEBs are mounted, two on its upper half, positioned at opposite sides of the detector, and two on its lower one, also facing each other. These FEBs were chosen, through the tests described in chapter \ref{febtests}, among the worst performing ones that could still be configured and operated. The choice of using poorly performing electronics is dictated by the will to recreate the worst possible operating conditions the final detector could face. For the same reason, the FEBs are not connected to a cooling system.

\begin{figure}[h]
	\centering
	\includegraphics[width=.6\textwidth, keepaspectratio]{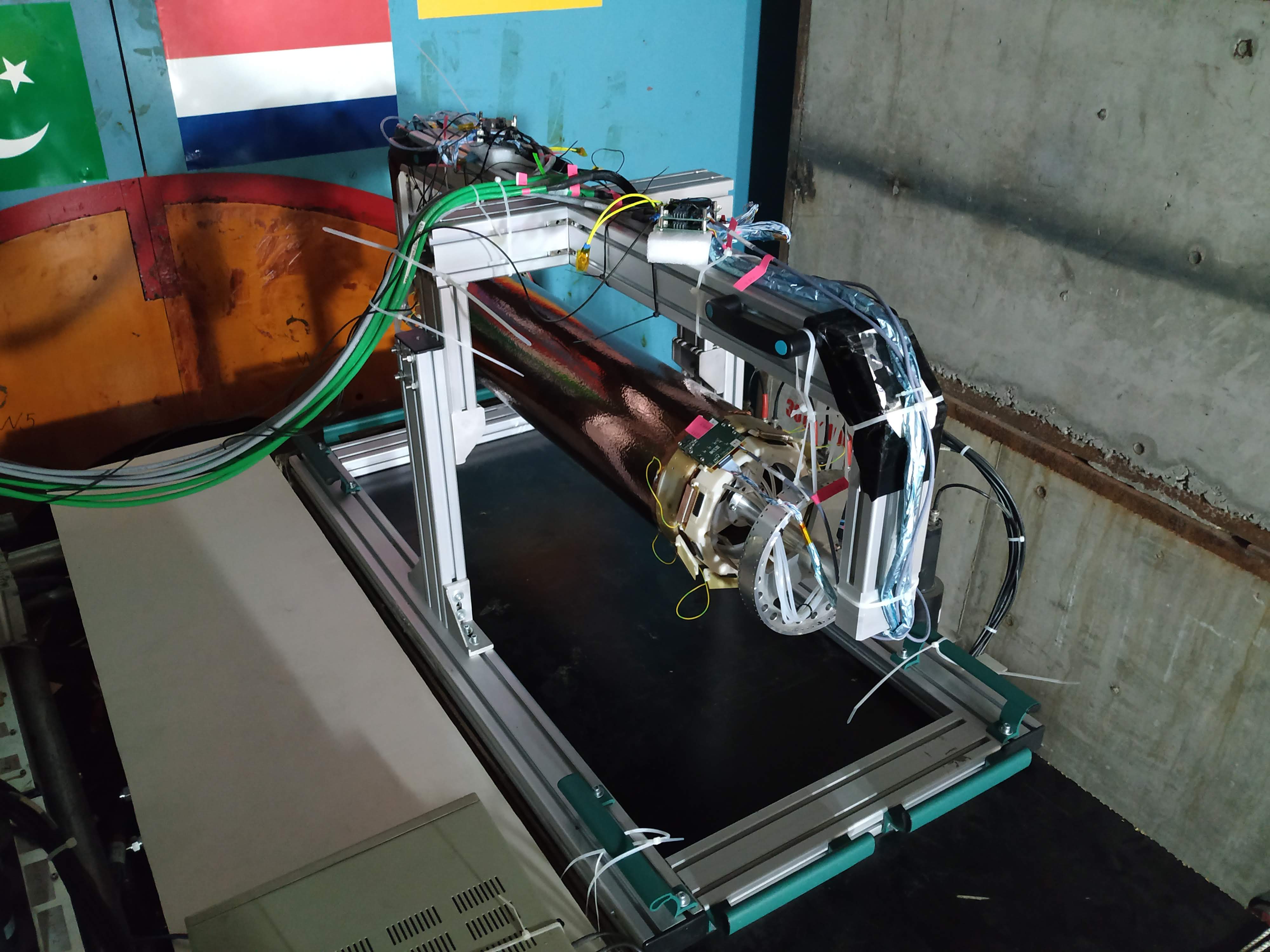}
	\caption[Photograph of the detector at the location chosen for the noise measurements.]{Photograph of the support holding the detector in the location chosen to perform the measurements.}
	\label{setuphoto}
\end{figure}

The detector is supported by an aluminum structure on top of which are fixed the four DLVPCs, stacked in couples. The support is secured on top of a scaffolding, directly above the beam line, to the east of BEPC-II southern collision point, where BESIII is located. Figure \ref{sipregion} in section \ref{bepici2} shows the instrumentation surrounding the setup. Figure \ref{setuphoto} instead shows the structure supporting the detector at the location chosen for the test. This choice was determined by its proximity to the collision point and therefore by its similarity with the CGEM-IT final location. The detector is parallel to the beam line so that one of its sides is closer to the collision point than the other one.

\begin{figure}[h]
	\centering
	\includegraphics[width=.6\textwidth, keepaspectratio]{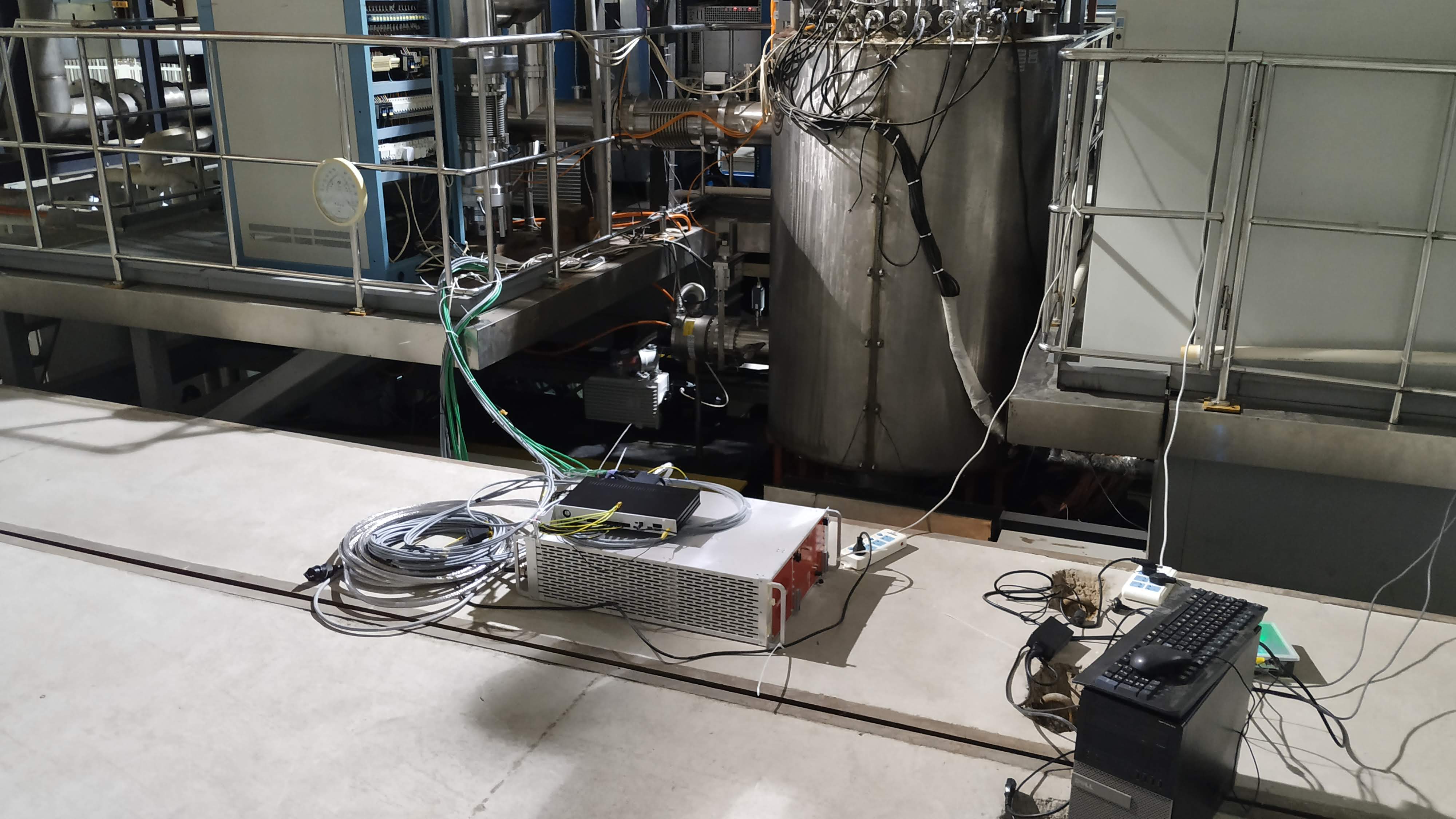}
	\caption[Photograph of the power supply system and GEMROC used in the noise measurements.]{Photograph of the power supply system and GEMROC used for the measurements. These were placed on top of the concrete bunker surrounding the collision point.}
	\label{onthebunker}
\end{figure}

A readout chain akin to the one described in section \ref{readout} was used to collect data with the setup. The power supply system and the GEMROC were placed on top of the concrete bunker that surrounds the interaction point for radiation shielding, as shown in Figure \ref{onthebunker}.

The LH cables, fastened to a railing, bridged the few meters separating the DLVPCs from the GEMROC and the power supply. The SH cables connecting FEBs and DLVPCs were first secured to the support as shown in figure \ref{setuphoto} for ease of transport, later they were separated from the aluminum structure of the support to prevent their shielding to come in contact with it.

The GEMROC and the power supply were connected to the BESIII LAN an Ethernet switch located on the platform above the collision point. A computer was placed inside the BESIII control room and connected to the same LAN through another Ethernet switch. The computer had access to the internet and allowed remote control of the data acquisition and of the power supply. 

\FloatBarrier

\subsection{Grounding}
\begin{figure}[h]
	\centering
	\includegraphics[width=.9\textwidth, keepaspectratio]{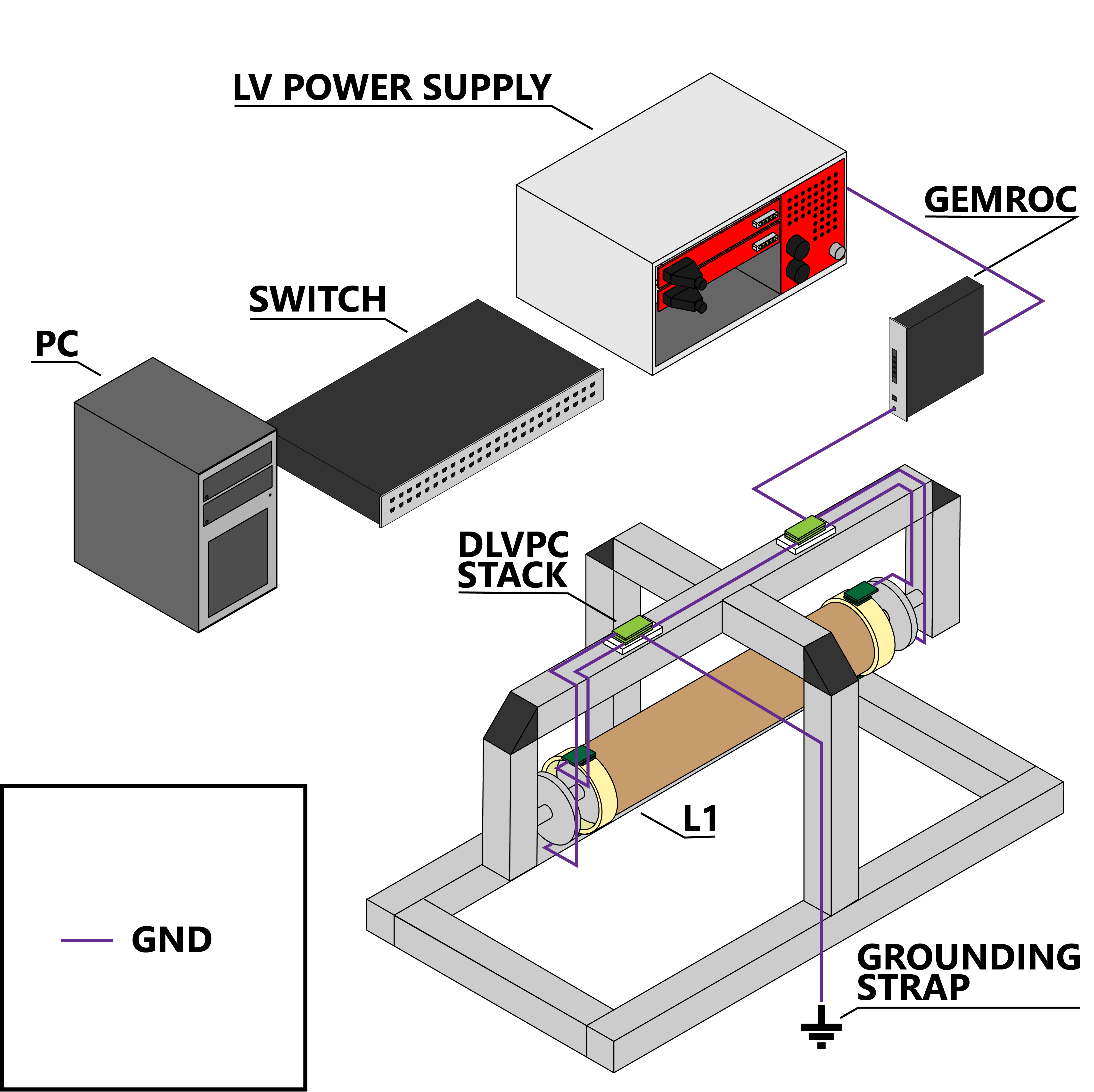}
	\caption[Grounding scheme used for the noise measurements.]{Schematic representation of the grounding scheme used for the measurements. SH and LH cables shielding connections are not depicted.}
	\label{grounding}
\end{figure}

The final grounding configuration of the setup is schematically represented in figure \ref{grounding}. The connections of the shielding of SH and LH cables are not represented.

The ground reference for the system is provided to the DLVPC stacks by one of the BESIII grounding straps present on site. From the DLVPCs, the reference is transferred to the GEMROC, through a dedicated cable, and to the FEBs, through one of the channels of the SH signal cables. The FEBs relay the reference potential to the detector ground, through the soldered copper band, and to the anode strips, through the connector. The strips that are not instrumented are connected to the detector ground through caps attached to the anode readout tails. The shielding of LH cables is connected on both sides, to the GEMROC and to the DLVPCs, while the ones of the SH cables are connected only on the DLVPC side.

In the final CGEM-IT design, the inner surface of L1 and the outer surface of L3 consist of copper clad Kapton foils that together enclose the detector in a Faraday cage.
As a single L1 detector was being used, it was necessary to build a conductive shell around it for completing the Faraday cage. This was done using a copper clad Kapton foil supported by a cylindrical Rohacell-Kapton sandwich. The shell was cut along its length; this made possible to install it without having to remove the detector from its support. Copper tape was used to patch the cut and to connect the inner and outer parts of the cage.

\section{Overview of the Collected Data}
\begin{sidewaysfigure}[h]
	\centering
	\includegraphics[width=.8\textwidth, keepaspectratio]{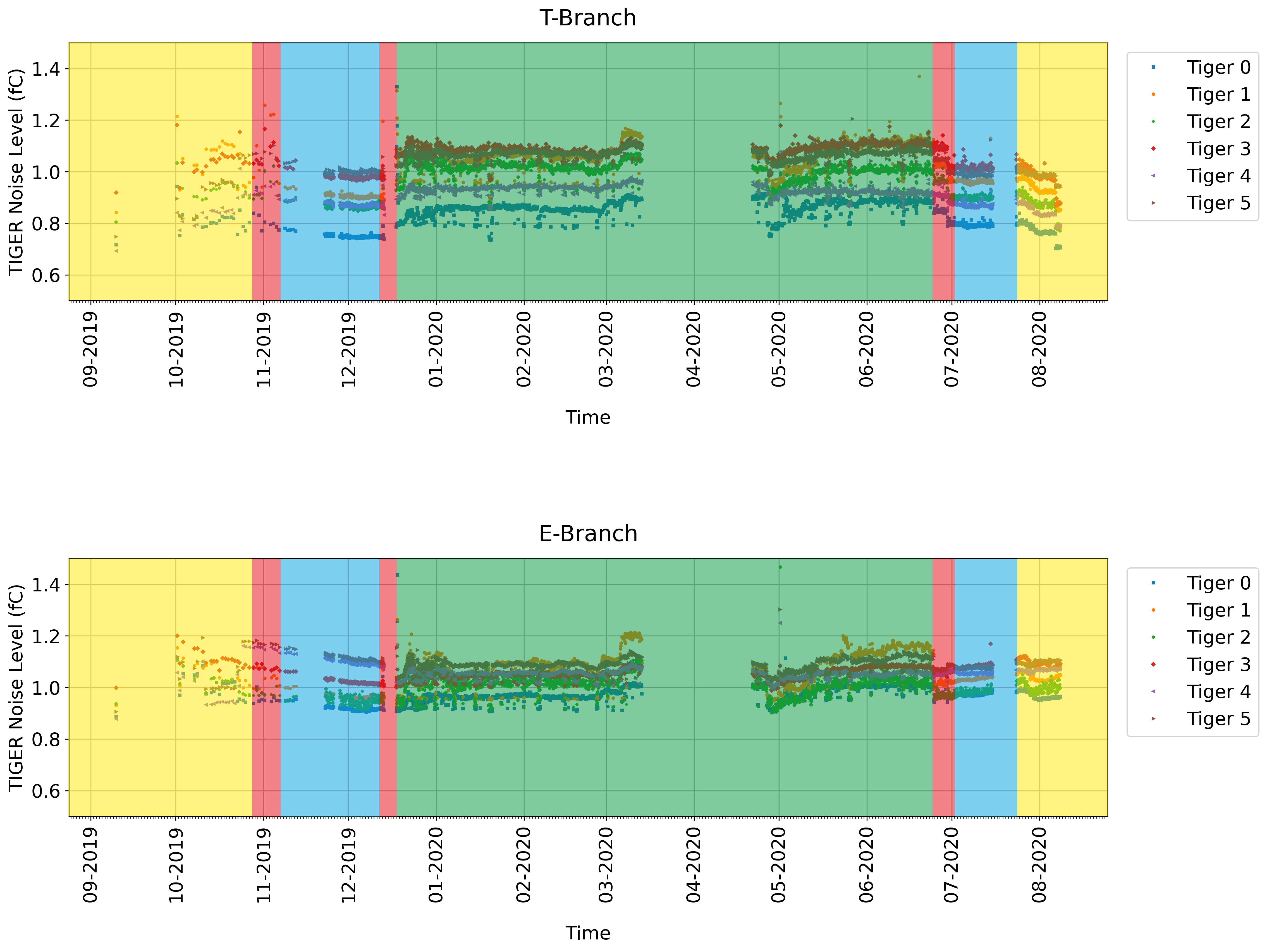}
	\caption[Overview of the data collected in the different BEPC-II operation phases.]{Overview of the data collected, during the whole measurement period, in relation to the BEPC-II phases. The shutdowns are represented in yellow, the SR production phases in blue and the BESIII data taking period in green. Data taken in the red regions, which correspond to machine recovery, machine study or to the switch between operation modes, were not used.}
	\label{phases}
\end{sidewaysfigure}

The operation schedule of the BEPC-II collider, in the period spanned by the noise measurements, is provided in table \ref{schedule}.
An overview of the data collected during these phases is shown in figure \ref{phases}.

\begin{table}[h]
	\centering
	\begin{tabular}{lllr} 	
		From		&	To			&	Task							&	Days	\\ \hline
		26/07/2019	&	23/10/2019	&	Summer Shutdown					&	90			\\
		24/10/2019	&	06/11/2019	&	Machine	recovery				&	14			\\
		07/11/2019	&	12/12/2019	&	SR operation					&	36			\\
		13/12/2019	&	19/12/2019	&	Switch to collision operation	&	7			\\
		20/12/2019	&	23/06/2020	&	Data Taking						&	186			\\
		24/06/2020	&	28/06/2020	&	Machine study					&	5			\\
		29/06/2020	&	01/07/220	&	Switch to SR operation			&	3			\\
		02/07/2020	&	23/07/2020	&	SR operation					&	22			\\
		24/07/2020	&	21/09/2020	&	Summer shutdown					&	60			\\
	\end{tabular}
	\caption[BEPC-II operation schedule in the period spanned by the noise measurements.]{BEPC-II operation schedule in the period spanned by the noise measurements.}
	\label{schedule}
\end{table}

The setup was installed in the BESIII experiment hall on the 1\textsuperscript{st} of October 2019. At that time, the ground configuration was different and relied on the power supply to provide the reference potential of the system. The definitive grounding scheme, relying on one of the BESIII grounding straps, was adopted on the 11\textsuperscript{th} of October 2019. The Faraday cage was installed between the 22\textsuperscript{nd} and the 23\textsuperscript{rd} of the same month, right before the beam was turned on.

Limited by the accelerator schedule, a test sample was taken on the same day the setup configuration was finalized, the day before the beginning of the SR production. This sample confirmed the operativity of the setup, reporting results analogous to the ones observed in the previous configurations. The reference samples were, instead, acquired during the 2020 summer shutdown.

The points to the left on the graph are sparser due to the fact that these measurements were acquired manually, launching a single threshold scan a day, for each electronic branch. On the 22\textsuperscript{nd} of November 2019 a script was implemented in GUFI allowing to continuously acquire data alternating between the electronic branches. This increases the statistic collected per day producing the denser data points visible throughout the rest of the measurement period.

\newpage

\FloatBarrier

To guarantee homogeneity, all the samples used for the analysis were collected in the final grounding configuration, with the Faraday cage installed and using the continuous scanning script. Apart from the measurements taken during the installation, which were launched while on site, the setup has always been remotely controlled.

The blanks observable in the graph are caused by crashes of the computer controlling the acquisition. These issues required someone to access the experiment control room to be addressed. After the recalling of the research team present on site, due to the COVID-19 pandemic, it became necessary to rely on the help of the BESIII personnel to solve issues that could not be fixed remotely.

The setup takes about 1 hour and 11 minutes to complete a threshold scan on all of its 512 instrumented channels. The scans are performed continuously, alternating between the two electronic branches. As the data from the two branches are treated separately, two measurements of the same channel, in the same branch, are separated by at least 2 hours and 22 minutes. This makes the setup inappropriate to discern the effects of rapidly varying parameters.

The SR production and the BESIII data taking phases offer consistent conditions, due to the beam current stability required. Data samples taken during machine recovery, machine study and the switch between operation modes, that is when the external conditions vary considerably, were not used in the analysis. This is because measurements taken on different channels within the same scan cycle may have been subject to very different conditions.

\begin{figure}[!h]
	\centering
	\includegraphics[width=\textwidth, keepaspectratio]{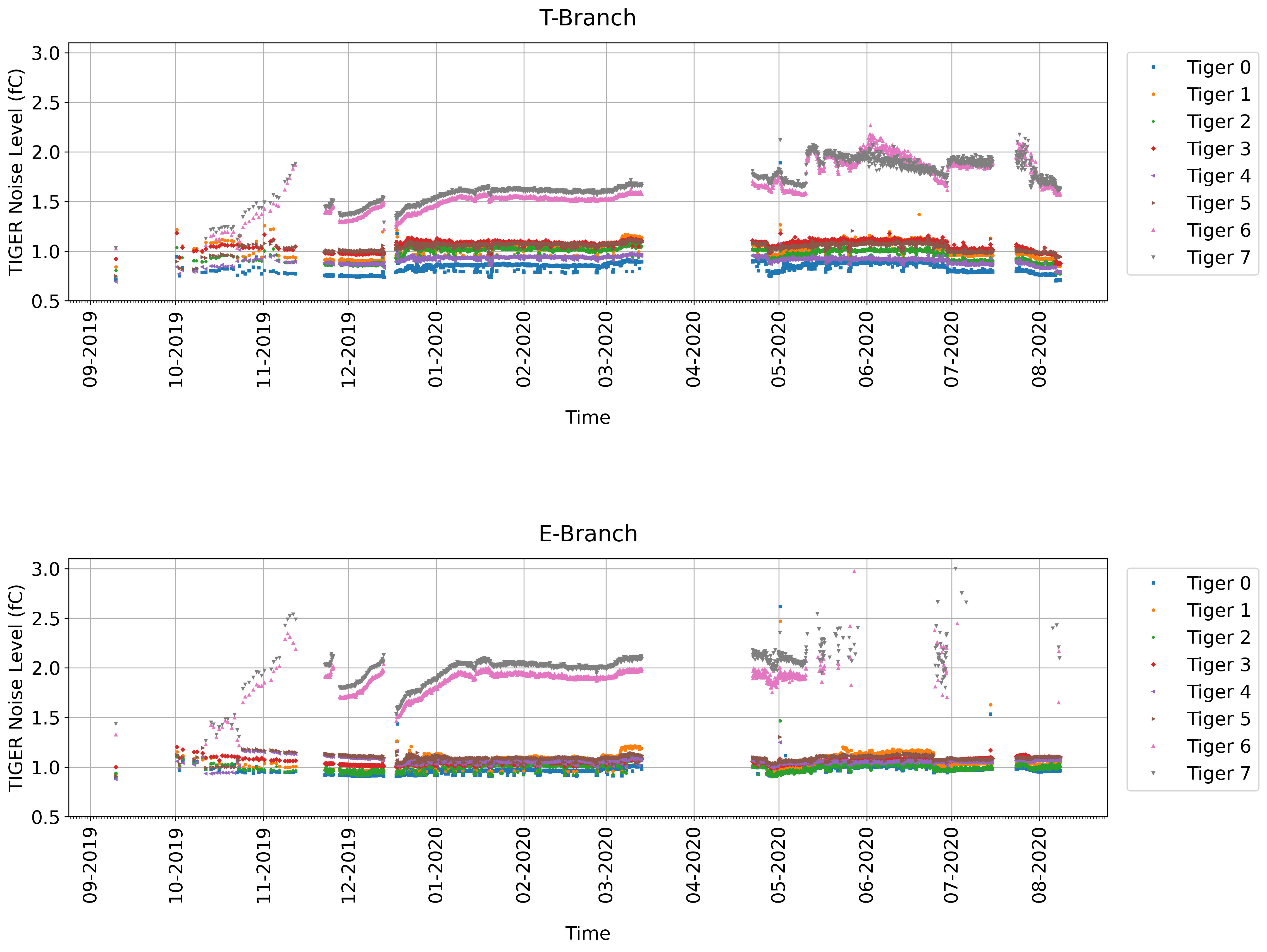}
	\caption[Erratic behavior of one of the FEBs in the setup.]{TIGERs 6 and 7, belonging to the same FEB, collected data that is incompatible with the others. Data from these two TIGERs are excluded from the analysis.}
	\label{erratic}
\end{figure}

Data collected from both TIGERs of one of the FEBs showed an erratic behavior when compared to the data collected by their counterparts, as can be observed in figure \ref{erratic}. This FEB was found to have a damaged connection to the detector ground.
Although this connection was patched during the operations to change the grounding scheme, the data it collected remained inconsistent and so it was decided to exclude it from the analysis.

\FloatBarrier

\section{Data Analysis}
The BEPC-II operation phases, on which the analysis focuses, are: the first SR production phase (SR1), the BESIII data taking phase (BESIII-DT), the second SR production phase (SR2) and the summer shutdown phase (SD). The number of data samples used for studying the noise level in each of them is not the same. This is because both their duration and the acquisition downtime, suffered by the setup, differ. Table \ref{samples} reports the number of samples used for each of the BEPC-II operation phases investigated.

\begin{table}[h]
	\centering
	\begin{tabular}{lllrr} 	
		Phase&From&To&Days&Samples\\\hline
		SR1&07/11/2019&12/12/2019&36&184\\
		BESIII-DT&20/12/2019&23/06/2020&186&1500\\
		SR2&02/07/2020&23/07/2020&22&140\\
		SD&24/07/2020&21/09/2020&60&156\\
	\end{tabular}
	\caption[Number of data samples used for the analysis during each of the BEPC-II operation phases on which it focuses.]{Number of data samples used for the analysis during each of the BEPC-II operation phases on which it focuses.}
	\label{samples}
\end{table}

The noise level of each TIGER for the studied phase is obtained by averaging together the measurements collected during its course. These do not seem to be distributed normally, so the error on the averages was obtained by halving their dispersion, after the removal of the five smallest and the five largest values. Discarding the extremes prevents the systematic errors to induce an overestimation of the statistical error. The errors of the single measurements are not shown in the plots for ease of readability, as they play no role in the determination of the results, or their errors, since they are obtained through an average over the whole period.

The data collected are qualitatively compared to the behavior of the beam parameters deemed relevant for each phase. The beam parameter data for the SR1 phase were provided by the personnel in charge of the accelerator. The ones used in the study of the BESIII-DT and SR2 phases were recovered from the web page of the BESIII slow control remote monitoring system \cite{paramsdata}.
In the first batch of data, the current of the electron beam is sampled every half a second. The slow control database, instead, contains a collection of beam parameters sampled every 30 seconds. For the study of the BESIII-DT phase the data related to both beam currents and both beam energies are used. For the SR2 phase, instead, only the electron beam current data are used, as in the database there are no energy data for the duration of this phase. Because the setup is not sensitive to such fine timescales the data were condensed averaging together the measurements contained in a 5 minutes time interval. For keeping track of the information lost when averaging together data presenting a sizable time variability, to each average was assigned an error equal to half of its dispersion.

\subsection{First SR Production Phase}
\begin{figure}[h]
	\centering
	\includegraphics[width=\textwidth, keepaspectratio]{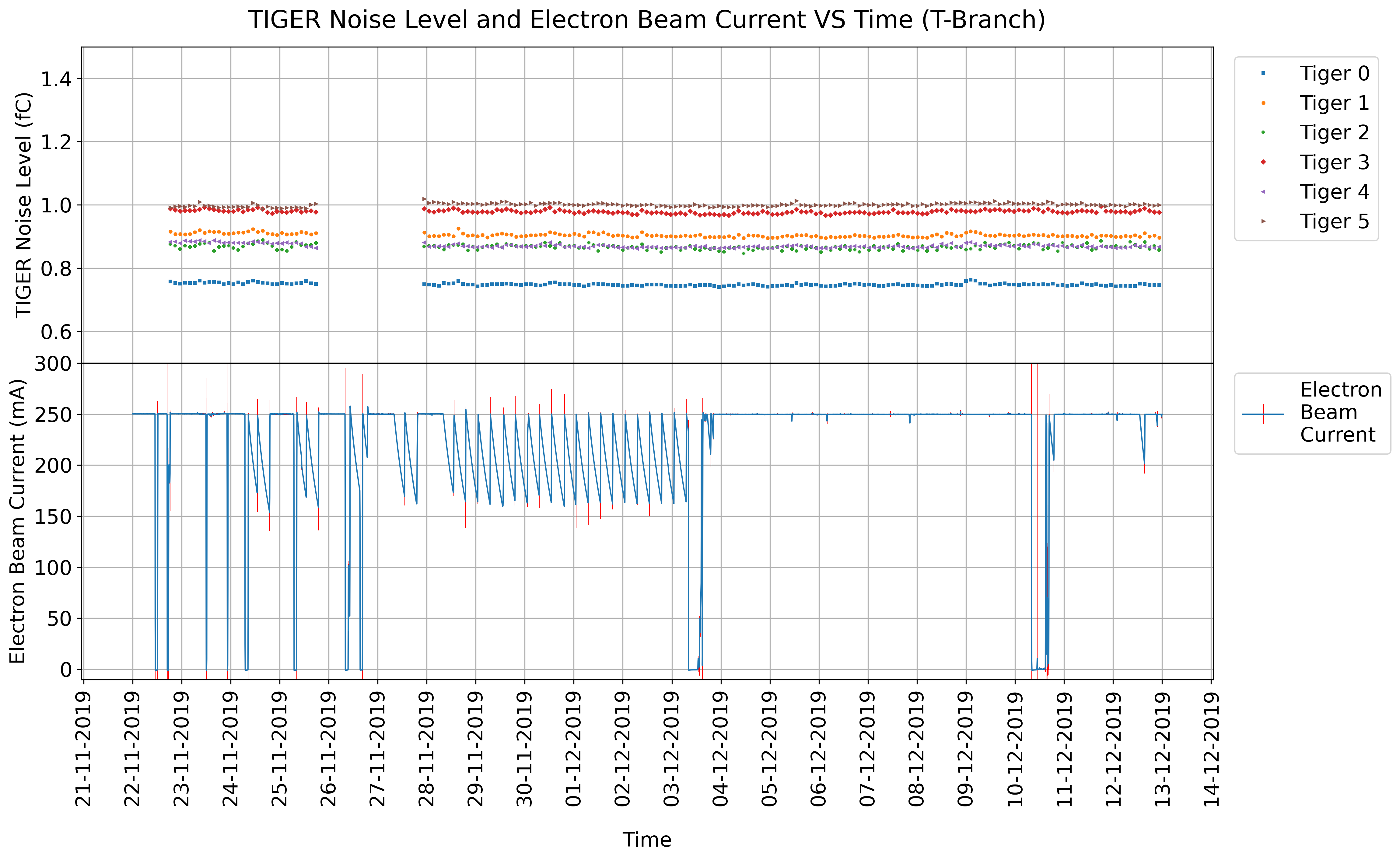}
	\vspace{1cm}
	\includegraphics[width=\textwidth, keepaspectratio]{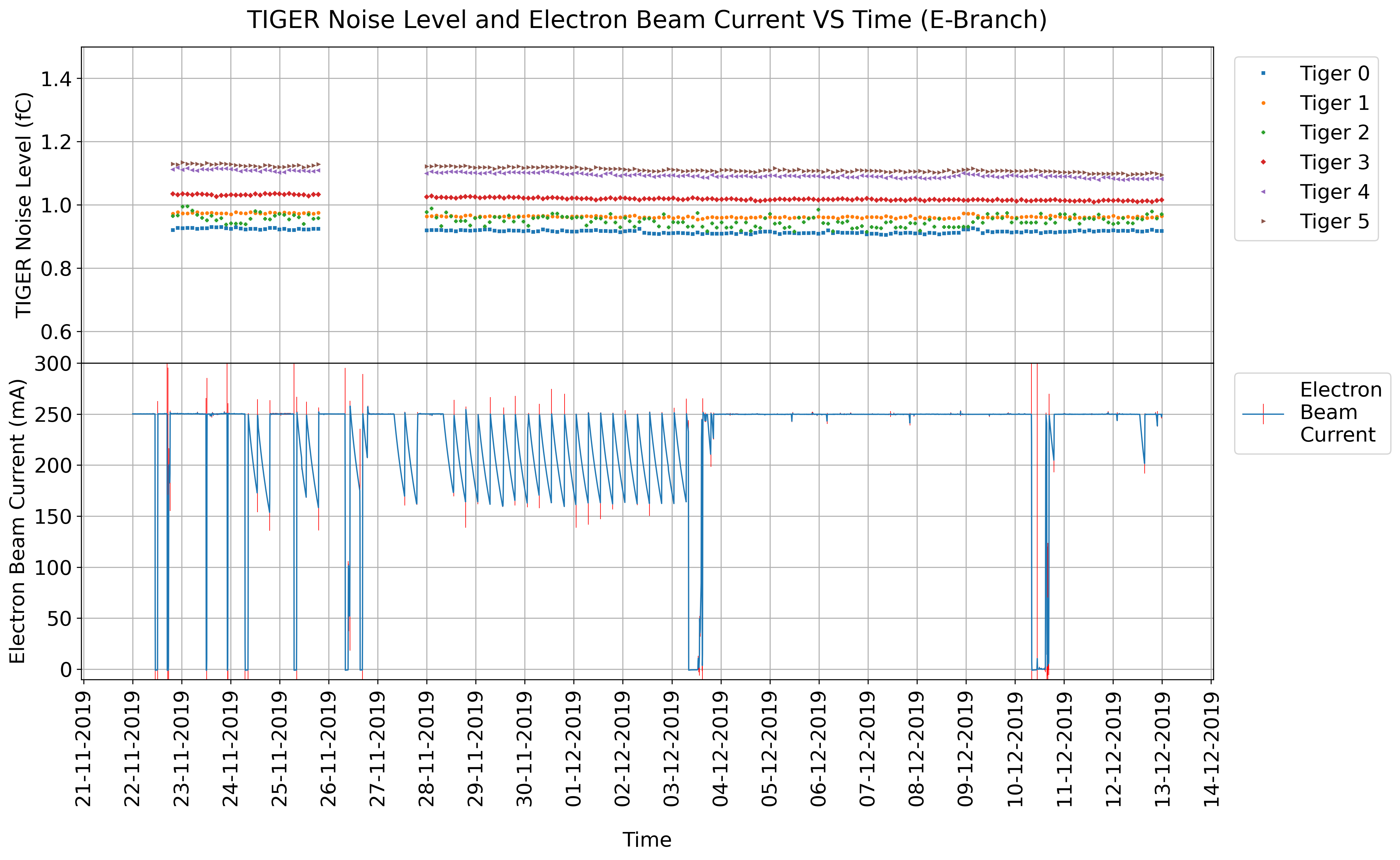}
	\caption[Comparison between the noise level of the TIGERs and the beam current in the first SR production phase.]{Comparison between the noise level of the TIGERs and the beam current in the first SR production phase. The data samples shown were collected between the 22/11/2019 and the 12/12/2019.}
	\label{sr1_1}
\end{figure}

Figure \ref{sr1_1} shows a comparison between the noise level of the 6 TIGERs and the beam current. The noise level of the TIGERs appears stable for the entire duration of the SR1 phase. Even when the beam current oscillates considerably or drops to 0 no corresponding variation is observed in the noise. This points towards the lack of a correlation between the two quantities.

\begin{figure}[h]
	\centering
	\includegraphics[width=\textwidth, keepaspectratio]{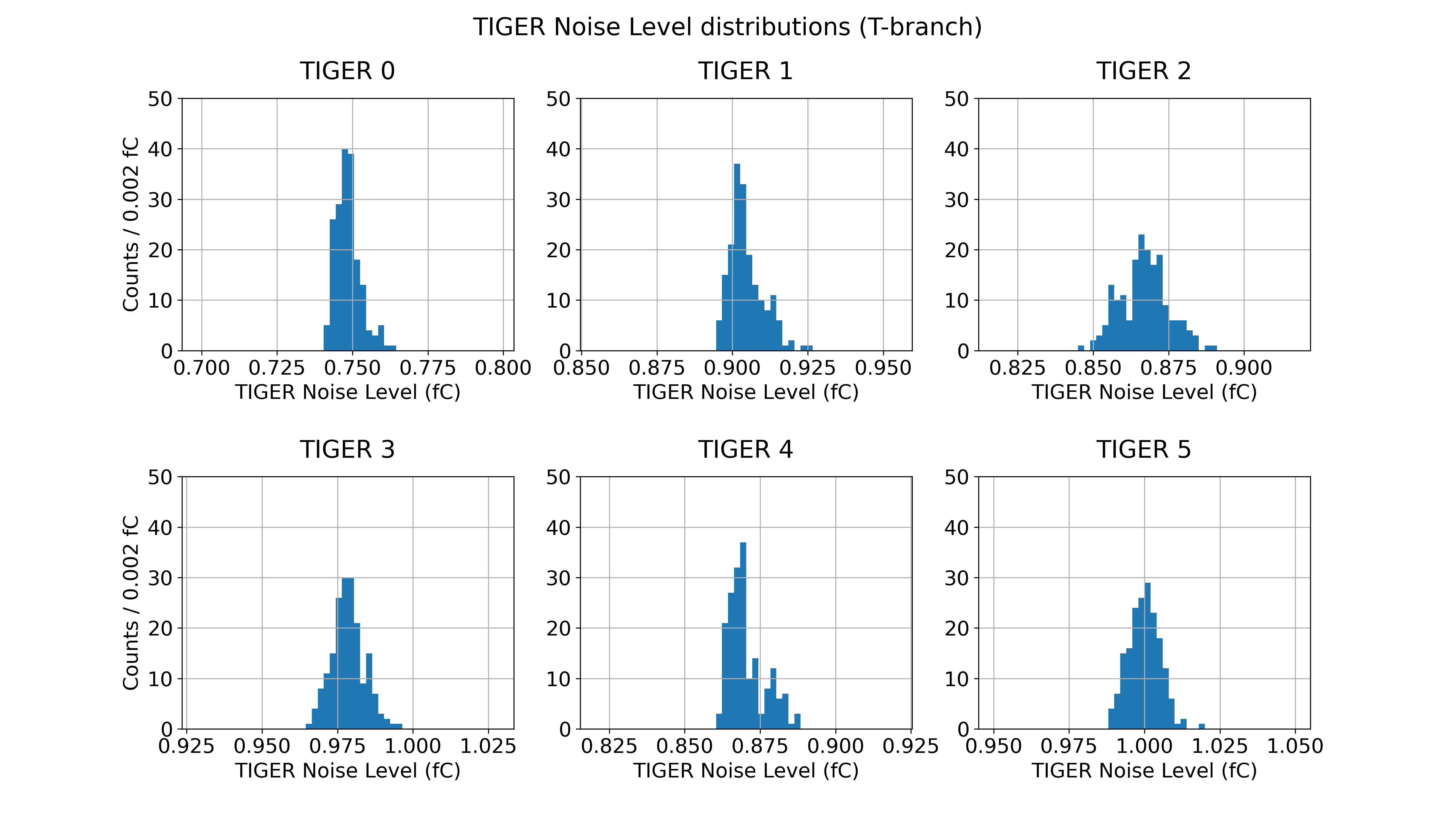}
	\vspace{1cm}
	\includegraphics[width=\textwidth, keepaspectratio]{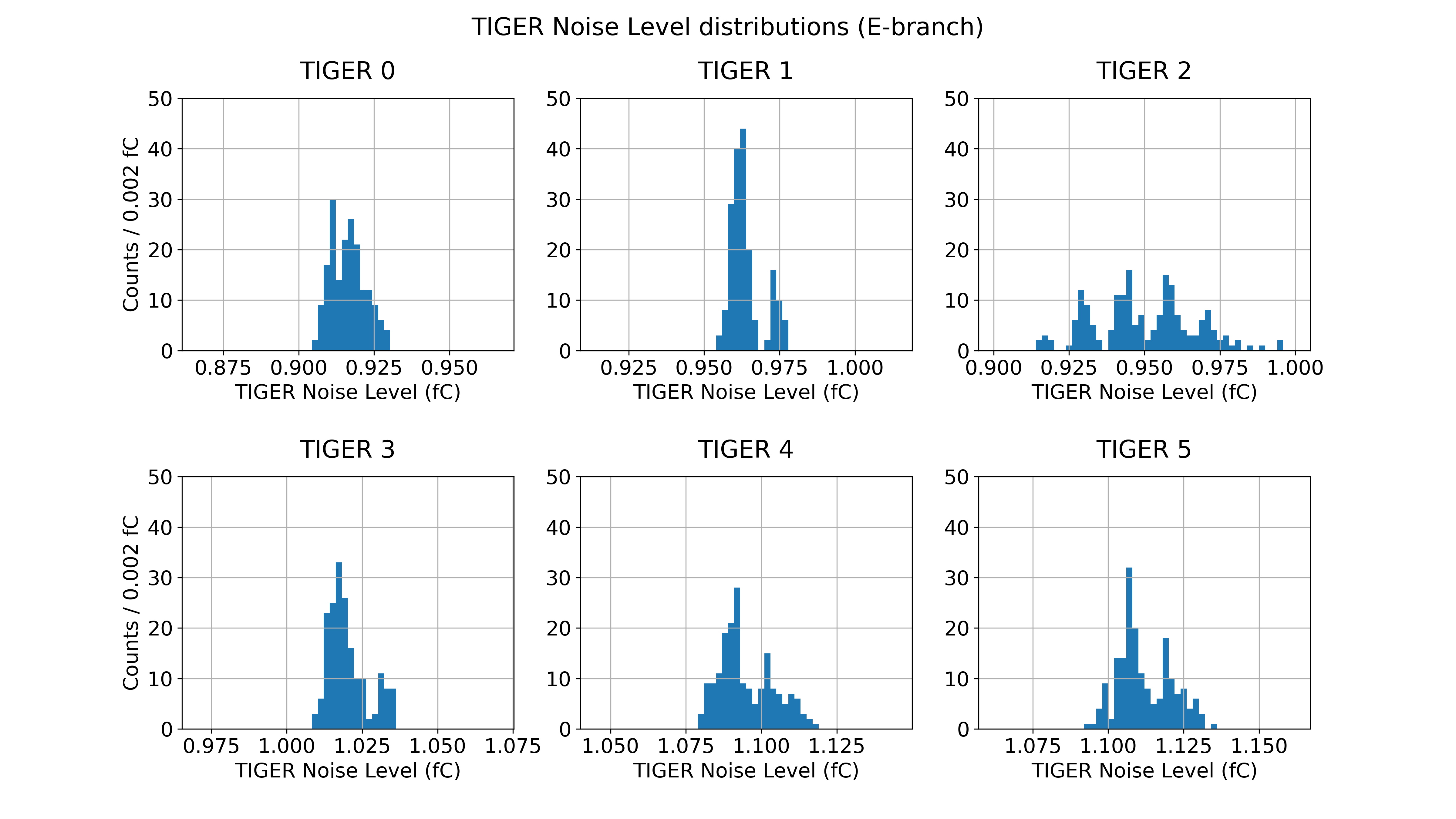}
	\caption[TIGER noise level distributions in the first SR production phase.]{TIGER noise level distributions in the first SR production phase. The The range is 0.1$\,$fC.}
	\label{sr1_2}
\end{figure}

Figure \ref{sr1_2} shows the distribution of the noise levels of each tiger for both electronic branches.
The T branch distributions mostly present a single peak while those obtained from data collected using the E branch mostly present two of them.
Tiger 2 experiences the largest oscillations and its distribution shows 4 separate peaks.
The presence of multiple peaks hints at a non-stochastic behavior. This could be a dependency on some variable that was not considered or an internal effect, due to the way the measurements are performed. The determination of the origin of this behavior was not within the initial goals and may therefore require a modification of the setup, of the experimental technique or  of both, to be investigated.
Nonetheless the fluctuations of the noise level observed remain below 10\% on values that are compatible with the expected performance of the TIGERs. These are also inferior to the systematic error that can be introduced, for example modifying the grounding or removing and then reinstalling the same FEB.

The time averaged noise levels for the first phase are reported in table$\,$\ref{sr1_3}. The measurements performed on the E branch are always higher than the ones taken on the T branch.

\begin{table}[h]
	\centering
	\begin{tabular}{rS[table-format=1.3]S[table-format=1.3]}
		\multicolumn{3}{c}{\textbf{T Branch}}\\\hline	
		TIGER&\multicolumn{1}{r}{Noise level (fC)} &\multicolumn{1}{r}{Error(fC)}\\\hline
		0&0.749&0.008\\
		1&0.905&0.009\\
		2&0.867&0.015\\
		3&0.978&0.010\\	
		4&0.870&0.011\\
		5&1.000&0.009\\
		&&\\
		\multicolumn{3}{c}{\textbf{E Branch}}\\\hline
		TIGER&\multicolumn{1}{r}{Noise level (fC)} &\multicolumn{1}{r}{Error(fC)}\\\hline
		0&0.916&0.010\\
		1&0.964&0.009\\
		2&0.95&0.03\\
		3&1.020&0.012\\	
		4&1.09&0.02\\
		5&1.11&0.02\\
	\end{tabular}
	\caption[Time averaged TIGER noise levels for the first SR production phase.]{Time averaged noise level of the TIGERs in the first SR production phase.}
	\label{sr1_3}
\end{table}

\FloatBarrier

\subsection{BESIII Data Taking Phase}
\begin{figure}[h]
	\centering
	\includegraphics[width=\textwidth, keepaspectratio]{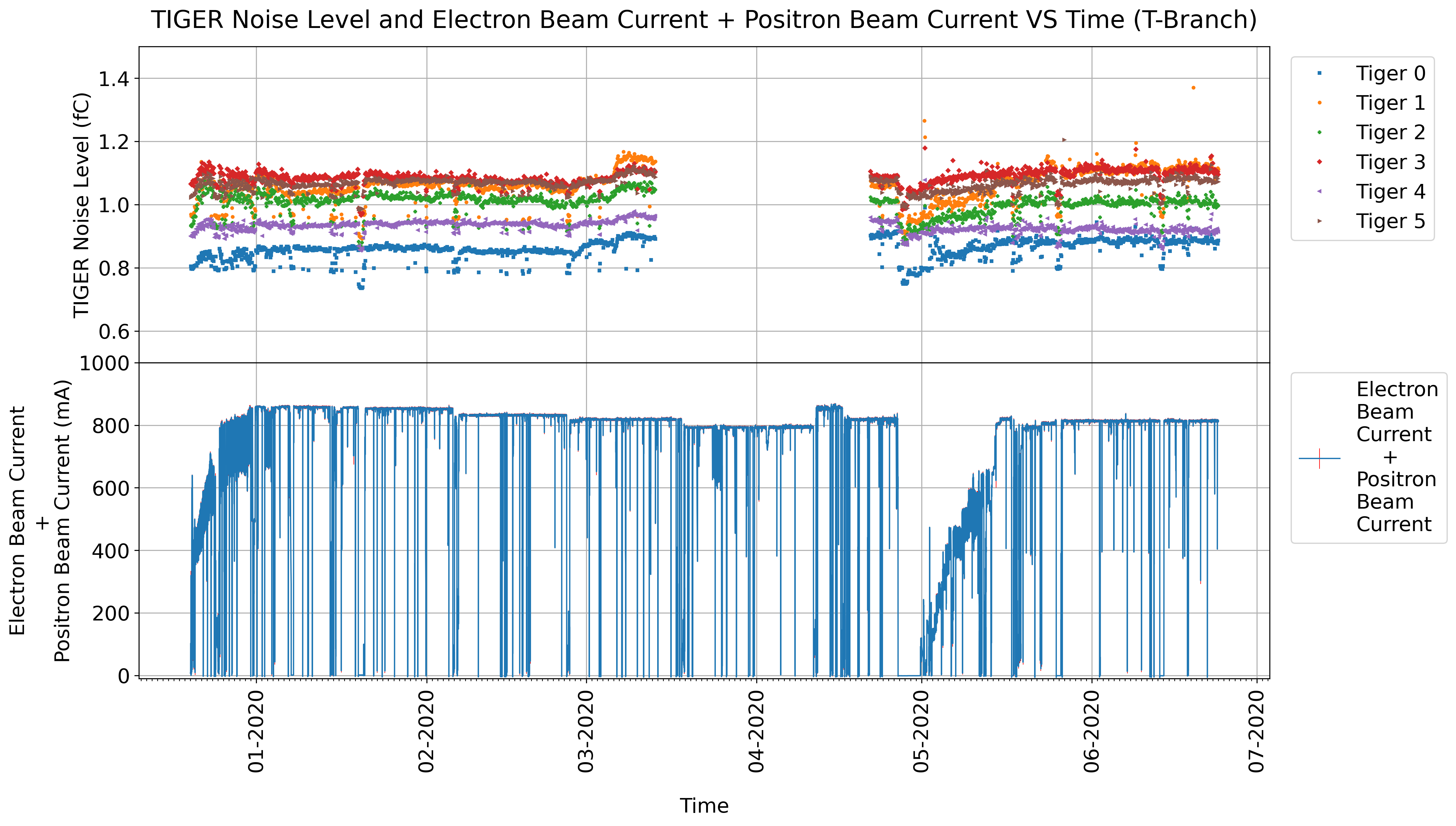}
	\vspace{1cm}
	\includegraphics[width=\textwidth, keepaspectratio]{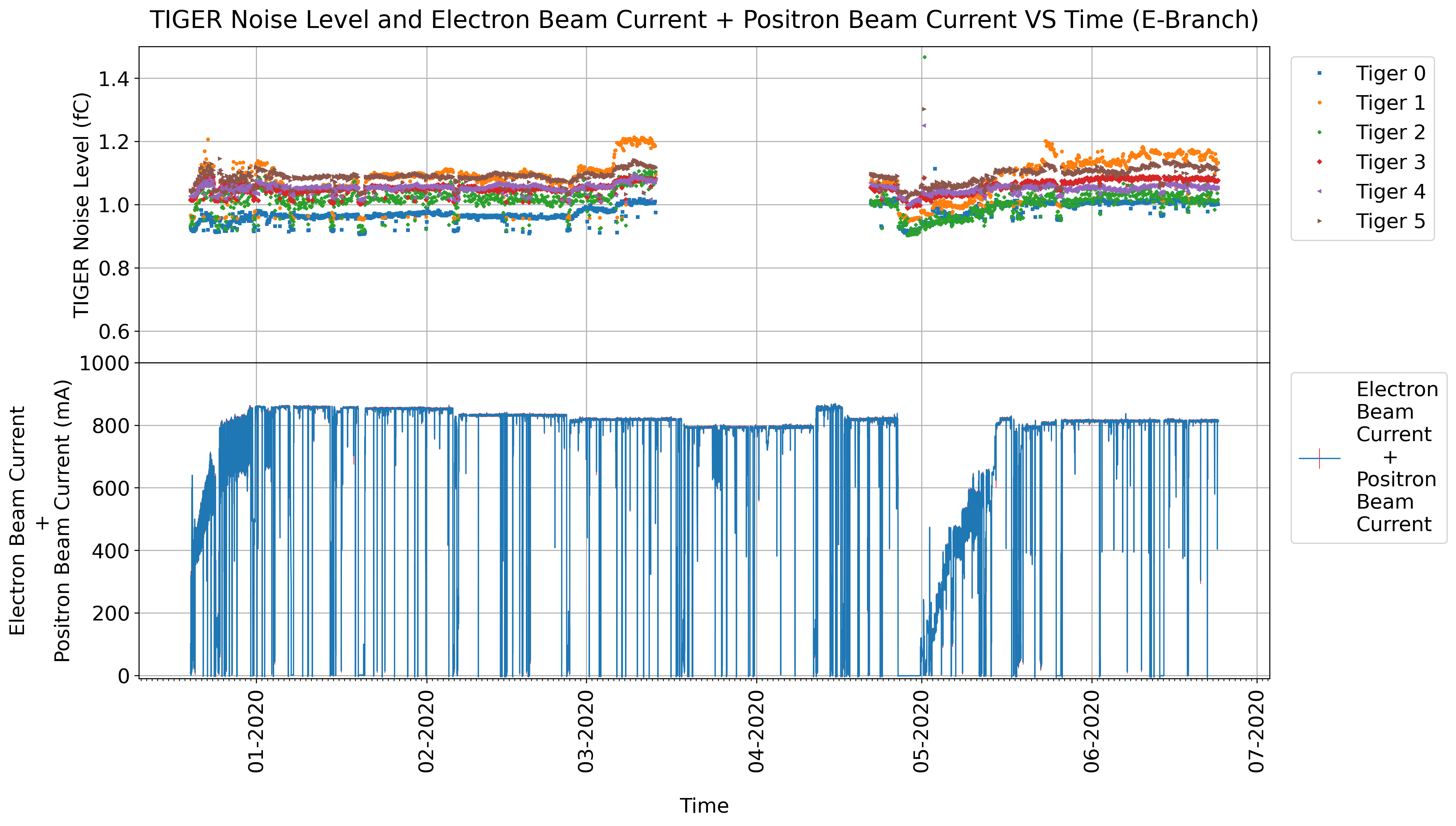}
	\caption[Comparison between the noise level of the TIGERs and the sum of the beam currents in the BESIII data taking phase.]{Comparison between the noise level of the TIGERs and the sum of the beam currents in the BESIII data taking phase. The data samples shown were collected between the 20/12/2019 and the 23/06/2020.}
	\label{daq1}
\end{figure}

\begin{figure}[h]
	\centering
	\includegraphics[width=\textwidth, keepaspectratio]{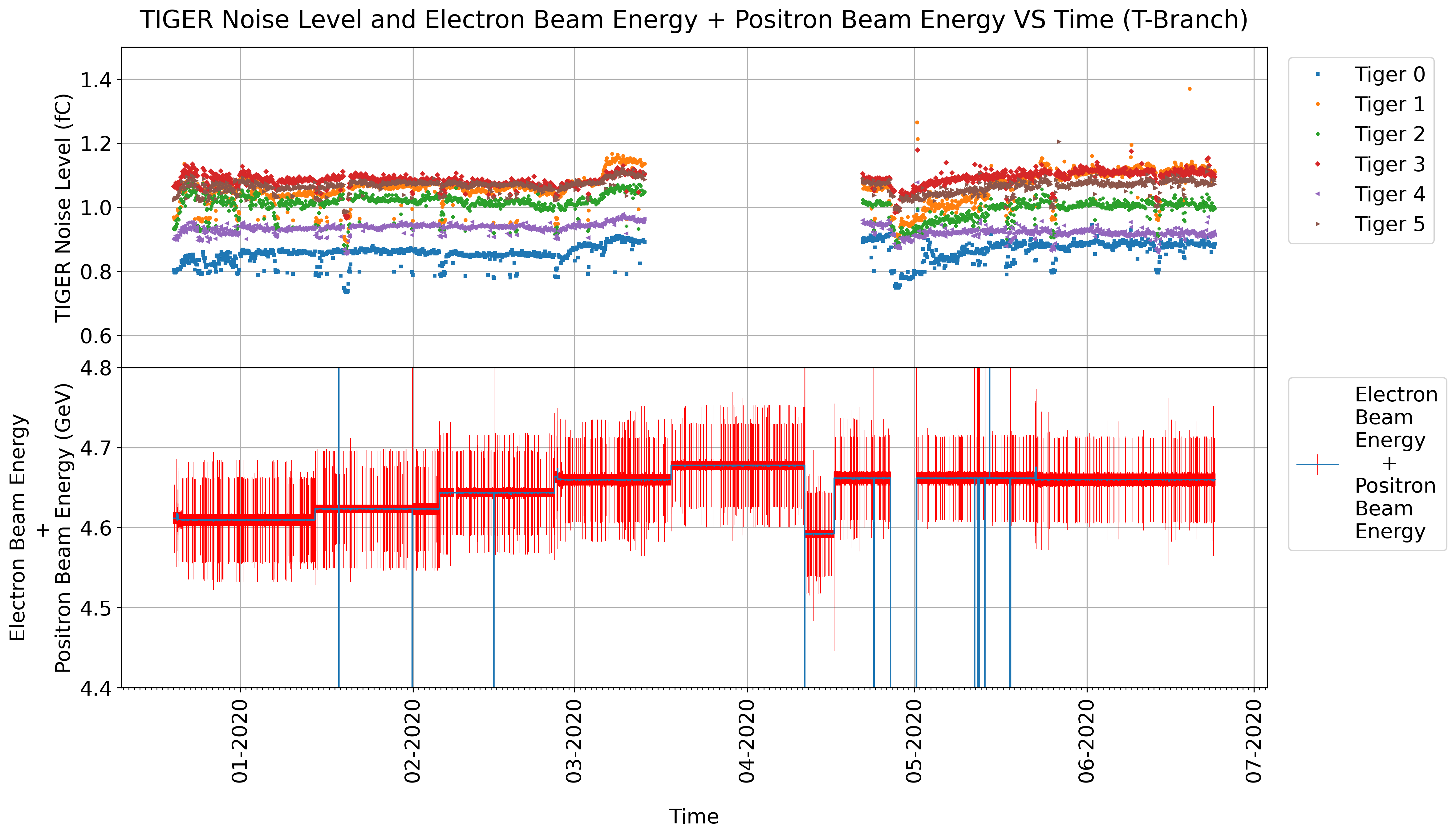}
	\vspace{1cm}
	\includegraphics[width=\textwidth, keepaspectratio]{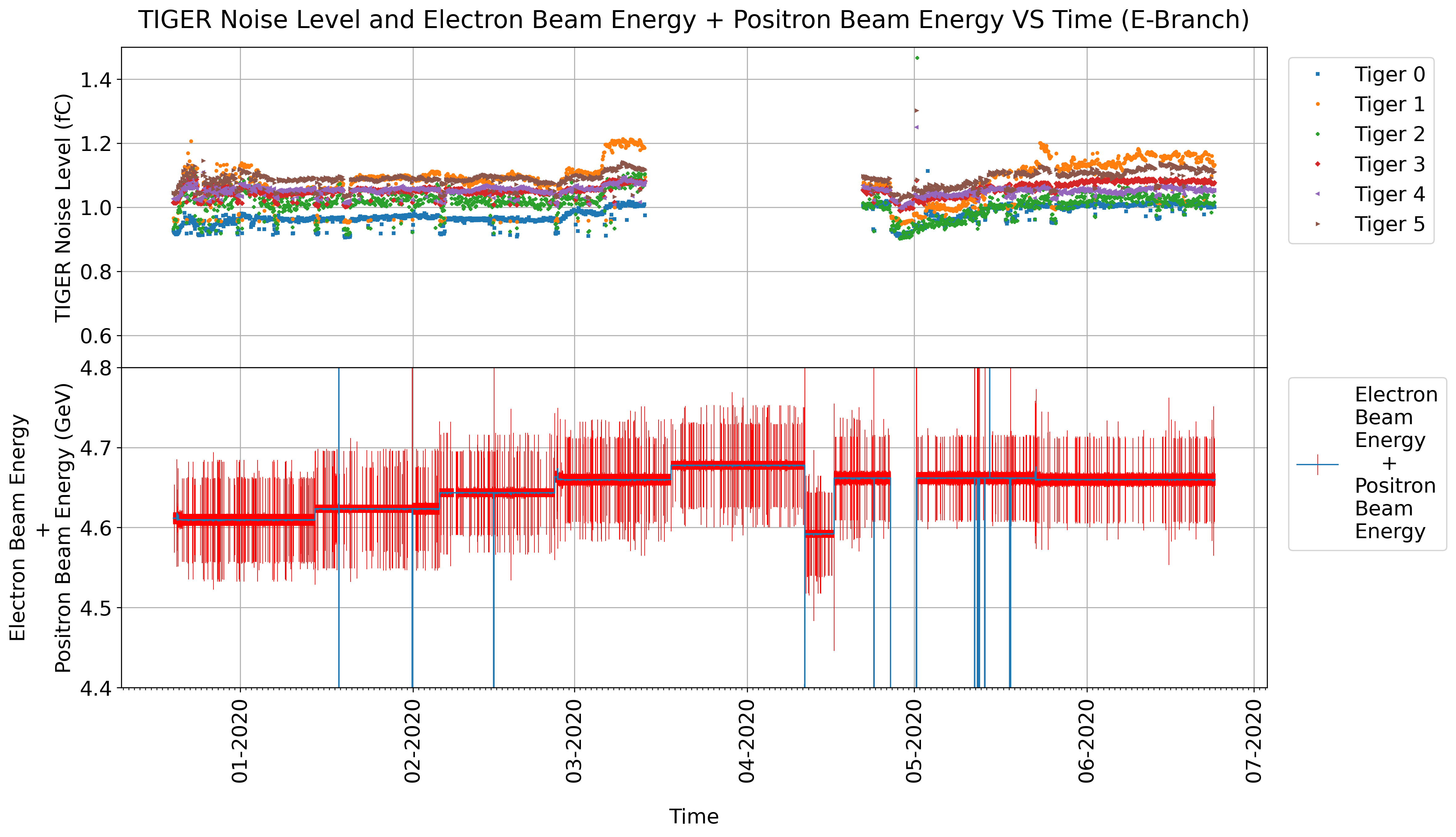}
	\caption[Comparison between the noise level of the TIGERs and the sum of the beam energies in the BESIII data taking phase.]{Comparison between the noise level of the TIGERs and the sum of the beam energies in the BESIII data taking phase. The data samples shown were collected between the 20/12/2019 and the 23/06/2020.}
	\label{daq2}
\end{figure}

\begin{figure}[h]
	\centering
	\includegraphics[width=\textwidth, keepaspectratio]{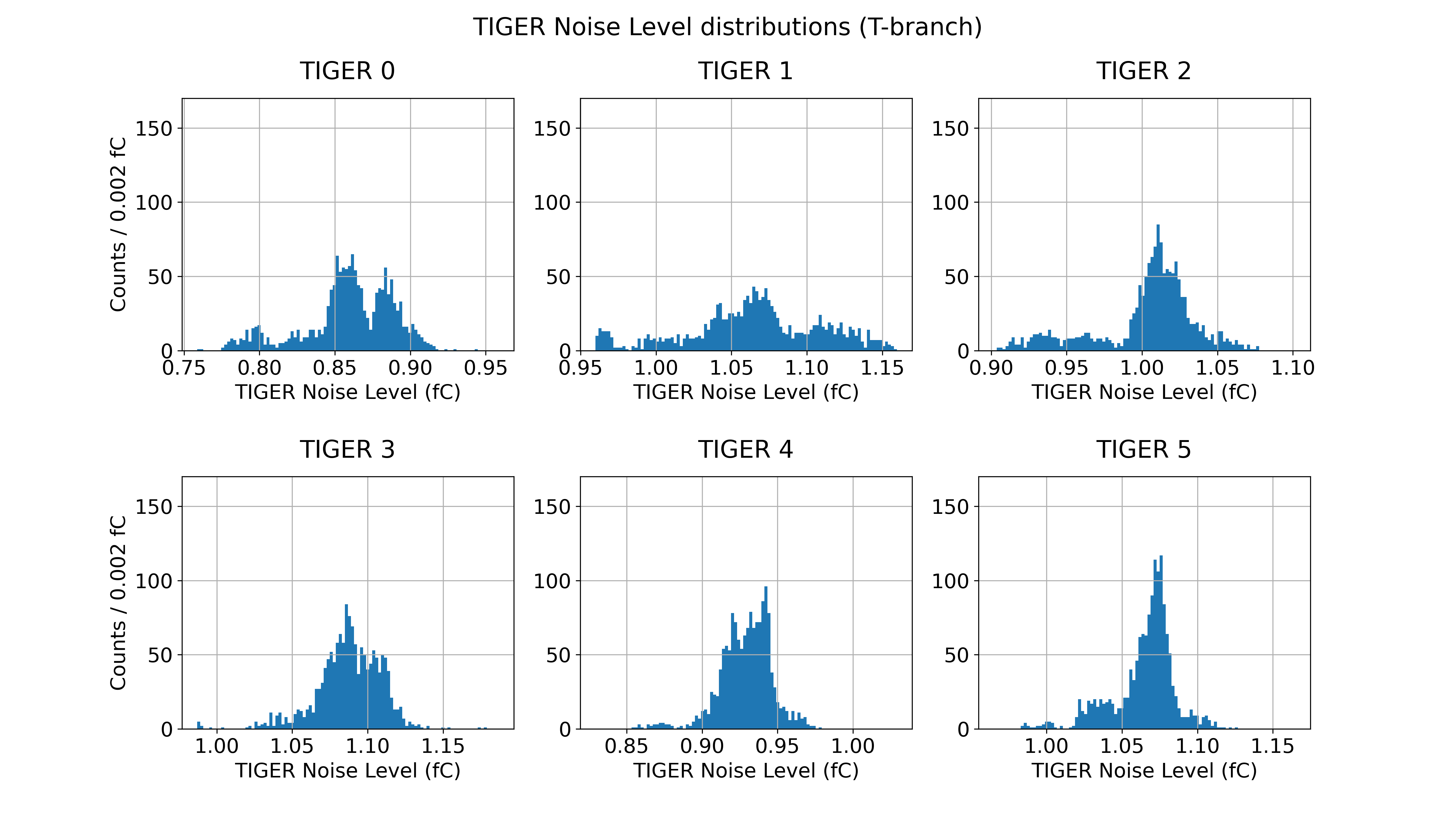}
	\vspace{1cm}
	\includegraphics[width=\textwidth, keepaspectratio]{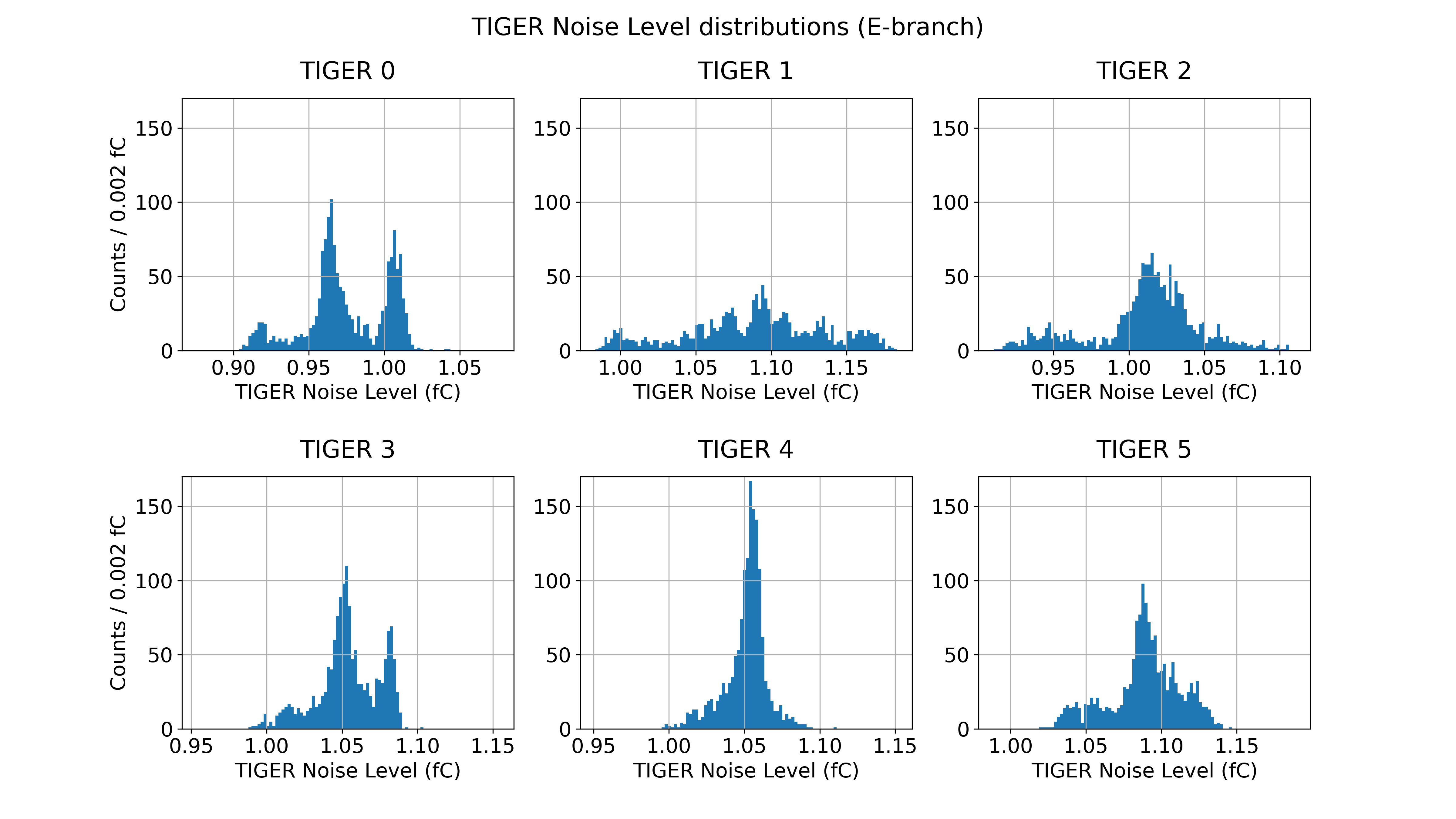}
	\caption[TIGER noise level distributions in the BESIII data taking phase.]{TIGER noise level distributions in the BESIII data taking phase. The range is 0.2$\,$fC.}
	\label{daq3}
\end{figure}

Figure \ref{daq1} shows a comparison between the noise level of the TIGERs and the sum of the two beam currents. The latter were summed assuming the relative contributions to the EM background would sum up according to the superposition principle.
In figure \ref{daq2} the data are instead compared to the sum of the beam energies, this may differ slightly from the center of mass energy as the readings are taken by instruments which are far from the collision point.
Figure \ref{daq3} shows distributions of the noise level of the TIGER for T and E branch, analogous to the ones presented for the SR1 phase.

Around the second week of March 2020, TIGER 1 registered a large increase in the noise level that is also matched by the other TIGERs, although to a much smaller degree. This variation is not stochastic, it lasts more than a week and it can be observed in both electronic branches. As it does not seem to be correlated to any of the monitored parameters, additional variables may be involved.

On the 26\textsuperscript{th} of April a beam injection failure caused the drop observable in both the current and the energy plots. In correspondence to this event, the setup registered a decrease in the overall level of noise. Other variations of these parameters do not seem to produce any relevant changes. This indicates that more than to the beam parameters studied, the drop in the noise may be related to some other factor, or factors, linked to the operation of the machines. The same considerations can be applied to the regions were the beam currents are building up and the stochastic oscillations in the noise level appear increased.

Other non stochastic variations visible in the plots present much shorter duration with respect to the two main events previously described.
The larger ones, like the one observed at the beginning of May, correspond to an increase of the noise level and involve a single data point.
The smaller ones, more frequent, cause instead a reduction in the noise level and may involve small groups of contiguous measurements.
While increases in the noise level localized to a single point also appear in the data collected during the SR2 phase, the smaller drops are unique to the BESIII-DT phase.

The frequency and duration of these smaller variations hints at a disappearance of an external contribution related to the data taking phase.
The larger isolated variations may instead be caused by communication errors between FEBs and GEMROCs. Being very rare and incompatible with the surrounding data, they were treated as systematic errors and, as such, they are excluded to the calculation of the error on the averages.

The sum of the beam energies, which varies in time according to the BESIII physics programme, shows a stepped behavior that is not replicated by the noise measurements. This points towards the lack of a strong correlation between the two quantities.

The noise level distributions are much more spread out than the ones related to the SR1 phase. This is due to both the presence of non stochastic variations and to a general increase in the amplitude of the stochastic oscillations. The time averaged noise levels for the BESIII data taking phase are reported in table \ref{daq4}.

\begin{table}[h]
	\centering
	\begin{tabular}{rS[table-format=1.3]S[table-format=1.3]}
		\multicolumn{3}{c}{\textbf{T Branch}}\\\hline	
		TIGER&\multicolumn{1}{r}{Noise level (fC)} &\multicolumn{1}{r}{Error(fC)}\\\hline
		0&0.86&0.09\\
		1&1.06&0.13\\
		2&1.00&0.10\\
		3&1.09&0.08\\	
		4&0.93&0.06\\
		5&1.06&0.06\\
		&&\\
		\multicolumn{3}{c}{\textbf{E Branch}}\\\hline
		TIGER&\multicolumn{1}{r}{Noise level (fC)} &\multicolumn{1}{r}{Error(fC)}\\\hline
		0&0.98&0.06\\
		1&1.08&0.13\\
		2&1.01&0.10\\
		3&1.05&0.05\\	
		4&1.05&0.04\\
		5&1.09&0.05\\
	\end{tabular}
	\caption[Time averaged results for the BESIII data taking phase.]{Time averaged noise level of the TIGERs in the BESIII data taking phase.}
	\label{daq4}
\end{table}

\FloatBarrier

\subsection{Second SR Production Phase}
\begin{figure}[h]
	\centering
	\includegraphics[width=\textwidth, keepaspectratio]{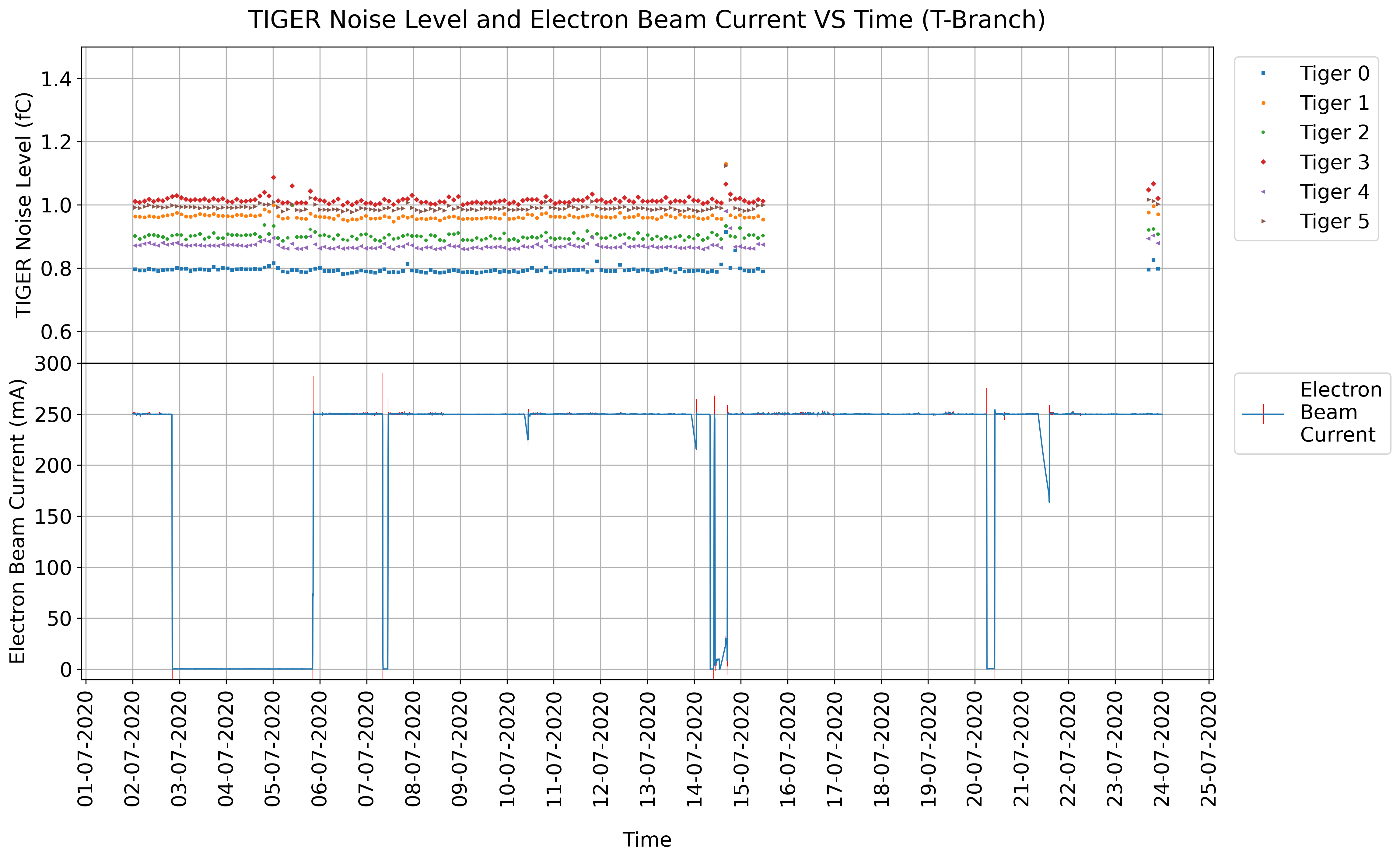}
	\vspace{1cm}
	\includegraphics[width=\textwidth, keepaspectratio]{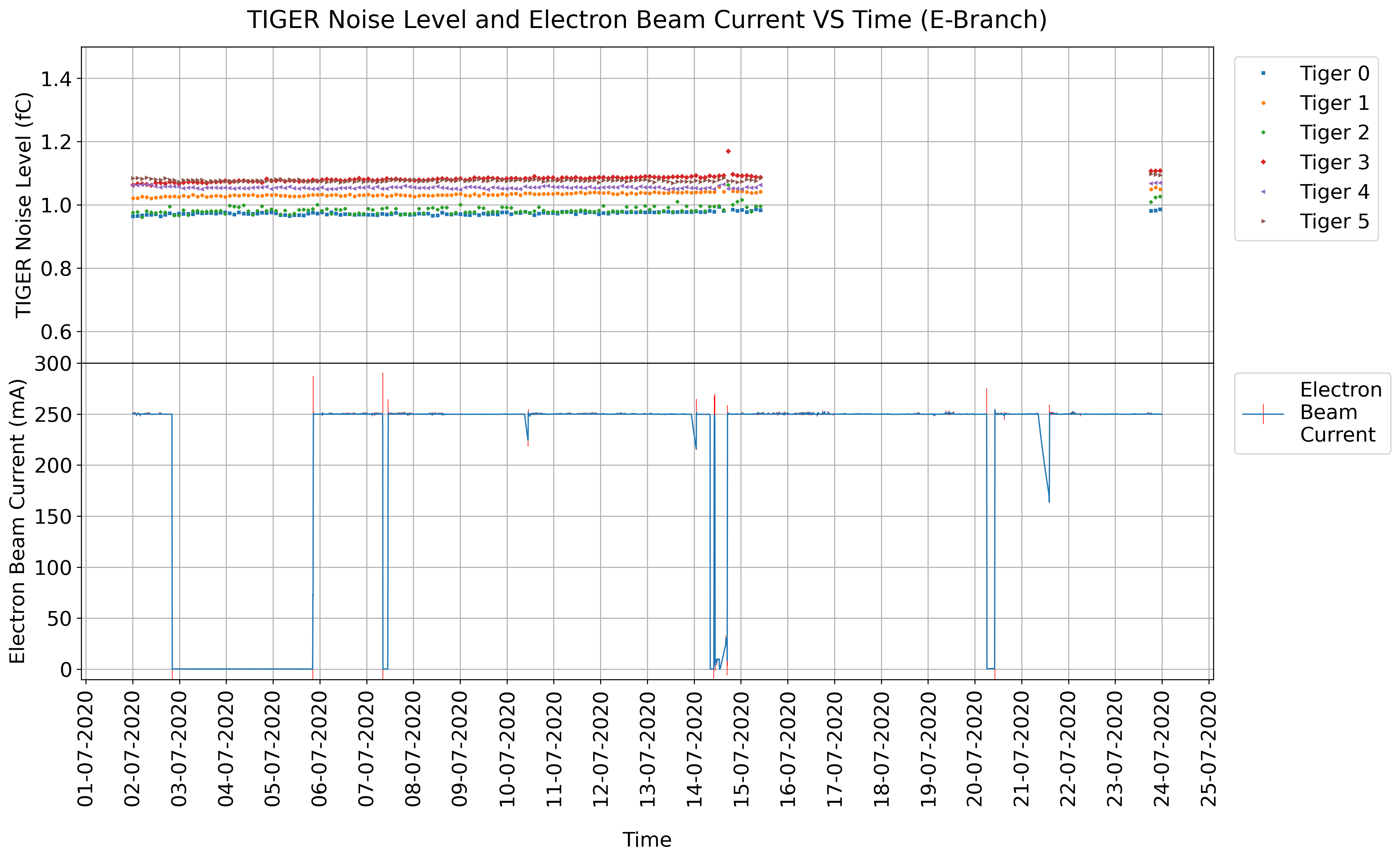}
	\caption[Comparison between the noise level of the TIGERs and the beam current in the second SR production phase.]{Comparison between the noise level of the TIGERs and the beam current in the second SR production phase. The data samples shown were collected between the 02/07/2020 and the 23/07/2020}
	\label{sr2_1}
\end{figure}

Figure \ref{sr2_1} shows a comparison between the noise level of the TIGERs and the electron beam current during the SR2 phase. The only relevant difference with the data taken during the SR1 phase is the appearance of sudden increases in the noise level limited to single measurements that are analogous to those observed in the BESIII-DT phase. These were treated as systematic errors in the same way.

\begin{figure}[h]
	\centering
	\includegraphics[width=\textwidth, keepaspectratio]{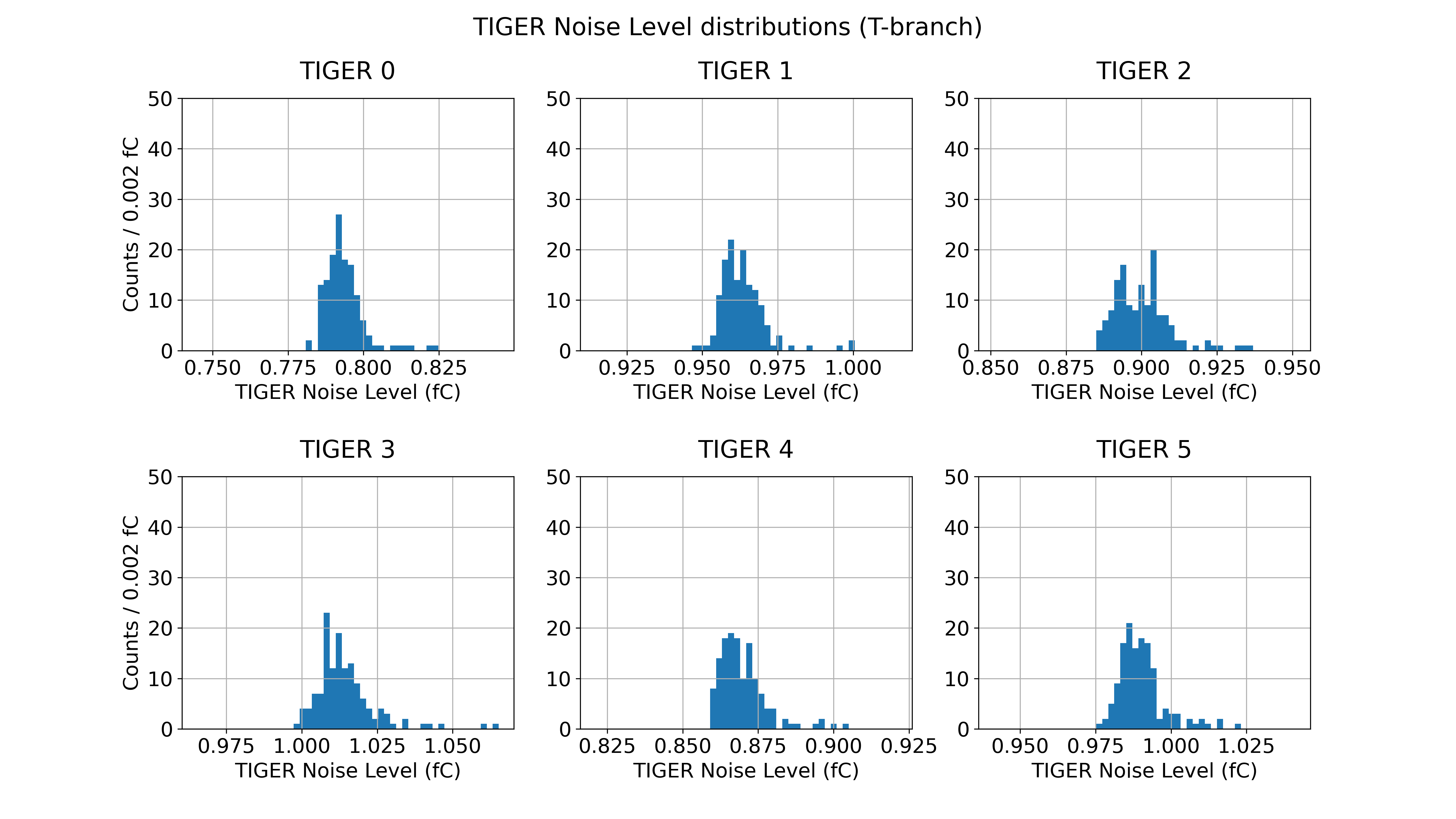}
	\vspace{1cm}
	\includegraphics[width=\textwidth, keepaspectratio]{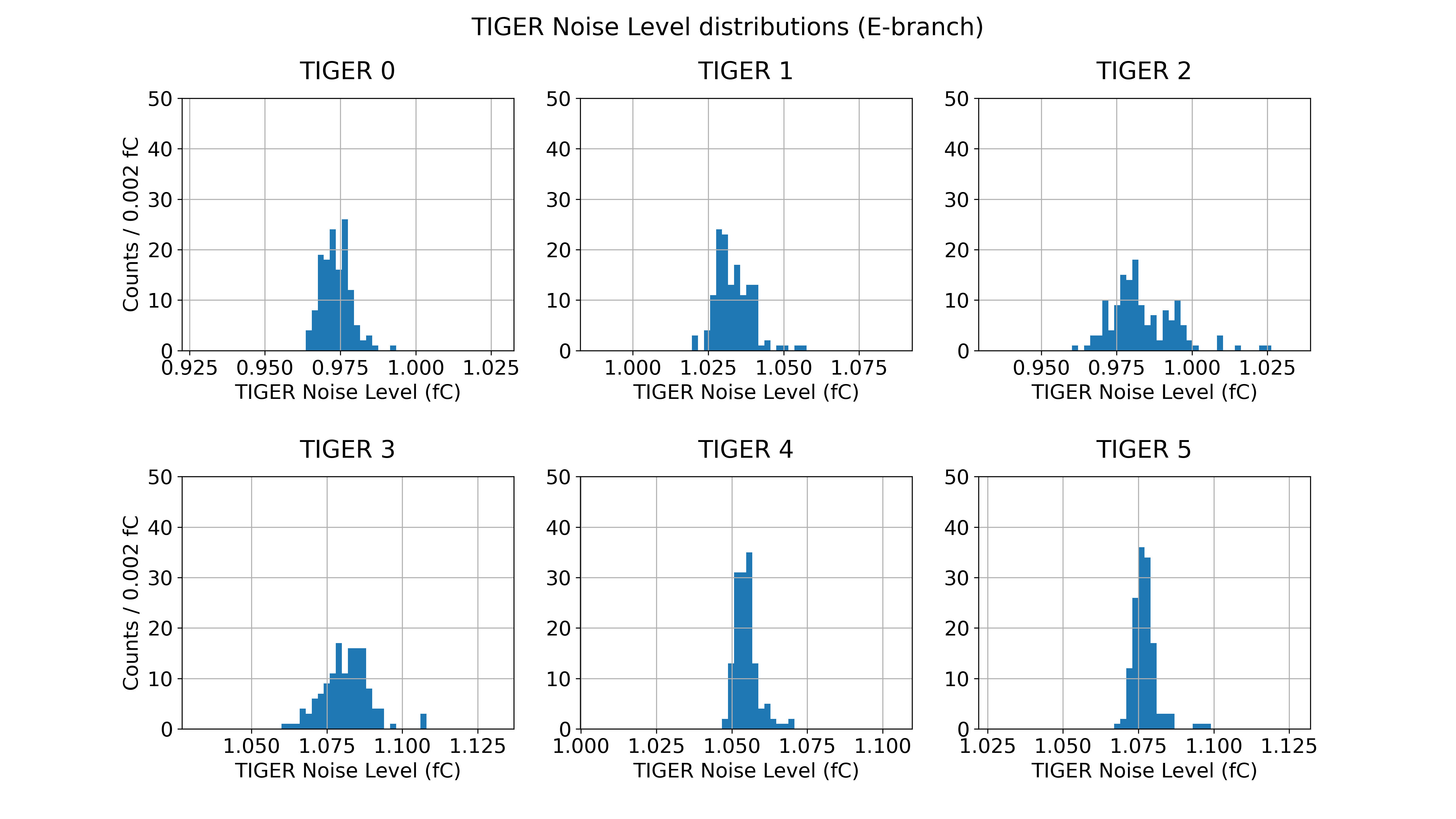}
	\caption[TIGER noise level distributions in the second SR production phase.]{TIGER noise level distributions in the second SR production phase. The range is 0.1$\,$fC.}
	\label{sr2_2}
\end{figure}

Figure \ref{sr2_2} shows the distributions of the noise level of the TIGERs during the second SR production phase while table \ref{sr2_3} reports the time averaged results.

\begin{table}[h]
	\centering
	\begin{tabular}{rS[table-format=1.3]S[table-format=1.3]}
		\multicolumn{3}{c}{\textbf{T Branch}}\\\hline	
		TIGER&\multicolumn{1}{r}{Noise level (fC)} &\multicolumn{1}{r}{Error(fC)}\\\hline
		0&0.795&0.014\\
		1&0.965&0.012\\
		2&0.90&0.02\\
		3&1.02&0.02\\	
		4&0.87&0.02\\
		5&0.99&0.02\\
		&&\\
		\multicolumn{3}{c}{\textbf{E Branch}}\\\hline
		TIGER&\multicolumn{1}{r}{Noise level (fC)} &\multicolumn{1}{r}{Error(fC)}\\\hline
		0&0.978&0.009\\
		1&1.038&0.010\\
		2&0.98&0.02\\
		3&1.082&0.013\\	
		4&1.055&0.007\\
		5&1.077&0.007\\
	\end{tabular}
	\caption[Time averaged results for the second SR production phase.]{Time averaged noise level of the TIGERs in the second SR production phase.}
	\label{sr2_3}
\end{table}

\FloatBarrier

\subsection{Summer Shutdown Phase} 
\begin{figure}[h]
	\centering
	\includegraphics[width=\textwidth, keepaspectratio]{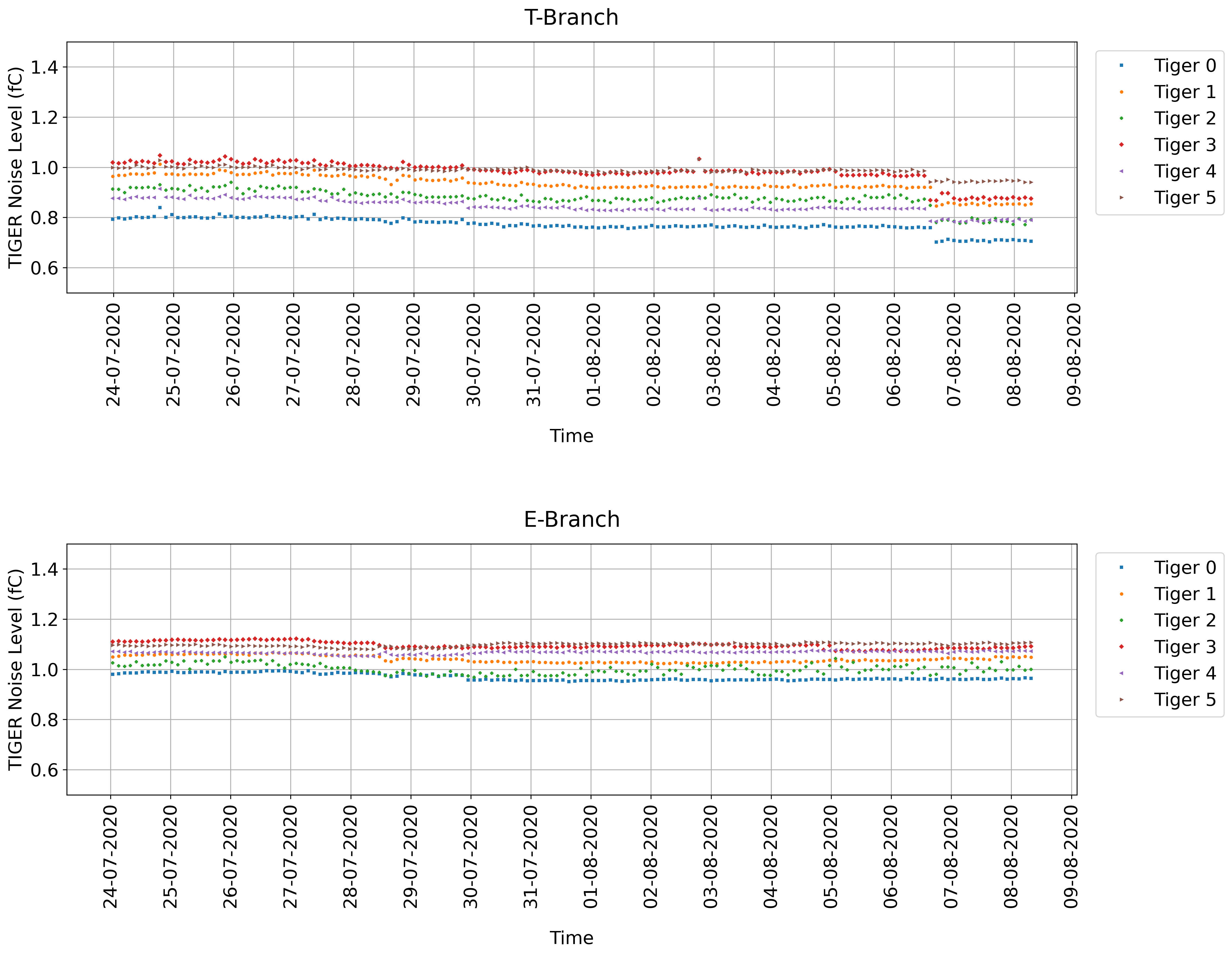}
	\caption[Noise level of the TIGERs in the summer 2020 shutdown.]{Noise level of the TIGERs during the shutdown of summer 2020. The data samples shown were collected between the 24/07/2020 and the 08/08/2020.}
	\label{shut1}
\end{figure}

\begin{figure}[h]
	\centering
	\includegraphics[width=\textwidth, keepaspectratio]{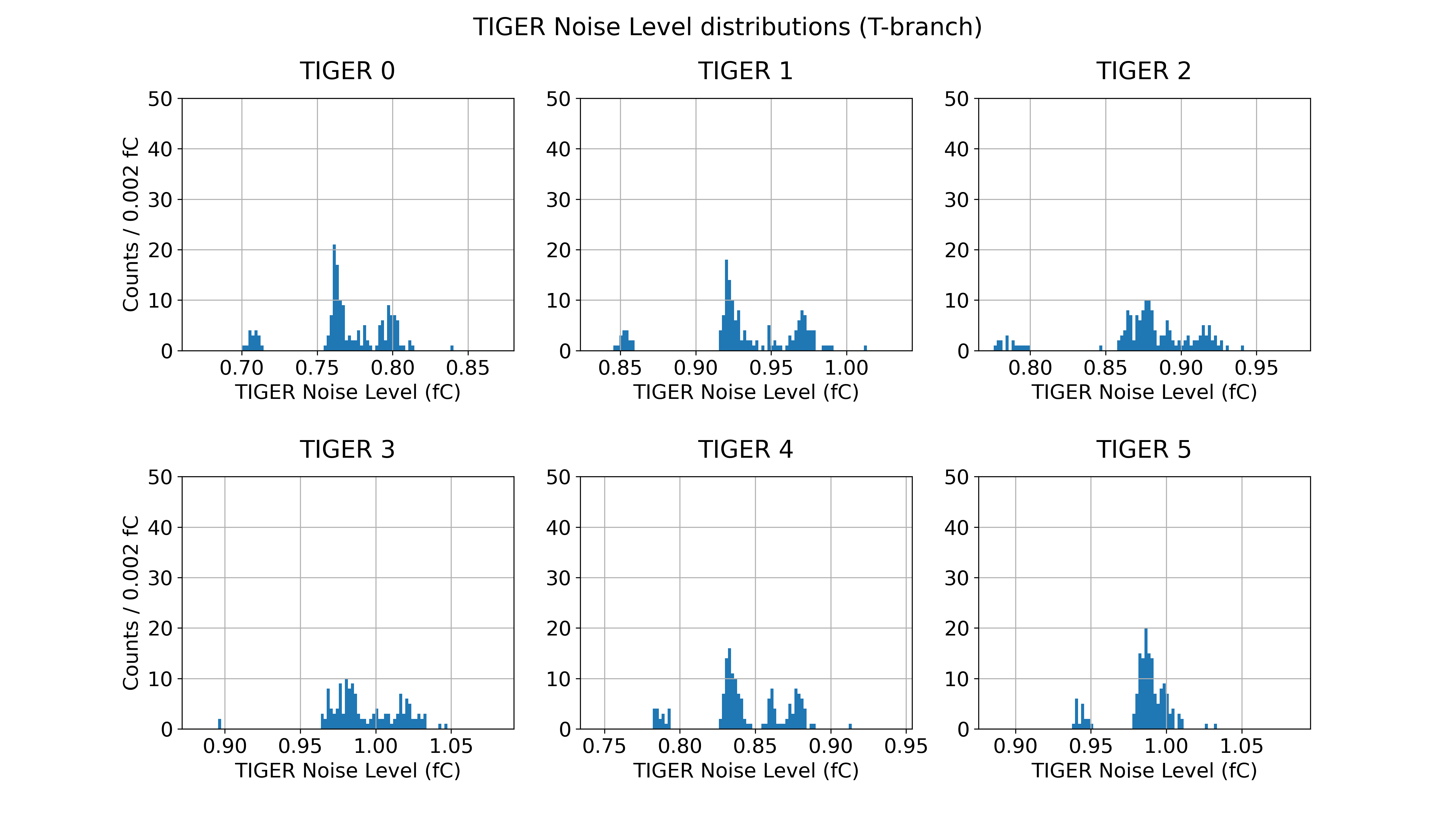}
	\vspace{1cm}
	\includegraphics[width=\textwidth, keepaspectratio]{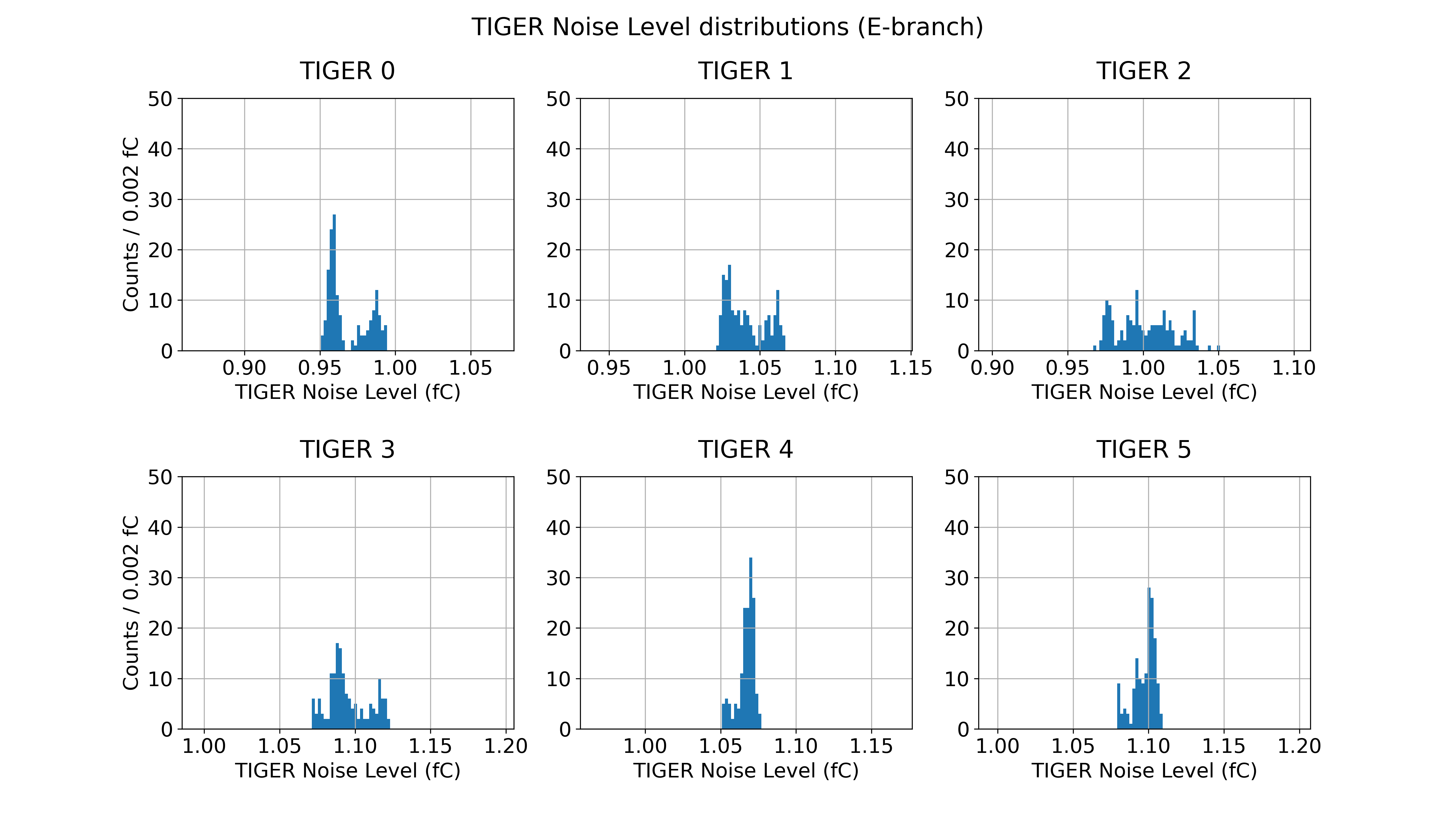}
	\caption[TIGER noise level distributions in the summer 2020 shutdown.]{TIGER noise level distributions in the summer 2020 shutdown. The range is 0.2$\,$fC.}
	\label{shut2}
\end{figure}

Figure \ref{shut1} shows the data collected as a reference for the other measurements during the 2020 summer shutdown. Measurements taken on the T branch in the last days before the failure of the computer controlling the acquisition show a clear decrease in the noise level that has no match on the E branch. As the phenomenon could not be investigated further the measurements taken in those days were excluded from both the distributions in figure \ref{shut2} and the averages in table \ref{shut3}.

The noise level seems to be progressively falling during the course of the shutdown, leading to wider distributions with multiple peaks. Either the EM background near the collision point is unstable even while the machines are not operating or the change is due to some other factor that was not considered. This behavior causes an increase of the error of the time averaged values, which are generally larger than those corresponding to measurements taken during the two SR phases.

\begin{table}[h]
	\centering
	\begin{tabular}{rS[table-format=1.3]S[table-format=1.3]}
		\multicolumn{3}{c}{\textbf{T Branch}}\\\hline	
		TIGER&\multicolumn{1}{r}{Noise level (fC)} &\multicolumn{1}{r}{Error(fC)}\\\hline
		0&0.78&0.02\\
		1&0.94&0.03\\
		2&0.89&0.03\\
		3&1.00&0.03\\	
		4&0.85&0.03\\
		5&0.991&0.014\\
		&&\\
		\multicolumn{3}{c}{\textbf{E Branch}}\\\hline
		TIGER&\multicolumn{1}{r}{Noise level (fC)} &\multicolumn{1}{r}{Error(fC)}\\\hline
		0&0.97&0.02\\
		1&1.04&0.02\\
		2&1.00&0.03\\
		3&1.10&0.02\\	
		4&1.066&0.010\\
		5&1.096&0.013\\
	\end{tabular}
	\caption[Time averaged results for the second SR production phase.]{Time averaged noise level of the TIGERs in the second SR production phase.}
	\label{shut3}
\end{table}

\FloatBarrier

\subsection{Secondary Variables}
\paragraph{Additional Beam Parameters}
Apart from beam currents and energies, the website of the BESIII slow control remote monitoring system \cite{paramsdata} provides a collection of other parameters related to the operation of BEPC-II. The ones that were investigated are the power supply settings of: the anti-solenoids used to compensate the BESIII SC solenoid; the SC quadrupoles; the SC skew quadrupoles; and the SC dipole correctors.
These parameters were chosen because they are related to components that are close to the setup but, as they showed little to no variation in the time frame spanned by the measurements, they added no relevant information to the analysis.

\paragraph{FEB Temperature}
Each FEB is equipped with a thermistor that allows to measure its temperature during operation. Due to the fact that these sensors were not calibrated, two measurements taken by the same FEB are only meaningful when compared with each other and not as absolute values. Moreover, the temperature readings taken by different FEBs are not comparable, as the values recorded may not have the same reference. Despite this considerations the temperatures were recorded to aid the debug of the setup in case any of the FEBs started to behave suspiciously.

\begin{figure}[h]
	\centering
	\includegraphics[width=\textwidth, keepaspectratio]{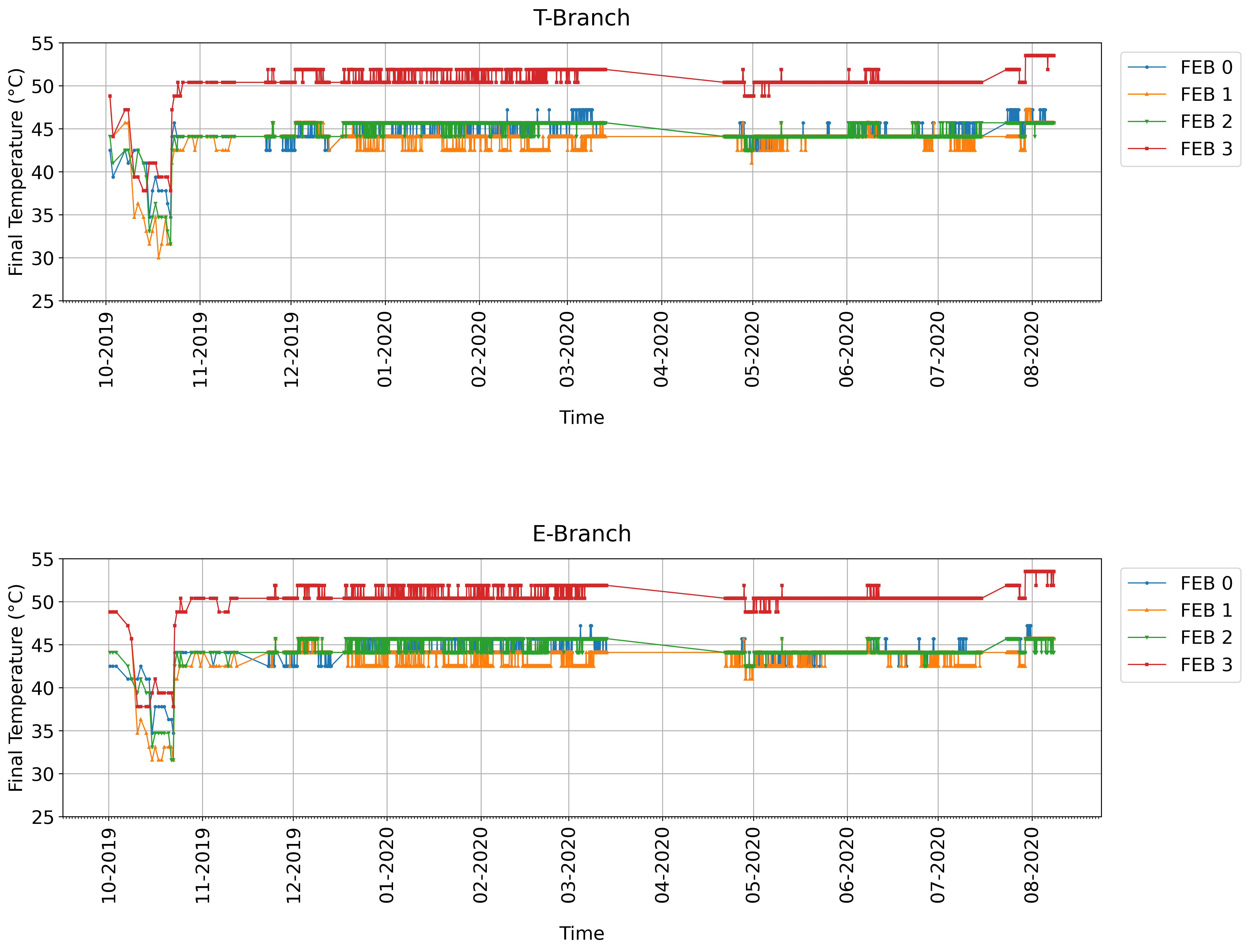}
	\caption[Temperature of the FEBs during the noise measurements]{Temperature of the FEBs during the noise measurements. The large variations observed for the first points are due to the fact that the readings were taken manually and consequently the FEBs had time to cool down.}
	\label{temperature}
\end{figure}

The continuous scanning script saves on a file the temperature of all four FEBs immediately before and after the scan of the 512 instrumented channels. Before the implementation of the script, the temperatures were recorded manually and, in the time required to register the values, the FEBs cooled down. This is the reason of the erratic variations observable at the beginning of the measurement period in the plot of figure \ref{temperature}. During operation in continuous scanning mode the FEBs do not have the time to cool down, so the initial and final values are close. In addition, the temperatures seemed to vary very little during the course of the data acquisition period, oscillating between values that are separated by the sensitivity of the thermistor. This sensitivity is substandard because the sensors are not meant to be precise thermometers but to monitor if the FEBs operate within a safe range of temperatures.

To investigate the behavior of the FEB temperature during a measurement cycle a simple test was performed.
The temperature of a FEB was recorded after the scan of progressively higher numbers of its channels.
It was observed that the temperature quickly reaches its final value after the scan of the first few channels and then remains stable until the end of the measurement.
It was consequently decided to adopt the final temperature, instead of the average of the two readings, as an estimate of the correct FEB temperature during the measurement cycle.

\paragraph{Environmental Temperature and Humidity}
Environmental variables, like humidity and temperature in the proximity of the detector, were not monitored. This is due to two interconnected reasons: the lack of equipment that could be remotely operated in the same way the setup was, at the time of installation, and the fact that the setup could not be accessed and upgraded while BEPC-II was operational. Equipment akin to the one that would have made this measurements possible is now being used to remotely monitor the cosmic ray telescope setup.

\paragraph{Ground Stability}
Another source of noise may have been the ground reference adopted, which was shared with other instrumentation present on site. This is not a problem because the grounding of the final detector will also be relying on the straps of the BESIII experiment and so the contribution due to an unstable reference potential is meant to be part of the test.

\section{Summary of the Results}
The time averaged noise levels obtained during the different BEPC-II operation phases are compared to the reference measurements taken during the summer shutdown to obtain the variations in table \ref{finalres}.
The errors on these variations are calculated by propagating the errors while considering independent measurements taken in different phases.

Most of the variations are compatible with 0 or very close to it when taking into account the related errors.
The largest variations are observed for the T branch during the BESIII data taking phase but these also are compatible with 0. This is due to the large errors caused by the spread of the distributions in both the SD and BESIII-DT phases.

The presence of negative values indicates that the contributions related to the operation of the accelerator and the experiment may not be the leading ones, against the assumption determining the choice of the reference measurements.

The data relative to the BESIII-DT and SD phases include also non stochastic contributions. In the former, frequent decreases of noise with similar magnitude and other behaviors related to the activity of BESIII or BEPC-II are observed. In the latter, the noise decreases progressively in time, despite the inactivity of the machines. 

These phenomena may be further investigated by including in the analysis new variables, for example the currents of the drift chamber, which depend from the particle background. The effects of environmental variables could be studied by adding more instrumentation to the setup. To study the effect of rapidly varying external conditions, it would be necessary to reduce the sample of channels scanned. This would narrow the interval between two scans of the same channel, making the setup more sensitive to short-duration phenomena.

\begin{table}[h]
	\centering		
		\begin{tabular}{r|S[table-format=2.3]S[table-format=2.3]|S[table-format=2.3]S[table-format=2.3]|S[table-format=2.3]S[table-format=2.3]}
		\multicolumn{7}{c}{\textbf{T Branch}}\\\hline	
		TIGER & \multicolumn{2}{c|}{$\Delta$SR1 \& Err (fC)} & \multicolumn{2}{c|}{$\Delta$SR2 \& Err (fC)} & \multicolumn{2}{c}{$\Delta$BESIII-DT \& Err (fC)}\\\hline
		0 & -0.03 & 0.02 & +0.02 & 0.03 & +0.08 & 0.09 \\
		1 & -0.04 & 0.03 & +0.02 & 0.03 & +0.12 & 0.14 \\
		2 & -0.02 & 0.03 & +0.01 & 0.04 & +0.11 & 0.11 \\
		3 & -0.02 & 0.03 & +0.02 & 0.04 & +0.09 & 0.09 \\
		4 & +0.02 & 0.03 & +0.02 & 0.03 & +0.08 & 0.06 \\
		5 & +0.01 & 0.02 & +0.00 & 0.02 & +0.07 & 0.06 \\
		\multicolumn{7}{c}{ }\\
		
		\multicolumn{7}{c}{\textbf{E Branch}}\\\hline
		TIGER & \multicolumn{2}{c|}{$\Delta$SR1 \& Err (fC)} & \multicolumn{2}{c|}{$\Delta$SR2 \& Err (fC)} & \multicolumn{2}{c}{$\Delta$BESIII-DT \& Err (fC)}\\\hline
		0 & -0.05 & 0.02 & +0.01  & 0.02  & +0.01 & 0.06 \\
		1 & -0.08 & 0.02 & -0.00  & 0.02  & +0.04 & 0.13 \\
		2 & -0.05 & 0.04 & -0.02  & 0.04  & +0.01 & 0.10 \\
		3 & -0.08 & 0.03 & -0.02  & 0.03  & -0.04 & 0.05 \\
		4 & +0.03 & 0.02 & -0.011 & 0.012 & -0.02 & 0.05 \\
		5 & +0.02 & 0.02 & -0.019 & 0.014 & -0.01 & 0.06
	\end{tabular}
	\caption[Variations of the noise level of the TIGERs with respect to the values obtained during the shutdown]{Variations of the noise level of the TIGERs with respect to the values obtained during the shutdown.}
	\label{finalres}
\end{table}
 	\pagestyle{empty}
 	\chapter*{Conclusions}
\addcontentsline{toc}{chapter}{Conclusions}
The performance decay of the inner tracker of the BESIII experiment led the BESIII collaboration to plan for its upgrade.
The Italian collaboration proposed the CGEM-IT, a new kind of GEM detector offering better rate capabilities and improved resistance to aging phenomena with respect to its wire-based counterparts.
The detector and its electronics are designed and built by INFN and University personnel for the Ferrara, LNF and Turin groups.

Thanks to the period spent at LNF and a three-months stay at IHEP, I was able to follow the path of the innermost layer of the CGEM-IT from the first moments of its construction to the operations that prepared it for being installed in a cosmic ray telescope setup, where it is currently collecting data.

While in Beijing, I performed the tests that allow to validate the quality of the front end boards before their installation on the detector. The experience acquired with the setup, the knowledge of the readout chain and the software I developed for the tests were fundamental when I joined the preparation of a setup for the collection of noise data inside the BESIII experiment hall.

These measurements aim to prove the robustness of the electronics and of the grounding scheme to the EM background present near the collision point of BEPC-II, where the CGEM-IT will be in operation. After the installation of the setup, I was responsible for the collection of the data and their analysis.

To study the contributions introduced by the activity of BESIII and BEPC-II, measurements were taken during different phases of the storage ring operation, which alternates periods of SR production and collider-mode running.

The analysis performed, based on the comparison between samples taken in the different BEPC-II operation modes and control samples collected during the summer shutdown, shows that the variation of the noise level is minimal or compatible with zero, with the largest variations measured remaining below 0.2$\,$fC. These values, although obtained through a comparative analysis, are about 20\% of the measured noise level, which is compatible with the design parameters.

The analysis does not show a strong correlation with the energy nor with the current of the beam but it hints at the possibility of a dependence from other variables related to the operation of either BESIII or BEPC-II that will be the subject of further investigations.

The test with a real CGEM detector in the experiment hall provides a platform for the integration of the CGEM-IT within the DAQ and Slow Control software of the BESIII experiment. While the setup is now incapable of remotely collecting data, due to the failure of the computer controlling the acquisition, more measurements are planned in the future to investigate the effects of rapidly varying beam parameters on a limited number of channels.

The setup in the experiment hall has now also been devoted to the first phases of the integration of the CGEM-IT with the DAQ and Slow Control software of the BESIII experiment. The same measurements presented in this thesis will also be repeated with this final configuration.
 	\appendix
 	\chapter{Complete Stratigraphy of the  CGEM-IT}
\label{stratos}
\begin{table}[!h]
	\centering
	\begin{tabular}{|l|l|l|r|}
		\multicolumn{4}{c}{\textbf{Layer 1}}                                                                            \\ \hline
		Sublayer                 &                                        & Material     & \multicolumn{1}{l|}{Thickness ($\upmu$m)} \\ \hline
		\multirow{9}{*}{Cathode} & \multirow{2}{*}{Faraday cage}          & Copper       & 3                              \\ \cline{3-4} 
		&                                        & Kapton       & 50                             \\ \cline{2-4} 
		& \multirow{5}{*}{Cylindrical Structure} & Epoxy        & 15                             \\ \cline{3-4} 
		&                                        & Honeycomb    & 1900                           \\ \cline{3-4} 
		&                                        & Epoxy        & 15                             \\ \cline{3-4} 
		&                                        & Kapton       & 25                             \\ \cline{3-4} 
		&                                        & Epoxy        & 15                             \\ \cline{2-4} 
		& \multirow{2}{*}{Cathodic plane}        & Kapton       & 50                             \\ \cline{3-4} 
		&                                        & Copper       & 3                              \\ \hline
		\multirow{3}{*}{GEM $\times$ 3}  & \multirow{3}{*}{GEM foil $\times$ 3}           & Copper       & 5                              \\ \cline{3-4} 
		&                                        & Kapton       & 50                             \\ \cline{3-4} 
		&                                        & Copper       & 5                              \\ \hline
		\multirow{14}{*}{Anode}  & \multirow{5}{*}{Anode readout plane}   & Copper       & 5                              \\ \cline{3-4} 
		&                                        & Kapton       & 50                             \\ \cline{3-4} 
		&                                        & Copper       & 5                              \\ \cline{3-4} 
		&                                        & Epoxy        & 25                             \\ \cline{3-4} 
		&                                        & Kapton       & 25                             \\ \cline{2-4} 
		& \multirow{7}{*}{Cylindrical structure} & Epoxy        & 15                             \\ \cline{3-4} 
		&                                        & Carbon fiber & 60                             \\ \cline{3-4} 
		&                                        & Epoxy        & 15                             \\ \cline{3-4} 
		&                                        & Honeycomb    & 3900                           \\ \cline{3-4} 
		&                                        & Epoxy        & 15                             \\ \cline{3-4} 
		&                                        & Carbon fiber & 60                             \\ \cline{3-4} 
		&                                        & Epoxy        & 15                             \\ \cline{2-4} 
		& \multirow{2}{*}{Ground plane}          & Copper       & 5                              \\ \cline{3-4} 
		&                                        & Kapton       & 50                             \\ \hline
	\end{tabular}        
	\caption[Complete stratigraphy of the first layer of the CGEM-IT.]{Complete stratigraphy of the first layer of the CGEM-IT.}
	\label{strato1}
\end{table}

\begin{table}[h]
	\centering
	\begin{tabular}{|l|l|l|r|}
		\multicolumn{4}{c}{\textbf{Layer 2}}                                                                         \\ \hline
		Sublayer                  &                                        & Material & \multicolumn{1}{l|}{Thickness ($\upmu$m)} \\ \hline
		\multirow{10}{*}{Cathode} & \multirow{8}{*}{Cylindrical structure} & Kapton   & 12,5                           \\ \cline{3-4} 
		&                                        & Epoxy    & 15                             \\ \cline{3-4} 
		&                                        & rohacel  & 1000                           \\ \cline{3-4} 
		&                                        & Epoxy    & 15                             \\ \cline{3-4} 
		&                                        & Kapton   & 12,5                           \\ \cline{3-4} 
		&                                        & Epoxy    & 15                             \\ \cline{3-4} 
		&                                        & rohacel  & 1000                           \\ \cline{3-4} 
		&                                        & Epoxy    & 15                             \\ \cline{2-4} 
		& \multirow{2}{*}{Cathodic plane}        & Kapton   & 50                             \\ \cline{3-4} 
		&                                        & Copper   & 5                              \\ \hline
		\multirow{3}{*}{GEM $\times$ 3}      & \multirow{3}{*}{GEM foil $\times$ 3}              & Copper   & 5                              \\ \cline{3-4} 
		&                                        & Kapton   & 50                             \\ \cline{3-4} 
		&                                        & Copper   & 5                              \\ \hline
		\multirow{14}{*}{Anode}   & \multirow{5}{*}{Anodic readout plane}  & Copper   & 5                              \\ \cline{3-4} 
		&                                        & Kapton   & 50                             \\ \cline{3-4} 
		&                                        & Copper   & 5                              \\ \cline{3-4} 
		&                                        & Epoxy    & 25                             \\ \cline{3-4} 
		&                                        & Kapton   & 25                             \\ \cline{2-4} 
		& \multirow{7}{*}{Cylindrical structure} & Epoxy    & 15                             \\ \cline{3-4} 
		&                                        & rohacel  & 2000                           \\ \cline{3-4} 
		&                                        & Epoxy    & 15                             \\ \cline{3-4} 
		&                                        & Kapton   & 12,5                           \\ \cline{3-4} 
		&                                        & Epoxy    & 15                             \\ \cline{3-4} 
		&                                        & rohacel  & 2000                           \\ \cline{3-4} 
		&                                        & Epoxy    & 15                             \\ \cline{2-4} 
		& \multirow{2}{*}{Ground plane}          & Copper   & 5                              \\ \cline{3-4} 
		&                                        & Kapton   & 50                             \\ \hline
	\end{tabular}
	\caption[Complete stratigraphy of the second layer of the CGEM-IT.]{Complete stratigraphy of the second layer of the CGEM-IT.}
	\label{strato2}
\end{table}

\begin{table}[h]
	\centering
	\begin{tabular}{|l|l|l|r|}
		\multicolumn{4}{c}{\textbf{Layer 3}}                                                                             \\ \hline
		Sublayer                  &                                        & Material     & \multicolumn{1}{l|}{Thickness ($\upmu$m)} \\ \hline
		\multirow{10}{*}{Cathode} & \multirow{8}{*}{Cylindrical structure} & Kapton       & 25                             \\ \cline{3-4} 
		&                                        & Epoxy        & 15                             \\ \cline{3-4} 
		&                                        & Carbon fiber & 60                             \\ \cline{3-4} 
		&                                        & Epoxy        & 15                             \\ \cline{3-4} 
		&                                        & Honeycomb    & 1900                           \\ \cline{3-4} 
		&                                        & Epoxy        & 15                             \\ \cline{3-4} 
		&                                        & Kapton       & 50                             \\ \cline{3-4} 
		&                                        & Epoxy        & 15                             \\ \cline{2-4} 
		& \multirow{2}{*}{Anodic readout plane}  & Kapton       & 50                             \\ \cline{3-4} 
		&                                        & Copper       & 3                              \\ \hline
		\multirow{3}{*}{GEM $\times$ 3}      & \multirow{3}{*}{GEM foil $\times$ 3}              & Copper       & 5                              \\ \cline{3-4} 
		&                                        & Kapton       & 50                             \\ \cline{3-4} 
		&                                        & Copper       & 5                              \\ \hline
		\multirow{16}{*}{Anode}   & \multirow{4}{*}{Anodic readout plane}  & Copper       & 5                              \\ \cline{3-4} 
		&                                        & Kapton       & 50                             \\ \cline{3-4} 
		&                                        & Copper       & 5                              \\ \cline{3-4} 
		&                                        & Epoxy        & 25                             \\ \cline{2-4} 
		& \multirow{8}{*}{Cylindrical structure} & Kapton       & 25                             \\ \cline{3-4} 
		&                                        & Epoxy        & 15                             \\ \cline{3-4} 
		&                                        & Kapton       & 125                            \\ \cline{3-4} 
		&                                        & Epoxy        & 15                             \\ \cline{3-4} 
		&                                        & Honeycomb    & 3900                           \\ \cline{3-4} 
		&                                        & Epoxy        & 15                             \\ \cline{3-4} 
		&                                        & Carbon fiber & 60                             \\ \cline{3-4} 
		&                                        & Epoxy        & 15                             \\ \cline{2-4} 
		& \multirow{2}{*}{Ground plane}          & Copper       & 5                              \\ \cline{3-4} 
		&                                        & Kapton       & 50                             \\ \cline{2-4} 
		& \multirow{2}{*}{Faraday cage}          & Kapton       & 50                             \\ \cline{3-4} 
		&                                        & Copper       & 5                              \\ \hline
	\end{tabular}
	\caption[Complete stratigraphy of the third layer of the CGEM-IT.]{Complete stratigraphy of the third layer of the CGEM-IT.}
	\label{strato3}
\end{table}

	\bibliographystyle{ieeetr}
	\bibliography{biblio}

\begin{thebibliography}{10}

\bibitem{bessite}
``Website of the {BESIII} experiment.''
  \url{http://bes3.ihep.ac.cn/index.html}.

\bibitem{record}
``Article on phys.org.''
  \url{https://phys.org/news/2016-04-bepcii-luminosity-world-11033cm2s.html}.

\bibitem{zcref}
{M. Ablikim \textit{et al.} (BESIII Collaboration)}, ``Observation of a charged
  charmoniumlike structure in
  ${e}^{\mathbf{+}}{e}^{\mathbf{\ensuremath{-}}}\ensuremath{\rightarrow}{\ensuremath{\pi}}^{\mathbf{+}}{\ensuremath{\pi}}^{\mathbf{\ensuremath{-}}}j/\ensuremath{\psi}$
  at $\sqrt{s}\mathbf{=}4.26\text{ }\text{ }\mathrm{GeV}$,'' {\em Phys. Rev.
  Lett.}, vol.~110, p.~252001, 2013.

\bibitem{Sauli_1997}
F.~Sauli, ``{GEM}: A new concept for electron amplification in gas detectors,''
  {\em Nucl. Instrum. Meth. A}, vol.~386, pp.~531--534, 1997.

\bibitem{BEPC-II_2009}
{C. ZHANG for BEPC-II Team}, ``{BEPC II}: construction and commissioning,''
  {\em Chinese Physics C}, vol.~33, no.~S2, pp.~60--64, 2009.

\bibitem{BESIII_2020}
{M. Ablikim \textit{et al.} (BESIII Collaboration)}, ``Future physics programme
  of {BESIII},'' {\em Chinese Physics C}, vol.~44, no.~4, 2020.

\bibitem{BESIII_2009}
{M. Ablikim \textit{et al.} (BESIII Collaboration)}, ``Design and construction
  of the {BESIII} detector,'' {\em Nucl. Instrum. Meth. A}, vol.~614,
  pp.~345--399, 2010.

\bibitem{mdcit}
{BESIII Collaboration}, ``The construction of the {BESIII} experiment,'' {\em
  Nucl. Instrum. Meth. A}, vol.~598, no.~1, pp.~7--11, 2009.

\bibitem{note}
Internal note.

\bibitem{Aging}
{DONG M.Y. \textit{et al.}}, ``Aging effect in the {BESIII} drift chamber,''
  {\em Chinese Physics C}, vol.~40, no.~1, 2016.

\bibitem{BES30}
{BESIII Collaboration}, {\em {30 Years of BES Physics, Proceedings of the
  Symposium}}, pp.~169--175.
\newblock World Scientific, 2019.

\bibitem{giulio}
G.~Mezzadri, {\em Search for hidden-strangeness pentaquark in
  $\mathrm{\lambda_c \, \rightarrow \, p \phi \pi^0}$ at BESIII}.
\newblock PhD thesis, University of Ferrara, 2017.

\bibitem{kloetdr}
{De Robertis G. \textit{et al.} (KLOE-2 Collaboration)}, ``Technical design
  report of the inner tracker for the {KLOE-2} experiment,'' 2010.

\bibitem{Sauli_2016}
F.~Sauli, ``The gas electron multiplier {(GEM)}: Operating principles and
  applications,'' {\em Nucl. Instrum. Meth. A}, vol.~805, no.~Supplement C,
  pp.~2--24, 2016.

\bibitem{riccardo}
R.~Farinelli, {\em Research and development in cylindrical triple-GEM detector
  with $\mathrm{\mu}$TPC readout for the BESIII experiment}.
\newblock PhD thesis, University of Ferrara, 2018.

\bibitem{mecreview}
I.~Balossino, ``Investigation and improvements of the mechanical structure of
  cylindrical {GEMs} for the {BESIII} experiment,'' {\em Journal of
  Instrumentation}, vol.~15, no.~08, pp.~C08013--C08013, 2020.

\bibitem{cdr}
``{CGEM-IT} conceptual design report.''
  \url{http://www.lnf.infn.it/esperimenti/bes3/bes3\_tdr\_cgem.pdf}.

\bibitem{tigerref}
A.~Rivetti {\em et~al.}, ``Tiger: A front-end asic for timing and energy
  measurements with radiation detectors,'' {\em Nucl. Instrum. Meth. A},
  vol.~924, pp.~181--186, 2019.

\bibitem{fabio}
F.~Cossio, {\em A mixed-signal ASIC for time and charge measurements with GEM
  detectors}.
\newblock PhD thesis, PoliTO, 2019.

\bibitem{cpad}
G.~Mezzadri, ``{The CGEM-IT of the BESIII experiment: preliminary results of
  the cosmic data taking}.'' {CPAD} 2021 Instrumentation Frontier Workshop,
  \url{https://indico.fnal.gov/event/46746/contributions/210384/}, March 2021.

\bibitem{Farinelli_2020}
R.~Farinelli {\em et~al.}, ``Preliminary results from the cosmic data taking of
  the besiii cylindrical gem detectors,'' {\em Journal of Instrumentation},
  vol.~15, no.~08, p.~C08004–C08004, 2020.

\bibitem{paramsdata}
``Web page of the {BESIII} slow control remote monitoring system.''
  \url{http://bes3db.ihep.ac.cn/SlowWeb/TableList.php?MainSys=BEPCII&SubSys=BEPCII_Major}.

\end{thebibliography}
	\chapter*{Special Thanks}
I would like to use these last pages to say thanks to all those who helped me in the writing of this thesis and in the work that led to it, and to those who were close to me during these university years.

First of all, I would like to thank my advisor, Dr. Gianluigi Cibinetto, for providing me the opportunity to experience the reality of scientific research and for guiding me over the course of my studies.

A special thanks to my co-advisor, Dr. Ilaria Balossino, who was my reference point when I was in China. It is thanks to her punctuality and organizational skills that I was able to complete this thesis within the deadline, while also being satisfied with the quality of the work done.

Thanks to Dr. Giulio Mezzadri, always willing to provide help and advice and to spend himself beyond duty.

Thanks to Eng. Michele Melchiorri with whom I have had the pleasure to work with and to discuss about the design of the detector and other topics as well, I learned something from each of these conversations.

Thanks to the remaining members of the BESIII group in Ferrara: Dr. Isabella Garzia, Dr. Diego Bettoni, Dr. Riccardo Farinelli and Dr. Marco Scodeggio, for always being willing to answer my questions and to put at my disposal their skills and knowledge.

In the weeks I spent in Frascati I had the opportunity to work alongside the researchers and technicians involved in the construction of the detector: Dr. Monica Bertani, Dr. Jing Dong, Dr. Alessandro Calcaterra and Dr. Emiliano Paoletti. Working with these people was an extremely educational and rewarding experience but above all a real pleasure. From the first day I have felt welcome and I have always been treated with care and respect despite my inexperience. A special thought goes to Dr. Stefano Cerioni, who taught me so much about the work of detector construction, which he faced with extreme passion and competence.

In China I met the researchers of the INFN section of Turin, who developed the front end electronics and the GUFI acquisition software I used to collect the noise data: Dr. Michela Greco, Dr. Alberto Bortone and Dr. Fabio Cossio. I wish to thank them for answering my many doubts about electronics and for adapting the acquisition system to my needs, allowing me to collect the data much more efficiently.

I want to thank my girlfriend Fahimeh, who was close to me and supported me even when I was far away but above all for putting up with me in the last months of desperate work.

Thanks to my parents for always supporting me and being the people I could always rely on in difficult times.

Thanks to my grandmother, to my aunts and uncles, for always taking an interest and for constantly encouraging me throughout my studies.

Finally, I would like to thank my friends: those from childhood, with whom I grew up, and those I met during my university years, with whom I shared the experiences of this journey.

\begin{otherlanguage}{italian}
\chapter*{Ringraziamenti}
Vorrei impiegare queste ultime pagine per ringraziare tutti coloro che mi hanno aiutato nella stesura di questa tesi e nel lavoro che ha portato alla sua scrittura, o che mi sono stati accanto in questi anni di università.

Anzitutto vorrei ringraziare il mio relatore, il Dott. Gianluigi Cibinetto, per avermi dato l'opportunità di vivere una vera e appassionante esperienza di ricerca e per avermi più volte consigliato e indirizzato nell'arco del mio percorso universitario.

Un ringraziamento speciale alla mia correlatrice, la Dott.ssa Ilaria Balossino, che è stata il mio punto di riferimento nei mesi passati in Cina e grazie alla cui puntualità e capacità organizzative ho potuto completare questa tesi entro le scadenze previste, rimanendo soddisfatto della qualità del lavoro svolto.

Un ringraziamento al Dott. Giulio Mezzadri, sempre disponibile a fornire aiuto e consiglio e a spendersi anche più del dovuto.

Un ringraziamento all'Ing. Michele Melchiorri con cui ho avuto più volte il piacere di lavorare e discutere del disegno del rivelatore e tanto altro ancora, da ciascuna di queste chiacchierate ho imparato qualcosa.

Grazie ai restanti membri del gruppo di BESIII di Ferrara: la Dott.ssa Isabella Garzia, il Dott. Diego Bettoni, il Dott. Riccardo Farinelli e il Dott. Marco Scodeggio, per essere sempre stati disponibili a rispondere alle mie domande e a mettere a mia disposizione le loro conoscenze.

Nelle settimane passate a Frascati ho avuto occasione di lavorare a fianco dei ricercatori e dei tecnici che si occupano della costruzione del rivelatore: la Dott.ssa Monica Bertani, la Dott.ssa Jing Dong, il Dott. Alessandro Calcaterra e il Dott. Emiliano Paoletti. Lavorare con queste persone è stata un'esperienza estremamente educativa e appagante ma soprattutto un vero piacere. Sin dal primo giorno mi sono sentito il benvenuto e sono sempre stato trattato con attenzione e rispetto nonostante la mia inesperienza. Un pensiero particolare va al Dott. Stefano Cerioni, che tanto mi ha insegnato sul lavoro di costruzione, che affrontava con estrema passione e competenza.

In Cina ho conosciuto i ricercatori della sezione INFN di Torino, i quali hanno sviluppato l'elettronica di front end e il software di acquisizione GUFI, che ho usato per la raccolta dei dati di rumore: la Dott.ssa Michela Greco, il Dott. Alberto Bortone e il Dott. Fabio Cossio. A loro va il mio ringraziamento per aver dato risposta ai miei tanti dubbi sull'elettronica e per aver adattato il sistema di acquisizione alle mie necessità, consentendomi di raccogliere dati in maniera più efficiente.

Voglio ringraziare la mia ragazza Fahimeh, che mi è stata vicino e che mi ha supportato anche quando ero lontano ma soprattutto per avermi sopportato negli ultimi mesi di disperato lavoro.

Un ringraziamento ai miei genitori per non avermi mai fatto mancare il loro sostegno ed essermi stati vicino nei momenti difficili.

Un grazie a mia nonna, alle zie e agli zii, per essersi sempre interessati ai miei studi e avermi costantemente incoraggiato.

Infine vorrei ringraziare gli amici: quelli di infanzia, con cui sono cresciuto, e quelli incontrati negli anni di università, con i quali ho condiviso le esperienze di questo percorso.
\end{otherlanguage}

\end{document}